%% file: main.tex
\documentclass[11pt,a4paper,listof=totoc,bibliography=totoc,openany]{scrbook} 

\usepackage{geometry}

\usepackage{amssymb,stmaryrd,mathrsfs,dsfont,mathtools}
\usepackage{afterpage}
\usepackage{epigraph}
\usepackage{amsthm}
\usepackage{soul}
\setul{0.5ex}{0.3ex}

\usepackage{thmtools, thm-restate}
\usepackage[dvipsnames]{xcolor} 			
\usepackage[hyphens]{url}
\usepackage[Conny]{fncychap} 

\usepackage[useregional]{datetime2}
\usepackage{lipsum}
\usepackage{xspace}
\usepackage{physics}
\usepackage{enumitem}
\usepackage{awesomebox}
\usepackage{tabu}
\usepackage{tablefootnote}
\usepackage{tabularx}
\usepackage{dcolumn}
\usepackage{xltabular}
\usepackage{colortbl}
\usepackage{wrapfig}
\usepackage[framemethod=TikZ]{mdframed}

\usepackage[export]{adjustbox}

\usepackage{graphics}
     \graphicspath{{fig/}}
\usepackage[style=altlist,acronym]{glossaries}

\usepackage{pgf}

\input{styling/tikz.tikzstyles}
\usepackage{tikz}
\usepackage{lib/tikzit}
\usepackage{lastpage}

\usetikzlibrary{positioning,chains,fit,shapes,calc,er,automata}

 \def\hmath$#1${\texorpdfstring{{\rmfamily\textit{#1}}}{#1}}
\ChTitleVar{\bfseries\Large\rmfamily}
\ChNameVar{\bfseries\Large\sffamily}
\definecolor{applegreen}{rgb}{0.55, 0.71, 0.0}
\definecolor{aqua}{rgb}{0.0, 1.0, 1.0}

\input xy
\xyoption{all}

\setacronymstyle{long-short-desc}

\input{settings/settings}

\usepackage{nameref}
\usepackage[capitalize,nameinlink]{cleveref}

\input{settings/commands}

\begin{document}


\frontmatter
\input{pages/layout/cover}
\input{pages/layout/title}

\input{pages/layout/disclaimer}
\input{pages/layout/acknowledgments}
\tableofcontents
\input{pages/util/abstract}

\mainmatter
    \input{pages/introduction/intro}

    \input{pages/theory/theory}
    \input{pages/content/whardness} 
    \input{pages/content/planarsds}
    \input{pages/content/outlook/open-q-further-research}

\backmatter
  \microtypesetup{protrusion=false}
  \printbibliography
   \listoffigures{}
 \listoftables{}
  \microtypesetup{protrusion=true}
\appendix

\end{document}

%% file: styling/tikz.tikzstyles
\tikzstyle{BLACK}=[draw=black, shape=circle, fill=black, inner sep=3pt]
\tikzstyle{DOM}=[fill={rgb,255: red,122; green,0; blue,42}, draw=black, shape=circle, inner sep=3pt]
\tikzstyle{NONE}=[fill={rgb,255: red,247; green,33; blue,212}, draw=black, shape=circle, inner sep=3pt]
\tikzstyle{NTWO}=[fill={rgb,255: red,0; green,177; blue,219}, draw=black, shape=circle, inner sep=3pt]
\tikzstyle{NTHR}=[fill={rgb,255: red,140; green,143; blue,14}, draw=black, shape=circle, inner sep=3pt]
\tikzstyle{GRAYN}=[fill={rgb,255: red,128; green,128; blue,128}, draw=black, shape=circle, inner sep=3pt]
\tikzstyle{BRGRAY}=[fill={rgb,255: red,236; green,236; blue,236}, draw=none, shape=circle, inner sep=3pt]
\tikzstyle{TEXTSTD}=[fill=white, draw=black, shape=rectangle, tikzit shape=rectangle, text width=3cm, rounded corners]
\tikzstyle{TEXTHIGH}=[fill={rgb,255: red,231; green,245; blue,255}, draw=black, shape=rectangle, tikzit shape=rectangle, text width=3cm, rounded corners]
\tikzstyle{TEXTHIGHGRAY}=[fill={rgb,255: red,220; green,220; blue,220}, draw=black, shape=rectangle, tikzit shape=rectangle, text width=3cm, rounded corners]
\tikzstyle{TEXTHIGHGREEN}=[fill={rgb,255: red,149; green,242; blue,145}, draw=black, shape=rectangle, tikzit shape=rectangle, text width=3.4cm, rounded corners]
\tikzstyle{TEXTHIGHYELLOW}=[fill={rgb,255: red,250; green,250; blue,160}, draw=black, shape=rectangle, tikzit shape=rectangle, text width=3cm, rounded corners]

\tikzstyle{EDGE}=[-, fill=none]
\tikzstyle{BLUE}=[-, draw={rgb,255: red,0; green,101; blue,189}]
\tikzstyle{GRAY}=[-, fill={rgb,255: red,128; green,128; blue,128}]
\tikzstyle{DARKGREEN}=[-, fill=none, draw={rgb,255: red,19; green,100; blue,13}, line width=1.25pt]
\tikzstyle{big dash}=[-, dashed, dash pattern=on 4mm off 2mm, fill={rgb,255: red,178; green,255; blue,253}]
\tikzstyle{big dash thick}=[-, thick, dashed, dash pattern=on 4mm off 2mm, fill={rgb,255: red,178; green,255; blue,253}]
\tikzstyle{new edge style 0}=[-, fill={rgb,255: red,224; green,254; blue,255}]
\tikzstyle{FILLGREEN}=[-, fill={rgb,255: red,232; green,255; blue,207}]
\tikzstyle{FILLBLUE}=[-, fill={rgb,255: red,231; green,255; blue,255}]
\tikzstyle{FILLPURPLE}=[-, fill={rgb,255: red,255; green,233; blue,239}]
\tikzstyle{FILLDARKBLUE}=[-, fill={rgb,255: red,238; green,230; blue,255}]
\tikzstyle{FILLDARKBLUEINVISIBLE}=[-, draw=none, fill={rgb,255: red,238; green,230; blue,255}]
\tikzstyle{FILLGREENINVISBLE}=[-, draw=none, fill={rgb,255: red,232; green,255; blue,207}]
\tikzstyle{ULTRAGRAY}=[-, draw={rgb,255: red,150; green,150; blue,150}]
\tikzstyle{EDGEDASHED}=[-, dashed, fill=none]
\tikzstyle{simple directed edge}=[->, draw=black, thick, line width=0.75mm]
\tikzstyle{simple dashed directed edge}=[->, draw=black, dashed]
\tikzstyle{SLIGHTLYDASHED}=[-, dotted, fill=none, draw={rgb,255: red,128; green,128; blue,128}]
\tikzstyle{simple dashed directed edge}=[->, draw=black, line width=0.75mm, dashed]
\tikzstyle{simple dashed directed edge not thick}=[->, draw=black, line width=0.45mm, dashed]
\tikzstyle{simple}=[->, draw=black, thick]
\tikzstyle{simple}=[draw=black, fill=none, tikzit draw=black, ->]
\tikzstyle{invis}=[-, draw=none]

%% file: settings/settings.tex
\PassOptionsToPackage{table,svgnames,dvipsnames}{xcolor}

\definecolor{TUMBlue}{HTML}{0065BD}
\definecolor{TUMSecondaryBlue}{HTML}{005293}
\definecolor{TUMSecondaryBlue2}{HTML}{003359}
\definecolor{TUMBlack}{HTML}{000000}
\definecolor{TUMWhite}{HTML}{FFFFFF}
\definecolor{TUMDarkGray}{HTML}{333333}
\definecolor{TUMGray}{HTML}{808080}
\definecolor{TUMLightGray}{HTML}{CCCCC6}
\definecolor{TUMAccentGray}{HTML}{DAD7CB}
\definecolor{TUMAccentOrange}{HTML}{E37222}
\definecolor{TUMAccentGreen}{HTML}{A2AD00}
\definecolor{TUMAccentLightBlue}{HTML}{98C6EA}
\definecolor{TUMAccentBlue}{HTML}{64A0C8}
\definecolor{VeryLightBlue}{HTML}{E8F8FF}

\definecolor{NONE}{HTML}{F721D4}
\definecolor{NTWO}{HTML}{00B1DB}
\definecolor{NTHREE}{HTML}{8C8F0E}

\definecolor{MATHAGREEN}{HTML}{B8E986}
\definecolor{MATHARED}{HTML}{E91111}
\definecolor{MATHABLUE}{HTML}{CBB3FF}
\definecolor{MATHAYELLOW}{HTML}{E6D830}

\definecolor{FILLGREEN}{HTML}{E8FFCF}
\definecolor{FILLPURPLE}{HTML}{E8FFCF}
\definecolor{FILLDARKBLUE}{HTML}{EEE6FF}

\definecolor{darkred}{HTML}{E80000}
\definecolor{darkgreen}{HTML}{006400} 
\definecolor{yellowgreen}{HTML}{9ACD32}

\usepackage[utf8]{inputenc}
\usepackage[T1]{fontenc}
\usepackage[sc]{mathpazo}
\usepackage[ngerman,american]{babel}

\usepackage[autostyle]{csquotes}

\usepackage[%
  backend=bibtex,
  url=false,
  sortcites=true,
]{biblatex}
\usepackage{hyperref} 
\hypersetup{
colorlinks=true,
linkcolor=darkgray,
citecolor=TUMBlue,
pdfsubject={Parameterized Complexity},
pdfauthor={Lukas Retschmeier},
pdfkeywords={ParameterizedComplexity, Algorithms, Kernelization}
}

\usepackage{graphicx}
\usepackage{scrhack} 
\usepackage{listings}
\usepackage{lstautogobble}
\usepackage{tikz}
\usepackage{pgfplots}
\usepackage{pgfplotstable}
\usepackage{booktabs}
\usepackage[final]{microtype}
\usepackage{caption}

\setkomafont{disposition}{\normalfont\bfseries} 
\BeforeTOCHead[toc]{{\cleardoublepage\pdfbookmark[0]{\contentsname}{toc}}}

\pgfplotsset{compat=newest}
\pgfplotsset{
  cycle list={TUMBlue\\TUMAccentOrange\\TUMAccentGreen\\TUMSecondaryBlue2\\TUMDarkGray\\},
}

\input{styling/tikz.tikzstyles}
\DeclareCaptionType{equ}[][]
\DeclareCaptionLabelFormat{blank}{}

\usepackage{etoolbox}
\makeatletter
\patchcmd{\@makechapterhead}{\vspace*{50\p@}}{\vspace*{-20\p@}}{}{}
\patchcmd{\@makeschapterhead}{\vspace*{50\p@}}{\vspace*{-20\p@}}{}{}
\patchcmd{\DOTI}{\vskip 80\p@}{\vskip 0\p@}{}{}
\patchcmd{\DOTIS}{\vskip 40\p@}{\vskip 0\p@}{}{}
\makeatother



%% file: settings/commands.tex
\newcommand*{\getUniversity}{Technical University Munich }
\newcommand*{\getSchool}{School of Computation}
\newcommand*{\getFaculty}{Information and Technology - Informatics}
\newcommand*{\getTitle}{On the Parameterized Complexity of Semitotal Domination on Graph Classes}
\newcommand*{\getTitleGer}{Über die Parametrisierte Komplexität des Problems der Halbtotalen Stabilen Menge auf Graphklassen}
\newcommand*{\getAuthor}{Lukas Retschmeier}
\newcommand*{\getDoctype}{Master's Thesis in Informatics}

\newcommand*{\getSupervisor}{Prof. Debarghya Ghoshdastidar}
\newcommand*{\getAdvisor}{Prof. Paloma Thomé de Lima}
\newcommand*{\getSubmissionDate}{\today}
\newcommand*{\getSubmissionLocation}{\textit{København S}}

\newcommand*\circled[1]{\tikz[baseline=(char.base)]{
            \node[shape=circle,draw,inner sep=2pt] (char) {#1};}}

\newtheorem{fact}{\protect\color{TUMSecondaryBlue} Fact}[section] 
\newtheorem{rgl}  {\protect\color{TUMSecondaryBlue} Rule} 	
\newtheorem{theorem}{\protect\color{TUMSecondaryBlue} Theorem}

\newtheorem{definition}{\protect\color{TUMSecondaryBlue} Definition}[section]
\newtheorem{graphclass}{\protect\color{TUMSecondaryBlue} Graph Class}

\newtheorem{corollary}{\protect\color{TUMSecondaryBlue} Corollary}[section]
\newtheorem{lemma}{\protect\color{TUMSecondaryBlue} Lemma}[section]
\newtheorem{proposition}{\protect\color{TUMSecondaryBlue} Proposition}[section]

\newcounter{casenum}
\newenvironment{caseof}{\setcounter{casenum}{0}}{\vskip.5\baselineskip}
\newcommand{\case}[2]{%
  \par\addvspace{.5\baselineskip}%
  \noindent \refstepcounter{casenum}\textbf{Case \thecasenum:}~#1\\*
  #2\ifhmode\unskip\fi
}

\crefname{case}{Case}{Cases}
\crefname{casenum}{\protect\textbf{Case}}{\protect\textbf{Cases}}
\Crefname{casenum}{\protect\textbf{Case}}{\protect\textbf{Cases}}
\creflabelformat{casenum}{#2\textbf{(#1)}#3}

\crefname{chapter}{Chapter}{Chapters}
\crefname{figure}{Figure}{Figures}
\crefname{section}{Section}{Section}

\newenvironment{caseofz}{\setcounter{casenum}{0}}{\vskip.5\baselineskip}
\newcommand{\casez}[2]{\vskip.5\baselineskip\par\noindent {\bfseries Case \arabic{casenum}:} #1\\#2\addtocounter{casenum}{1}}

\crefname{rgl}{Rule}{Rules}
\crefname{lemma}{Lemma}{Lemmas}
\crefname{fact}{Fact}{Facts}
\crefname{theorem}{Theorem}{Theorems}
\crefname{definition}{Definition}{Definitions}
\crefname{graphclass}{Graph Class}{Graph Classes}
\crefname{corollary}{Corollary}{Corollaries}
\crefname{proposition}{Proposition}{Propositions}
\crefname{figure}{Figure}{Figure}

\Crefname{rgl}{Rule}{Rules}
\Crefname{fact}{Fact}{Facts}
\Crefname{theorem}{Theorem}{Theorems}
\Crefname{definition}{Definition}{Definitions}
\Crefname{graphclass}{Graph Class}{Graph Classes}
\Crefname{corollary}{Corollary}{Corollaries}
\Crefname{figure}{Figure}{Figure}

\newacronym[description={Problem},%
    name={Semitotal Dominating Set}]{sds}{sds}{Semitotal Dominating Set}

\newcommand\kernelsize{358}

\newenvironment{abstract}[1]{%
\begin{center}\normalfont\textbf{Abstract}\end{center}
\begin{quotation} #1 \end{quotation}
}{%
\vspace{1cm}
}

\input{settings/abbreviations.tex}

\input{settings/mdframed-environments.tex}

%% file: settings/abbreviations.tex
\newcommand{\name}[1]{\textsc{#1}}
\newcommand{\dom}	{\name{Dominating Set}\xspace}
\newcommand{\pdomp}	{\name{Planar Dominating Set with Property P}\xspace}
\newcommand{\DOM}	{\name{Dominating Set (dom)}\xspace}
\newcommand{\doms}	{\name{dom}\xspace}
\newcommand{\tdoms}	{\name{tdom}\xspace}
\newcommand{\sdoms}	{\name{sdom}\xspace}
\newcommand{\pdom}	{\name{Planar Dominating Set}\xspace}

\newcommand{\is}{\name{Independent Set}\xspace}

\newcommand{\tdom}{\name{Total Dominating Set}\xspace}
\newcommand{\tdomin}{\name{Total Domination}\xspace}
\newcommand{\TDOM}{\name{Total Dominating Set (tdom)}\xspace}
\newcommand{\ptdom}{\name{Planar Total Dominating Set}\xspace}

\newcommand{\sdom}{\name{Semitotal Dominating Set}\xspace}
\newcommand{\sdomDE}{\name{Halbtotale Stabile Menge}\xspace}
\newcommand{\SDOM}{\name{Semitotal Dominating Set (sdom)}\xspace}
\newcommand{\sdomin}{\name{Semitotal Domination}\xspace}
\newcommand{\msdom}{\name{Minimum Semitotal Dominating Set}\xspace}
\newcommand{\sdomd}{\name{Semitotal Domination Decision}\xspace}

\newcommand{\msdomDE}{\name{Kleinste Halbtotale Stabile Menge}\xspace}
\newcommand{\psdom}{\name{Planar Semitotal Dominating Set}\xspace}

\newcommand{\cdom}{\name{Connected Dominating Set}\xspace}
\newcommand{\pcdom}{\name{Planar Connected Dominating Set}\xspace}

\newcommand{\prbdom}{\name{Planar Red-Blue Dominating Set}\xspace}
\newcommand{\rbdom}{\name{Planar Red-Blue Dominating Set}\xspace}
\newcommand{\dreg}{\textit{$D$-region decomposition}\xspace}

\newcommand{\efdom}{\name{Efficient Dominating Set}\xspace}
\newcommand{\pefdom}{\name{Planar Efficient Dominating Set}\xspace}

\newcommand{\eddom}{\name{Edge Dominating Set}\xspace}
\newcommand{\peddom}{\name{Planar Edge Dominating Set}\xspace}

\newcommand{\dirdom}{\name{Directed Dominating Set}\xspace}
\newcommand{\pdirdom}{\name{Planar Directed Dominating Set}\xspace}

\newcommand{\idom}{\name{Independent Dominating Set}\xspace}

\newcommand{\NPc}{\name{\textbf{NP}-Complete}\xspace}
\newcommand{\NPcs}{\name{NPc}\xspace}
\newcommand{\NPcn}{\name{\textbf{NP}-Completeness}\xspace}

\newcommand{\SAT}{\name{Boolean Satisfiability}\xspace}
\newcommand{\SATs}{\name{SAT}\xspace}
\newcommand{\NPh}{\name{\textbf{NP}-hard}\xspace}
\newcommand{\NP}{\name{\textbf{NP}}\xspace}
\newcommand{\Pt}{\name{\textbf{P}}\xspace}
\newcommand{\Ptt}{\name{P}\xspace}
\newcommand{\FPT}{\name{\textbf{FPT}}\xspace}
\newcommand{\FPTt}{\name{FPT}\xspace}
\newcommand{\FPTl}{\name{Fixed-Parameter Tractable}\xspace}

\newcommand{\WONE}{\ensuremath{\mathbf{W[1]}}\xspace}
\newcommand{\WTWO}{\ensuremath{\mathbf{W[2]}}\xspace}
\newcommand{\WTWOhs}{\ensuremath{W[2]}\xspace}
\newcommand{\WONEhs}{\ensuremath{W[1]}\xspace}
\newcommand{\WHIERARCHY}{\ensuremath{\mathbf{W}}-hierarchy\xspace}
\newcommand{\Wt}{\ensuremath{\mathbf{W[t]}}\xspace}

\newcommand{\G}		{\ensuremath{G=(V,E)}\xspace}
\newcommand{\GB}	{\ensuremath{G'=(V',E')}\xspace}

\newcommand{\sr}	{simple region}

\newcommand{\dGR} {\ensuremath{d_{G_{\mathfrak{R}}}}}
\newcommand{\GR} {\ensuremath{{G_{\mathfrak{R}}}}}
\newcommand{\VR} {\ensuremath{{V(\mathfrak{R})} } }

\newcommand{\ntwi}{\ensuremath{\abs{N_2(v,w) \cap V(R)}}}
\newcommand{\nthi}{\ensuremath{\abs{N_3(v,w) \cap V(R)}}}

\newcommand{\mc}{\mathcal}

\newcommand{\zerotext}[2][0pt]{\makebox[#1][l]{\qquad#2}}

\newcommand{\Dvw}   {\ensuremath{\mc{D}}\xspace}
\newcommand{\Dv}    {\ensuremath{\mc{D}_{v}}\xspace}
\newcommand{\Dw}    {\ensuremath{\mc{D}_{w}}\xspace}

\newcommand{\Nthreev}    {\ensuremath{N_3(v)}}

%% file: settings/mdframed-environments.tex
\ifstrempty{#1}%

\newcounter{theo}[section] 
\renewcommand{\thetheo}{\arabic{section}.\arabic{theo}}

\newcounter{lem}[section] 
\renewcommand{\thelem}{\arabic{section}.\arabic{lem}}

\newcounter{prf}[section]
\renewcommand{\theprf}{\arabic{section}.\arabic{prf}}

\newcounter{prb}[section]
\renewcommand{\theprf}{\arabic{prb}}
\newenvironment{prb}[2][]{%
\refstepcounter{prb}%
\ifstrempty{#1}%
{\mdfsetup{%
frametitle={%
\tikz[baseline=(current bounding box.east),outer sep=0pt]
\node[anchor=east,rectangle,fill=VeryLightBlue]
{\strut PROBLEM~\theprf};}}
}%
{\mdfsetup{%
frametitle={%
\tikz[baseline=(current bounding box.east),outer sep=0pt]
\node[anchor=east,rectangle,fill=VeryLightBlue]
{\strut ~#1};}}%
}%
\mdfsetup{innertopmargin=5pt,skipabove=20pt,skipbelow=20pt,linecolor=TUMBlue,%
linewidth=2pt,topline=true,%
frametitleaboveskip=\dimexpr-\ht\strutbox\relax
}
\begin{mdframed}[nobreak=true]\relax%
\label{#2}}{\end{mdframed}}

\newcounter{cc}[section]
\renewcommand{\thecc}{\arabic{cc}}
\newenvironment{cc}[2][]{%
\refstepcounter{cc}%
\ifstrempty{#1}%
{\mdfsetup{%
frametitle={%
\tikz[baseline=(current bounding box.east),outer sep=0pt]
\node[anchor=east,rectangle,fill=VeryLightBlue]
{\strut Complexity Class~\thecc};}}
}%
{\mdfsetup{%
frametitle={%
\tikz[baseline=(current bounding box.east),outer sep=0pt]
\node[anchor=east,rectangle,fill=VeryLightBlue]
{\strut ~#1};}}%
}%
\mdfsetup{innertopmargin=5pt,
linecolor=TUMBlue,%
linewidth=2pt,
topline=true,%
frametitleaboveskip=\dimexpr-\ht\strutbox\relax
}
\begin{mdframed}[nobreak=true]\relax%
\label{#2}}{\end{mdframed}}

%% file: pages/layout/cover.tex
\afterpage{%
\newgeometry{
top=1.25in,
bottom=1.25in,
left=1.25in,
right=1.25in,
bindingoffset=0.25in,
heightrounded
}
\begin{titlepage}
  \vspace*{\fill}
  \begin{center}
  \IfFileExists{logos/tum.pdf}{%
    \includegraphics[height=20mm]{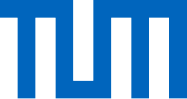}
  }{%
    \vspace*{20mm}
  }

  \vspace{5mm}
  {\huge\MakeUppercase{\getSchool{}}}\\

  \vspace{5mm}
  {\large\MakeUppercase{\getFaculty{}}}\\

  \vspace{7mm}
  {\large\MakeUppercase{\getUniversity{}}}\\

  \vspace{20mm}
  {\huge \getDoctype{}}

  \vspace{15mm}
  {\huge\bfseries \getTitle{}}

  \vspace{15mm}
  {\LARGE \getAuthor{}}

  \IfFileExists{logos/faculty.png}{%
    \vfill{}
    \includegraphics[height=20mm]{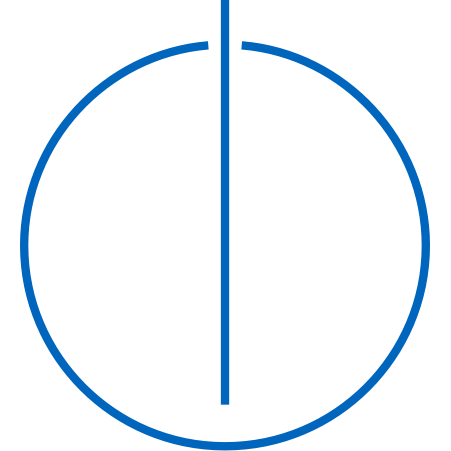}
  }{}
\end{center}
\end{titlepage}
\clearpage
\restoregeometry}

%% file: pages/layout/title.tex
\afterpage{%
\newgeometry{
top=1.25in,
bottom=1.25in,
left=1.25in,
right=1.25in,
bindingoffset=0.25in,
heightrounded
}
\begin{titlepage}
  \centering
  \IfFileExists{logos/tum.pdf}{%
    \includegraphics[height=20mm]{logos/tum.pdf}
  }{%
    \vspace*{20mm}
  }

  \vspace{5mm}
  {\huge\MakeUppercase{\getSchool{}}}\\

  \vspace{5mm}
  {\large\MakeUppercase{\getFaculty{}}}\\

  \vspace{7mm}
  {\large\MakeUppercase{\getUniversity{}}}\\

  \vspace{20mm}
  {\Large \getDoctype{}}

  \vspace{15mm}
  {\huge\bfseries \getTitle{}}

  \vspace{15mm}
  {\huge\bfseries \foreignlanguage{ngerman}{\getTitleGer{}}}

  \vspace{15mm}
  \begin{tabular}{l l}
    Author:          & \getAuthor{} \\

    Supervisors:                      & \getAdvisor{} \\
                       &  \textit{IT University Copenhagen} \\[1ex]
                    & \getSupervisor{} \\
                      &  \textit{Technical University Munich} \\[1ex]
    Submission Date: & February 15th, 2023 \\
  \end{tabular}

  \IfFileExists{logos/faculty.png}{%
    \vfill{}
    \includegraphics[height=20mm]{logos/faculty.png}
  }{}
\end{titlepage}

\clearpage
\restoregeometry}

%% file: pages/layout/disclaimer.tex
\afterpage{%
\newgeometry{
top=1.25in,
bottom=1.25in,
left=1.25in,
right=1.25in,
bindingoffset=0.25in,
heightrounded
}
\vspace*{0.70\textheight}
\noindent
I confirm that this \MakeLowercase{\getDoctype{}} is my own work and I have documented all sources and material used.

\vspace{35mm}
\noindent

\begin{tabu} to \textwidth {X[l]X[r]}
 \getSubmissionLocation & \getAuthor{}  \\ 
 \getSubmissionDate &   \\
\end{tabu}
\cleardoublepage{}
\clearpage
\restoregeometry}

%% file: pages/layout/acknowledgments.tex
\afterpage{%
\newgeometry{
top=1.25in,
bottom=1.25in,
left=1.25in,
right=1.25in,
bindingoffset=0.25in,
heightrounded
}
\addcontentsline{toc}{chapter}{Acknowledgments}
\thispagestyle{empty}

\vspace*{20mm}

\begin{center}
{\usekomafont{section} Acknowledgments}
\end{center}
\vspace{10mm}

I have many people to thank for their support, but first and foremost, I would like to thank my supervisor \textit{Paloma}!
Without your valuable ideas, patience and feedback, I would not have been able to write this thesis and have stayed motivated for such a long time. 
Our weekly discussions were really helpful and I could not have wished for better guidance. 
Thank you for always having time for my questions and ideas and supporting me wherever I needed help. 
Apart from this, without your initial crusade against the university's bureaucracy and all the effort you put into your side project of ``Finding him a Ph.D.'', I might not have arrived so well in Copenhagen!

I would like to extend my gratitude to my professor at TUM, \textit{Debarghya}, for instantly agreeing to formally supervise my thesis. 
Your immediate willingness to take on this role and the uncomplicated and instant communication was highly appreciated!

Moreover, I would like to thank the whole \textit{Algorithms Group} at ITU for letting me be a part of the group.
From the moment I joined, you all made me feel welcome and integrated, and I am grateful for the opportunity to have written my thesis in such a friendly, helpful and productive environment.
Thank you, \textit{Thore}, for ``getting the initial ball rolling'' and enabling this stay.

Special thanks to \textit{Jan Arne} in Bergen for his introductory course into parameterized complexity. 
Without this course and your motivating feedback afterward, I might not have tried to write my thesis in this area.

Thank you to my \textit{family} for always being there for me and supporting me, regardless of how far I am away.
Furthermore, I would like to acknowledge the assistance of my \textit{befriended ducks} and their quacky influence on some parts of the thesis.

Lastly, I would like to thank openAI's \textit{ChatGPT Dec 15} for supporting me in writing this acknowledgments section and \textit{Dall-E 2} for the immense amount of creativity that went into the chapter's illustrations!
What a time to be alive!

\cleardoublepage{}

\clearpage
\restoregeometry}

%% file: pages/util/abstract.tex
\chapter{Abstract}

\begin{abstract}
{\sffamily
For a graph \G, a set $D$ is called a \textit{semitotal dominating set}, if $D$ is a dominating set and every vertex $v \in D$ is within distance two to another witness $v' \in D$. 
The \msdom problem is to find a semitotal dominating set of minimum cardinality. 
The semitotal domination number $\gamma_{t2}(G)$ is the minimum cardinality of a semitotal dominating set and is squeezed between the domination number $\gamma(G)$ and the total domination number $\gamma_t(G)$.
 Given \G and a positive integer $k$, the \sdomd problem asks if $\gamma_{t2} \leq k$.

After introduced by Goddard, Henning and McPillan~\cite{Goddard2014}, \NP-completeness of the problem was shown for various graph classes like general graphs, \emph{split}, \emph{planar}, \emph{chordal bipartite} and \emph{circle} graphs~\cite{Henning2019, Kloks2021}.
Contrary, there exist polynomial-time algorithms for \emph{block} and \emph{interval} graphs as well as for \emph{graphs of bounded mim-width}, \emph{graphs of bounded clique-width}~\cite{Kloks2021, Galby2020,Courcelle1990,Henning2022,Henning2019}.
After giving a status about the complexity of the problem, we start a systematic look through the lens of \textit{parameterized complexity} by showing that \sdom is $\WTWOhs$-hard for bipartite graphs and split graphs when parameterized by solution size.
On the positive side, we extend a technique proposed by Alber et al.~\cite{Alber2004} for \dom to construct a linear kernel of size $\kernelsize \cdot k$ for \sdom on planar graphs.
 
 This result complements known linear kernels for other domination problems like \cdom, \rbdom, \efdom, \eddom, \idom, and \dirdom on planar graphs \cite{Diekert2005,Garnero2018,Guo2007,Garnero2017,Luo2013,Alber2006}

\textbf{Keywords: }Domination; Semitotal Domination; Parameterized Complexity; Planar Graphs; Linear Kernel; Problem Reduction; Graph Theory
} 
\end{abstract}

\newpage

\begin{abstract}
{\sffamily

Ein Graph \G und eine Menge $D$ wird als \textit{halbtotale stabile Menge} bezeichnet, falls $D$ eine stabile Menge ist und jeder Knoten $v \in D$ maximal einen Abstand von zwei zu einem anderen Zeugen $v' \in D$ besitzt. 
Das \msdomDE Problem frägt nach einer halbtotalen stabilen Menge von minimaler Kardinalität. 
Sei $\gamma_{t2}(G)$ die minimale Kardinalität einer halbtotalen stabilen Menge. Diese ist zwischen der minimalen stabilen Menge $\gamma(G)$ and der minimalen total stabilen Menge $\gamma_t(G)$ eingezwängt.
Gegeben \G und ein positives $k$, das \sdomDE Problem frägt, ob $\gamma_{t2} \leq k $ ist.

Nachdem das Problem von Goddard, Henning und McPillan~\cite{Goddard2014} eingeführt wurde, konnte \NP-vollständigkeit für viele Graphklassen wie \emph{split}, \emph{plättbare}, \emph{chordal bipartite} und \emph{zirkuläre} Graphen bereits gezeigt werden~\cite{Henning2019, Kloks2021}.
Andererseits existieren polynomialzeit Algorithmen sowohl für \emph{block} und \emph{interval} Graphen, als auch für \emph{Graphen mit beschränkter mim-width} und \emph{Graphen mit beschränkter clique-width}~\cite{Kloks2021, Galby2020,Courcelle1990,Henning2022,Henning2019}.

Nach einer umfassenden Analyse zum Stand des Problems, beginnen wir eine systematische Analyse aus Sicht der \textit{parametrisierten Komplexität} und zeigen, dass \sdomDE bei einer parameterisierung durch die Größe der Lösungsmenge $\WTWOhs$-hart für \textit{bipartite} und \textit{split} Graphen ist.
Basierend auf vorangegangener Arbeiten~\cite{Alber2004,Garnero2018} war es uns möglich Reduktionsregeln anzugeben, die einen linearen Problemkern der Größe $\kernelsize \cdot k$ für plättbare Graphen erzeugt.
Dies vervollständigt existierende lineare Problemkerne ähnlicher Problem wie \cdom, \rbdom, \efdom, \eddom, \idom oder \dirdom \cite{Diekert2005,Garnero2018,Guo2007,Garnero2017,Luo2013,Alber2006}.

\textbf{Schlagworte}: Stabile Menge; Halbtotale Stabile Menge; Parameterisierte Komplexität; Plättbare Graphen; Linearer Problemkern; Problemreduktion; Graph Theorie
}
\end{abstract}

%% file: pages/introduction/intro.tex
\chapter{Introduction}\label{ch:introduction}

\vspace*{-50pt}

\begin{figure}[ht]
        \includegraphics[width=0.35\textwidth, right]{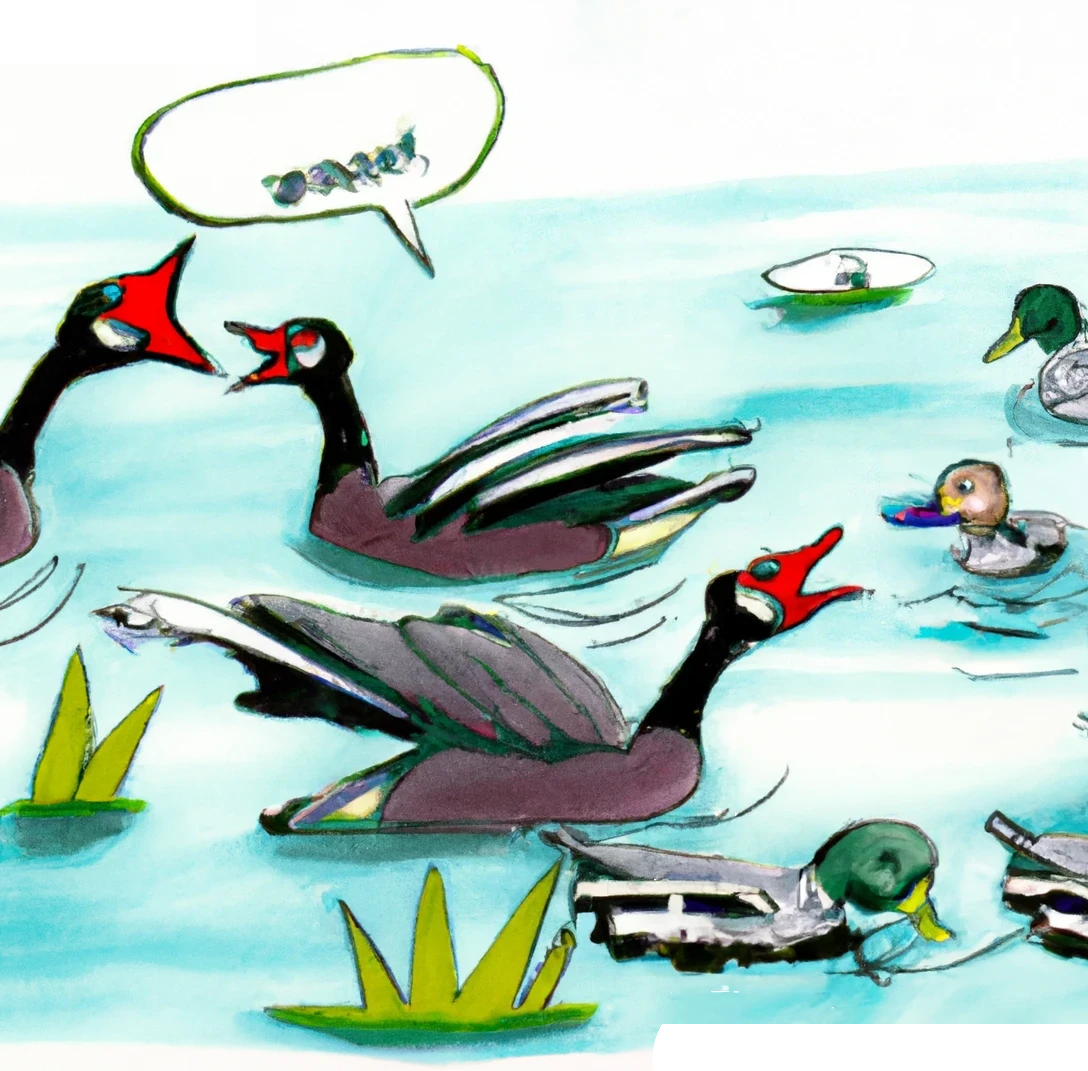}
        \captionsetup{textformat=empty,labelformat=blank}
        \caption[Generated with Dalle-E. Knowledge Cutoff 09-2022]{Generated with Dall-E. \url{https://labs.openai.com/}. ``A duck dominating sitting on a sea rose''}
\end{figure}

\epigraph{\itshape We have seen [\ldots], which is to say, all meaning comes from analogies.}{Douglas Richard Hofstadter, \textit{I am A Strange Loop}}

\begin{figure}[t]
    \centering
    \includegraphics[width=0.9\columnwidth]{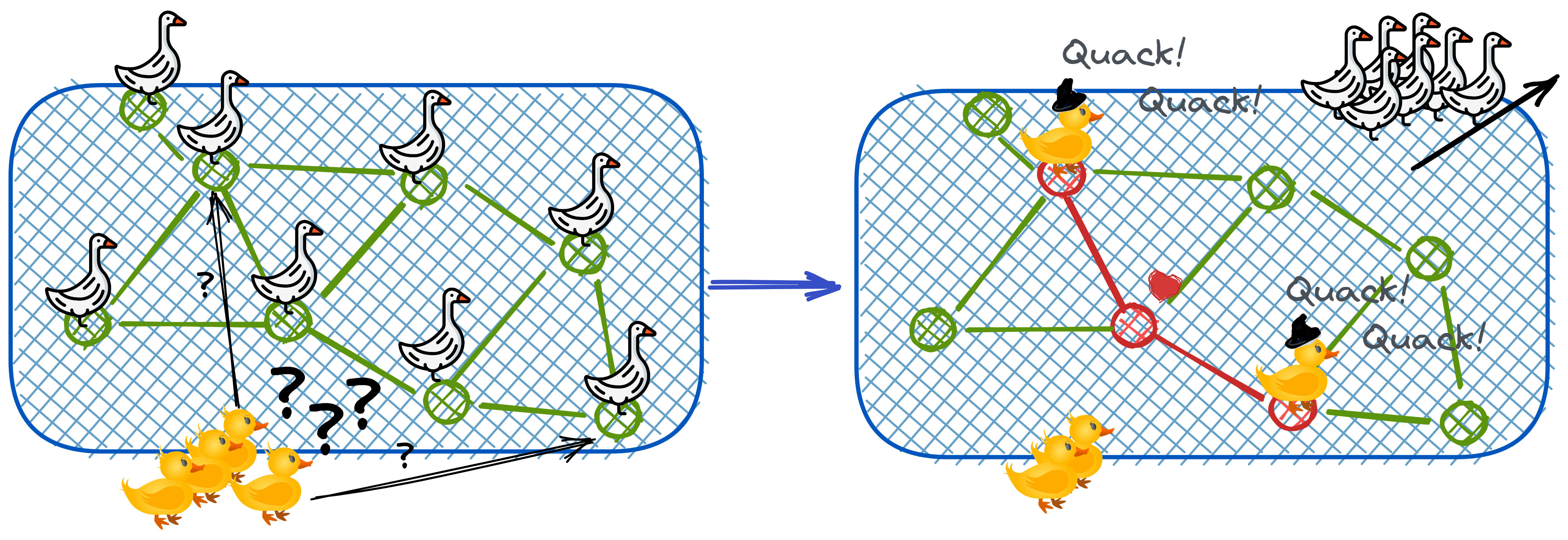}
    \caption[Introductions: Merganser Lake. Own Drawing. Embedded icons under public domain from {\href{https://creazilla.com/}{https://creazilla.com/}}]{\textit{Left: All water lilies ({\setulcolor{green}\ul{green}}) are occupied by members of the \textit{geesiosi} clan!
    An edge is drawn between two lilies if they are less than ten meters away such that one can hear the neighboring duck's quacking. 
    The handwritten arrows have been your first solution proposal which was refused by the \textit{Supreme Duck Decision Board}.
    Right: Your second and final solution: Two ducks are enough to make all \textit{geesiosi}s flee. Furthermore, they are only two water lilies apart ({\setulcolor{red}\ul{red edge}}) and therefore have someone to quack together with!}}
    \label{fig:duck-lake}
\end{figure}

Quack! Quack! They were careless for a second, and immediately the \textit{dreaded geesiosi} clan took the opportunity to conquer the  \textit{Merganser Lake}, which belongs to your befriended ducks!
Now they are sitting on all the beautiful water lilies and refuse to give them back. The desperate ducks rely on your assistance!
They have given you a map of the lake (see the left side of \cref{fig:duck-lake}) and marked all the water lilies in green.
You instantly assured of helping and started to analyze their situation!

You see that the \textit{geesiosi} members are terrified of the ducks' quacking, and you assume that one single duck could free up an entire water lily by sitting there and driving away all the geese on neighboring plants that are within a radius of ten meters! 
After thinking about this for a while, you realized that this might be the critical observation to regaining the lake!
After some more deep contemplating, you came up with a good assignment of ducks to water lilies, where only a minimum number of ducks is required to liberate the whole territory again.

Happy with your first idea, you present it to the \textit{Supreme Duck Decision Board}, but the \textit{Chief Strategy Duck} shared her worries with you: 
``We have to hold the fort and protect the lake against another future rush of the \textit{geesiosi}!'', they said, ``and it is a tedious task to sit alone on a water lily waiting the whole day! They would rather want to have another duck not too far away to have someone around to quack with together!''

After revising your solution, you came up with a new one where there is always another friend sitting at most two water lilies away. (see the right side of \cref{fig:duck-lake}). 
Now they should be close enough to dispel boredom, and your ducks were fully satisfied with your suggestion. 
Fantastic!

While you saw the chosen two ducks being sent out over the water's surface, you were still thinking about the problem.
It looked so easy at first, but in the end, one had to try all the possible configurations (and of course, you did not tell the ducks that it was that simple because they think you are a wizard!).
You wonder whether there is a way where you do not have to check all the possible configurations.

Back in your library, you learn from some ancient scrolls that this problem has already been formalized by a professor called Henning~\cite{Henning2019} as the \sdom problem, which is a variant of the intensively studied \dom problem. 
You read that both problems are \NPc~\cite{Garey1979,Henning2019}, and they are probably tough to solve in the general case efficiently, but there might still be hope if additional information is known. 
You look back into the map (\cref{fig:duck-lake}) and observe that none of the edges cross with each other, and you are getting curious if it can be used for something\ldots

\section*{Content of the Thesis}

Emerged during the last two decades, \textit{parameterized complexity} is a modern branch of computer science that showed many practical implications. 
This thesis systematically analyzes the \sdom problem through the lens of \textit{parameterized complexity}. 

\paragraph{Our contributions}

While many authors have already stated positive results - for example, there exist polynomial-time algorithms for \emph{AT-free}, \emph{block} and \emph{interval} graphs as well as for \emph{graphs of bounded mim-width} and \emph{graphs of bounded clique-width}~\cite{Kloks2021, Galby2020,Courcelle1990,Henning2022,Henning2019} - \NP-completeness was shown for various graph classes like general graphs, \emph{split}, \emph{planar}, \emph{chordal bipartite} and \emph{circle} graphs~\cite{Henning2019, Kloks2021}.

We will further investigate these \NPc cases by applying the framework of \textit{parameterized complexity}. 
We showed $\WTWOhs$-intractability for \textit{general}, \textit{bipartite} and \textit{split} using parameterized reductions from \dom when parameterized by solution size.

In a groundbreaking paper, Alber, Fellows, and Niedermeier~\cite{Alber2004} first gave a linear kernel for \pdom. 
They showed that a planar graph can be decomposed into a linear number of smaller regions. 
This motivated the introduction of local reduction rules that shrink the number of vertices in such a region to a constant size. 
Following up on this result, a plethora of other explicit linear kernels for domination problems on planar graphs were found~\cite{Guo2007, Garnero2017, Luo2013, Alber2006} and it made us believe we can also transfer them to \psdom.
Our hunch turned out to be true, and by applying similar techniques as by Garnero and Sau~\cite{Garnero2018}\footnote{We will rely on two different versions of this paper throughout the thesis. The \textit{arXiv} versions are explicitly marked.} for \ptdom, we were able to give an explicit kernel for \psdom of size $\kernelsize k$. 
More precisely, we are going to prove the following theorem:

\begin{restatable}[]{theorem}{centraltheo}\label{thm:central}
    The \psdom problem parameterized by solution size admits a linear kernel on planar graphs. 
    There exists a polynomial-time algorithm that, given a planar graph $(G, k)$, either correctly reports that $(G, k)$ is a NO-instance or returns an equivalent instance $(G', k)$ such that $\abs{V(G')} \leq \kernelsize \cdot k$.
\end{restatable}

This thesis is organized into the following chapters:

\begin{itemize}
    \item \Cref{ch:prelim} will give the necessary definitions in the fields of \textit{graph theory} and \textit{parameterized complexity}.
    \item In \cref{ch:semitotal-domination}, we will discuss the \sdom problem and its relation to \dom and \tdom in more detail. 
    As they are closely related, we will gather the complexity status for various graph classes and compare them with each other in \cref{ch:complexity-status}. 
    We will then show $\WTWO$-intractability for general, bipartite, split (and chordal) graphs.
    \item \Cref{ch:linkern} is the mainstay of this thesis. 
    We are going to construct a linear kernel for \psdom following an approach first suggested by Alber, Fellows and Niedermeier~\cite{Alber2004}. 
    \item In \cref{ch:closing}, we will give further ideas on how to improve the kernel and an outlook about interesting open problems for \sdom in general.
\end{itemize}

%% file: pages/theory/theory.tex
\chapter{Terminology and Preliminaries}\label{ch:prelim}

\vspace*{-50pt}

\begin{figure}[ht]
        \includegraphics[width=0.35\textwidth, right]{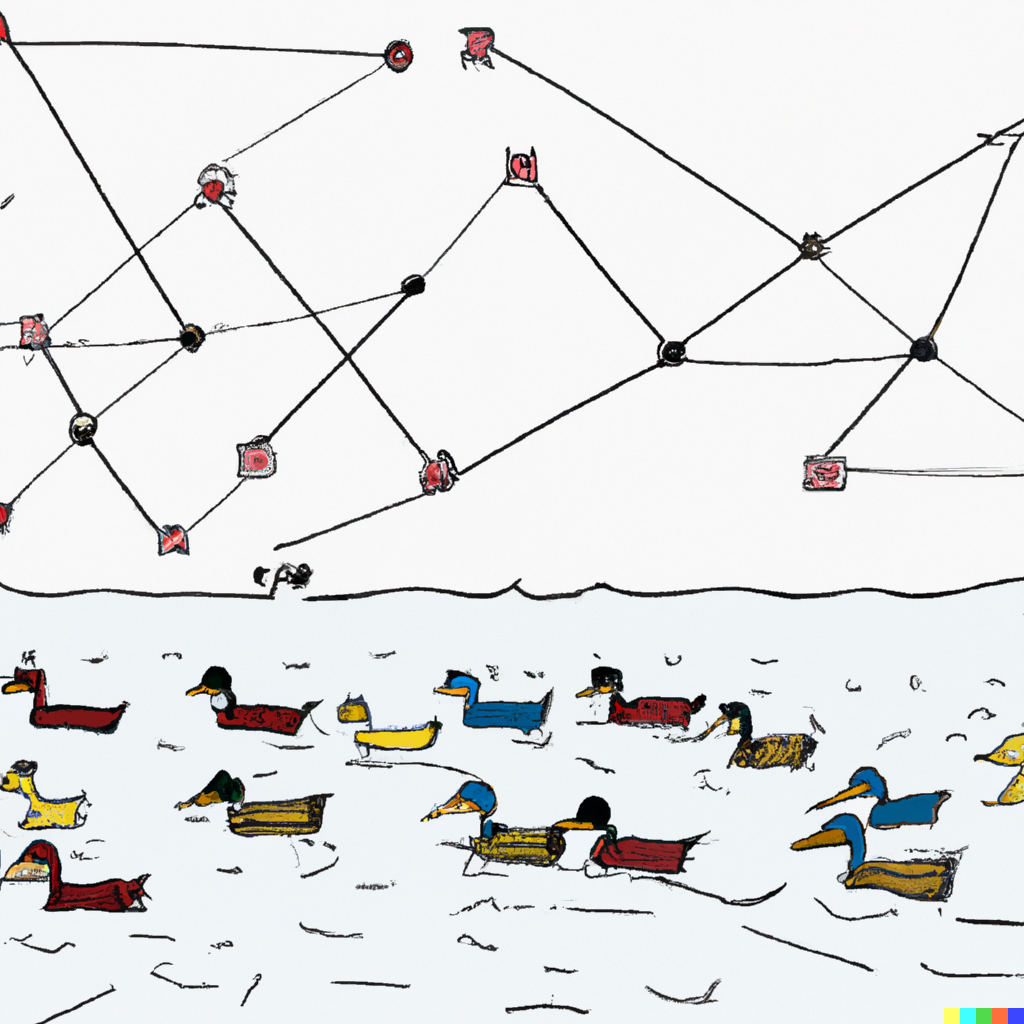}
        \captionsetup{textformat=empty,labelformat=blank}
        \caption[Generated with Dalle-E. Knowledge Cutoff 09-2022]{Generated with Dall-E. \url{https://labs.openai.com/}. ``Ducks learning graph theory while swimming on a sea sketched in color complex''}
\end{figure}

\epigraph{\itshape ``All we have to decide is what to do with the time that is given to us.''}{J. R. R. Tolkien, \textit{Gandalf} in \textit{Lord of the Rings}}

This chapter will introduce the core definitions used throughout this thesis. 
Most of the graph theory definitions are taken from~\cite{Diestel2010}. 
For definitions in the area of \textit{parameterized complexity}, the book by Cygan et al.~\citeauthor{Cygan2015} gives an excellent introduction.
For standard mathematical notation, we refer the reader to any introductory textbook into discrete mathematics, for instance~\cite{Rosen2012}.
\section{Graph Theory}
We will start by first giving basis definitions of graph theory and then define different graph classes necessary for this thesis.

\begin{definition}[Graph]
    A simple graph is a pair $G = (V, E)$ of two sets where $V$ denotes the vertices and $E \subseteq V \times V$ the edges of the graph.  A vertex $v \in V$ is incident with an edge $e \in E$ if $v \in e$. Two vertices $x, y$ are adjacent, or neighbors, if $\{x,y \} \in E$. By this definition, graph loops and multiple edges are excluded.
    
    A multigraph is a pair $(V, E)$ of disjoint sets together with a map $E \rightarrow V \cup [V]^2$ assigning to every edge either one or two vertices, its ends. Multigraphs can have loops and multiple edges.
    
    We usually denote the vertex set by $V(G)$ and its edge set by $E(G)$.
\end{definition}

Unless stated otherwise, we usually consider only \textit{simple graphs}, but the notion of \textit{multigraphs} gets essential when we later talk about the \textit{underlying multigraph} of a \dreg. 

\begin{definition}[Subgraph and Induced Subgraph]
    Let \G and $G' = (V', E')$ be two graphs. If $V' \subseteq V$ and $E' \subseteq E$ then $G'$ is a \underline{subgraph} of $G$. 
    If $G$ is a subgraph of $G'$ and $G'$ contains all the edges to $G$ with both endpoints in $V(G')$, then $G'$ is an \underline{induced subgraph} of $G$, and we write $G' = G[V(G')]$.
\end{definition}

\begin{definition}[Degrees]
    Let \G be a graph. The \textit{degree} $d_G(v)$ (shortly $d(v)$ if $G$ is clear from the context) of a vertex $v \in V$ is the number of neighbors of v. 
    We call a vertex of degree $0$ \underline{isolated}, and one of degree $1$ a \underline{pendant}. 
    If all the vertices of $G$ have the same degree $k$, then $g$ is $k$-regular.
\end{definition}

\begin{definition}[Closed and Open Neighborhoods {\cite{Balakrishnan2012}}]
    Let \G be a (non-empty) graph. 
    The set of all neighbors of $v$ is the \underline{open neighborhood} of $v$ and denoted by $N(v)$; the set $N[v] = N(v) \cup \{v\}$ is the \underline{closed neighborhood} f $v$ in $G$. When G needs to be made explicit, those open and closed neighborhoods are denoted by $N_G(v)$ and $N_G[v]$. 
\end{definition}

\begin{definition}[Isomorphic Graphs]
Let \G and $G' = (V', E')$ be two graphs. We call $G$ and $G'$ \underline{isomorphic}, if there exists a bijection $\phi: V \rightarrow V'$ with $\{x, y\} \in E \Leftrightarrow \phi(x)\phi(y) \in E'$ for all  $x,y \in V$. Such a map $\phi$ is called \underline{isomorphism}.

If a graph $G$ is isomorphic to another graph $h$, we denote $G \cong H$. 
\end{definition}

\begin{definition}[Paths and Cycles]
    A path is a non-empty graph $P = (V,E)$ of the form $V = \bigcup_{i  \in [k]} \{x_i\}$ and $E = \bigcup_{i \in  [k-1]} \{x_ix_{i+1}\}$ where all the $x_i$'s are distinct. 
    The vertices $x_0$ and $x_k$ are \underline{linked} by $P$ and are called the \textit{ends} of $P$. The \underline{length} of a path is its number of edges, and the path on $n$ vertices is denoted by  $P_n$. 
    We refer to a path $P$ by a natural sequence of its vertices: $P = x_0x_1...x_k$. Such a path $P$ is a path between $x_0$ and $x_k$, or a $x_0,x_k$-path.
    If $P = x_0...x_k$ is a path and $k \geq 2$, the graph with vertex set $V(P)$ and edge set $E(P) \cup \{x_kx_0\}$ is a \underline{cycle}. The cycle on $n$ vertices is denoted as $C_n$.
    The \underline{distance} $d_G(v,w)$ from a vertex $v$ to a vertex $w$ in a graph $g$ is the length of the shortest path between $v$ and $w$. If $v$ and $w$ are not linked by any path in $G$, we set $d_G(v,w) = \infty$. Again, if $G$ is clear from the context, we omit the subscripted $G$ and just write $d(v,w)$ instead.
\end{definition}

\subsection*{Graph Classes}

A \textit{graph class} is a set of graphs sharing a common structural property.

\begin{definition}[Graph Parameters]
Let \G be a graph.
An  \underline{independent set} of $G$ is a set of pairwise non-adjacent vertices. 
A \underline{clique} of $G$ is a set of pairwise adjacent vertices. 
A \underline{vertex cover} of $G$ is a subset of vertices containing at least one endpoint of every edge. 
A \underline{dominating set} is a subset $D$ of vertices such that all vertices not contained in are adjacent to some vertex in $D$.
A \underline{dominating set} is a subset $D$ of vertices such that all vertices not contained in are adjacent to some vertex in $D$.
The \underline{chromatic number}, $\chi(G)$, of a graph \G is the smallest number of colors, such that adjacent vertices in $V$ are colored differently.

\end{definition}

\begin{graphclass}[Perfect graph]
If the chromatic number equals the size of the maximum clique, \G is called \underline{perfect}. 
\end{graphclass}

\begin{graphclass}[r-partite]
    Let $r \geq 2$ be an integer. A Graph $G = (V, E)$ is called \underline{r-partite} if $V$ admits a partition into $r$ classes such that every edge has its ends in different classes: Vertices in the same partition class must not be adjacent. 
    A \textit{$2$-partite} graph is called \underline{bipartite}. 
    
    An $r$-partite graph in which every two vertices from different partition classes are adjacent is called \underline{complete}. For the \underline{complete bipartite graph} on bipartitions $X \uplus Y$ of size $m$ and $n$, we shortly write $K_{m,n}$. 
\end{graphclass}

\begin{graphclass}[Complete]
If all vertices of a graph \G are pairwise adjacent, we say that $G$ is \underline{complete}. 
A complete graph on $n$ vertices is a $K_n$. A $K_3$ is called a \underline{triangle}.
\end{graphclass}


\begin{graphclass}[Chordal]
For a graph \G, an edge that joins two vertices of a cycle, but is not itself an edge of the cycle is a \underline{chord} of that cycle.

Furthermore, we say $G$ is \underline{chordal} (or \textit{triangulated}) if each of its cycles of length at least four has a chord. In other words, it contains no induced cycle other than triangles.

\end{graphclass}

\begin{graphclass}[Split]
A \underline{split graph} is a graph \G whose vertices can be partitioned into a clique and an independent set.    
\end{graphclass}


    

\begin{graphclass}[Plane and Planar Graphs]

A \textit{plane graph} is a graph \G that can be drawn in the plane $\mathbb{R}^2$ in such a way that no edges cross each other.
A \underline{plane} graph is a planar graph together with a concrete drawing in the plane, the \underline{plane embedding} of $G$. 

\end{graphclass}

\input{pages/theory/pc/pc.tex}

\chapter{On Parameterized Semitotal Domination}\label{ch:semitotal-domination}

\vspace*{-50pt}

\begin{figure}[ht]
        \includegraphics[width=0.35\textwidth, right]{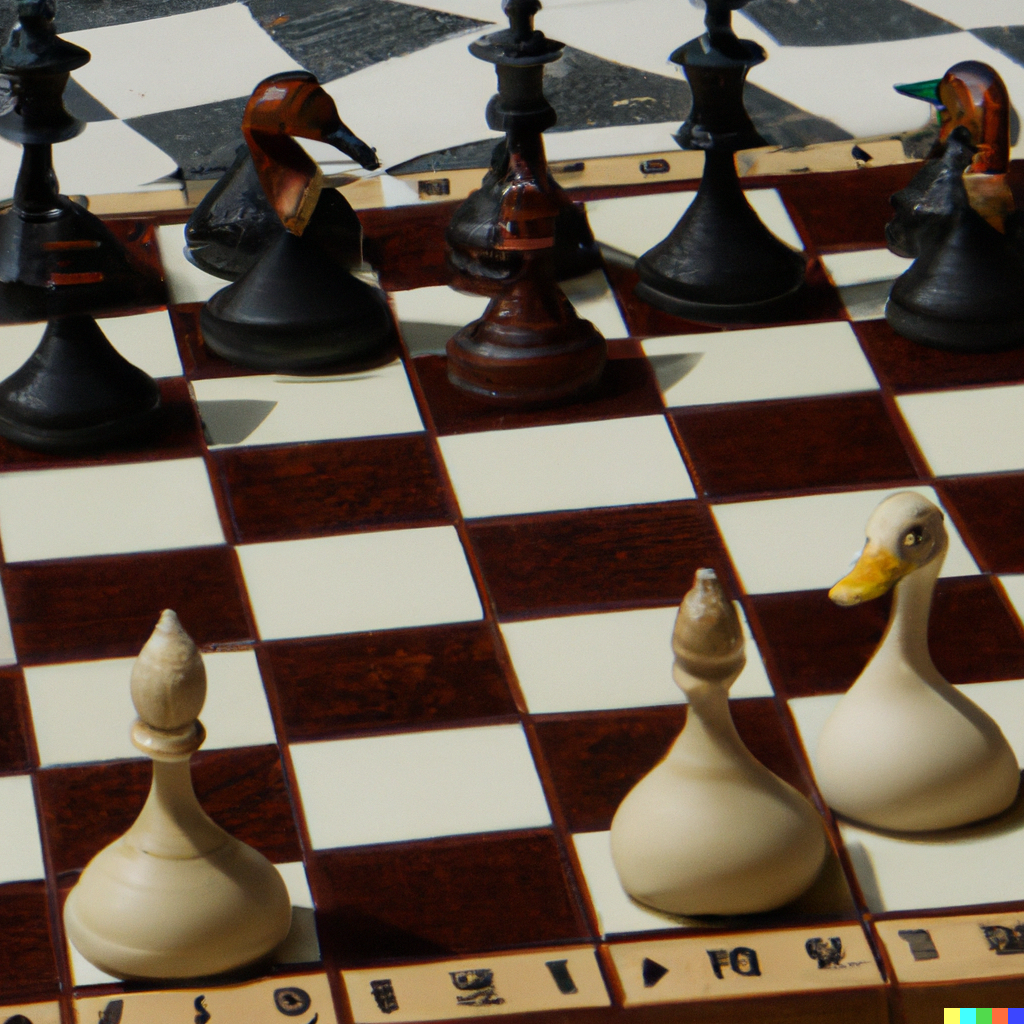}
        \captionsetup{textformat=empty,labelformat=blank}
        \caption[Generated with Dalle-E. Knowledge Cutoff 09-2022]{Generated with Dall-E. \url{https://labs.openai.com/}. ``Duck playing chess''}
\end{figure}

\epigraph{\itshape ``This set must be small, but also complete, \\ 
to conquer the
graph with ease and fleet, \\
with vertices chosen so carefully, \\
the solution is found, so elegantly.''
}{ChatGPT, Generated by \emph{OpenAI Language Model}, \\
Knowledge Cutoff: 2021-09}

In connection with various chessboard problems, the concept of domination can be traced back to the mid-1800s.
For example, De Jaenosch attempted in 1862 to find the minimum number of queens required to fully cover an $n \times n$-chessboard~\cite{Jaenisch1862}.
Because of the immense amount of publications related to domination that followed, Haynes, Hedetniemi, and Slater started a comprehensive survey of the literature in 1998~\cite{Haynes1998, Haynes1998b}. 
Twenty years later, through a series of three more books, Haynes, Henning and Hedetniemi complemented the survey with the latest developments~\cite{Haynes2020, Haynes2021, Haynes2022}.

We are now introducing the problems of \dom, \sdom, and \tdom and dedicate the rest of the chapter to giving a current status about the complexity on various graph classes. 

\section{Domination Problems}

We will define this work's three most important domination problems: \dom, \sdom, and \tdom.
Recall that a dominating set of a graph \G is a subset $D \subseteq V$, such that every vertex from $V \setminus D$ is adjacent to some vertex in $D$.
We say that a vertex $d$ is a \textit{dominating vertex} or \textit{dominator} if $d \in D$ and that $d$ \textit{dominates} all of its neighbors.
Before defining the following problems, we give the following terminology: 

\begin{definition}[Domination Numbers]
   The \underline{domination number} in a graph $G$ is the minimum cardinality of a dominating set (ds) of $G$, denoted as $\gamma(G)$. 
   The \underline{total domination number} is the minimum cardinality of a total dominating set (tds) of $G$, denoted by $\gamma_t(G)$.
   The \underline{semitotal domination number} is the minimum cardinality of a semitotal dominating set (sds) of $G$, denoted by $\gamma_{t2}(G)$.

   We say that a ds $D$ is \underline{minimal} if no proper subset $S' \subset S$ is a ds and that $D$ is a \underline{minimum} if it is the smallest ds.
\end{definition}

\begin{prb}[\MakeUppercase{\DOM}~{\cite[p. 586]{Cygan2015}}]{prb:ds}
    \begin{tabularx}{1.0\textwidth}{>{\hsize=0.30\hsize}X>{\hsize=0.8\hsize}X}
        \textbf{Input} & Graph \G, $k \in \mathbb{N}$\\
        \textbf{Question} & Is there a set {$D \subseteq V$} of size at most $k$ such that ${N[D] = V}$? \\
    \end{tabularx}
\end{prb}

The \dom problem asks for a subset $D$ of size at most $k$ whose set of neighbors are adjacent to all the other remaining vertices in a graph.
Assume that we have an arbitrary dominating set $D$ for some connected graph $G$ with at least two vertices.
For each $v \in D$, there must be at least one other dominating vertex $v' \in D$ that is at most three steps away.
Proven in \cref{fact:distance}, this motivates the definition of the following problem variants.

\begin{fact}\label{fact:distance} Let \G be a connected graph and $D$ a ds with $\abs{D} > 1$. 
For any $v_1 \in D$ there exists at least one other $v_2 \in D$ with $d(v_1,v_2) \leq 3$.
\end{fact}
\begin{proof}

Assume a ds $D$ and $v_1 \in D$ for which no other dominating vertex is in a distance less than three.
Then exists $v_2$ with $d(v_1,v_2) > 3$. 
By connectivity there is a path $p = (v_1, p_1, p_2, ...,  p_i, v_2)$ from $v_1$ to $v_2$ and no $p_i \in D $. 
For $D$ to be a valid ds, $p_2,...p_{i-1}$ have to be dominated by a neighbor as well, and therefore, there must exist a $u \in N(p_2) \cap D$ with $d(v_1, u) \leq 2$ giving a contradiction.
\end{proof}
We obtain the following variants if we lower the maximal allowed distance between two vertices in $D$.

Requiring every vertex $v \in V$ to have another direct neighbor in $D$ leads to the \tdom problem. 
We denote this neighbor as a \textit{witness} of $v$.

\begin{prb}[\MakeUppercase{\TDOM}~{\cite[p. 596]{Cygan2015}}]{prb:sds}
    \begin{tabularx}{1.0\textwidth}{>{\hsize=0.30\hsize}X>{\hsize=0.8\hsize}X}
        \textbf{Input} & Graph \G, $k \in \mathbb{N}$\\
        \textbf{Question} & Is there a dominating set $D \subseteq V$ of size at most $k$, such that for every $u \in V(G)$ there exists $v \in D$ with $\{u,v\} \in E$? \\
    \end{tabularx}
\end{prb}

It is natural to ask what happens if we restrict this distance property to be at most two, which leads us straight to the idea of \sdomin.
For a semitotal dominating set $D$, we say that $v$ \emph{witnesses} $v'$ if $v, v' \in D$ and $d(v,v') \leq 2$.
\sdomin was introduced by Goddard, Henning and McPillan~\cite{Goddard2014} as a relaxation of \tdomin. 

\begin{prb}[\MakeUppercase{\SDOM}~{\cite{Goddard2014}}]{prb:tsds}
    \begin{tabularx}{1.0\textwidth}{>{\hsize=0.30\hsize}X>{\hsize=0.8\hsize}X}
        \textbf{Input} & Graph \G, $k \in \mathbb{N}$\\
        \textbf{Question} & Is there a dominating set $D \subseteq V$ of size at most $k$, such that ${N[D] = V}$ and for all $d_1 \in D$ there exists another $d_2 \in D$ such that ${d(d_1, d_2) \leq 2}$?\\
    \end{tabularx}
\end{prb}

\begin{figure}
     \begin{equation*}
         \input{fig/tikz/ds-examples.tikz}
     \end{equation*}
    \caption[An example for various dominating sets]{\textit{An example for a minimum dominating set, semitotal dominating set and a total dominating set, where $\gamma(G) < \gamma_{t2}(G) < \gamma_t(G)$ are strict. 
    In the first case, only two vertices suffice to dominate all others.
    In the second one, we need a witness between $d_1$ and $d_2$ at most distance two. 
    In the last case, $d_1$ and $d_2$ both need a neighbor in the total dominating set.}}
    \label{fig:dsexamples}
\end{figure}
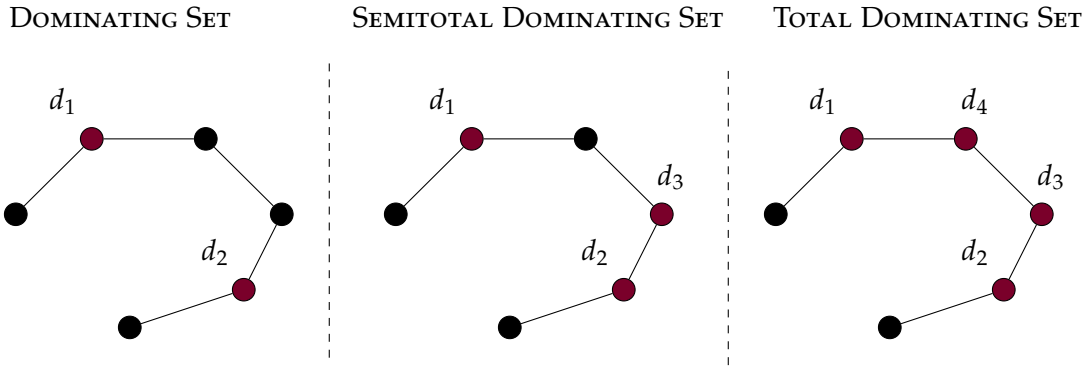

Because any tds is also an sds and every sds is also a ds, $\gamma_{t2}$ is squeezed between $\gamma$ and $\gamma_t$.
The  following fact was first observed by Goddard and Henning~\cite{Goddard2014}:

\begin{fact}
For every graph $G$ with no isolated vertex, $\gamma(G) \leq \gamma_{t2}(G) \leq \gamma_t(G)$.
\end{fact}

There are graphs where this inequality is strict.
\Cref{fig:dsexamples} demonstrates that the minimum ds, sds, and tds can strictly differ in size on a fixed graph \G. 
A ds does not need witnesses, and we can choose two vertices $d_1$ and $d_2$ to dominate the graph; this is the only minimum solution.
As $d(d_1, d_2) = 3$, we have to introduce additional dominating vertices for a minimum sds (middle) and tds (right). 
Their only purpose is to bridge the gap between $d_1$ and $d_2$, but they have no function as dominators.

\section{Complexity Status of \sdom}\label{ch:complexity-status}

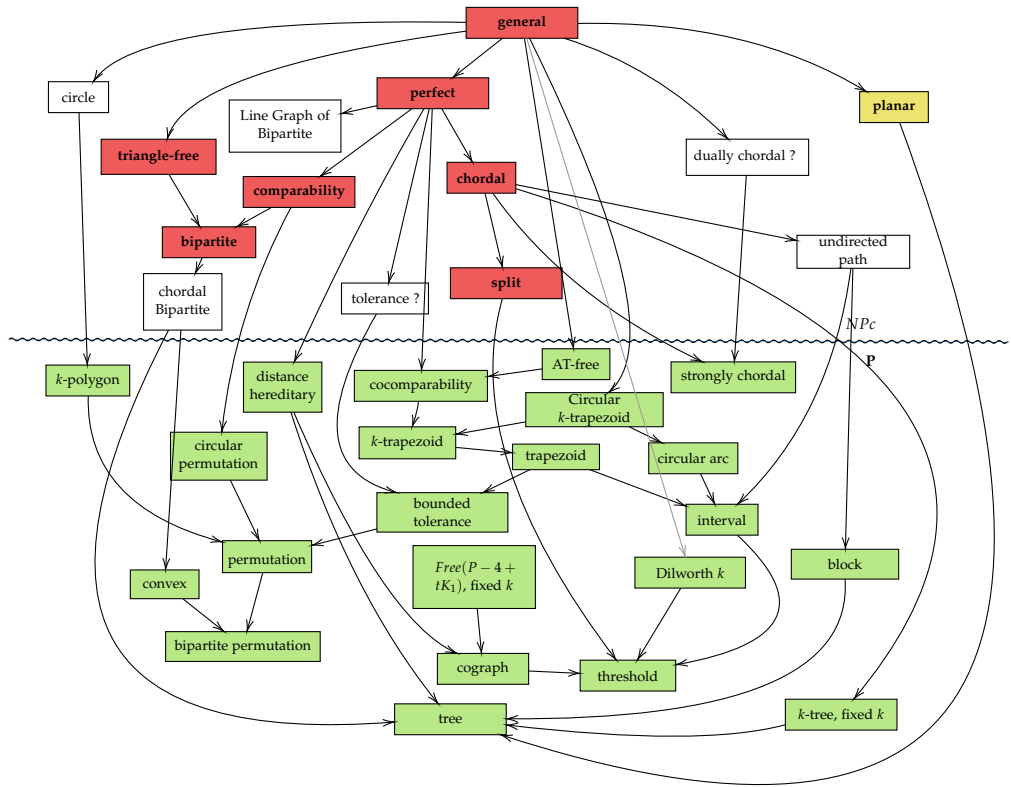
\begin{figure}
    \centering
    \resizebox{1\textwidth}{!}{
        \input{fig/mathcha/classes.tikz}
    }
    \caption[Graph inclusions]{\textit{Computational complexity of \sdom. Graph classes that admit a polynomial time algorithm are marked in {\setulcolor{MATHAGREEN}\ul{green}}, those with an fpt algorithm when parameterized by solution size in {\setulcolor{MATHAYELLOW}\ul{yellow}}, those that are not fpt by solution size in {\setulcolor{MATHARED}\ul{red}}, and those that are unknown are left \textit{white}.
    An \textbf{?} denotes problems where the classical complexity is also unknown.\\
    \underline{convex}, \underline{interval}, \underline{(circular k-)trapezoid}, \underline{circular permutation}, \underline{Dilworth-k}, \underline{$k$-polygon} and \underline{convex bipartite} graphs have bounded mim-width, and a branch decomposition can be found in polynomial time \cite{Belmonte2011} and therefore can be solved in polynomial time by using the algorithm proposed by Galby et al. \cite{Galby2020}.
    }}
    
    \label{fig:bigpicture}
\end{figure}

Goddard et al.~\cite{Goddard2014} first showed that \SDOM is \NPc in the general case and initiated the systematic analysis of the problem.  
\cref{tab:complexities} complements the complexity status for \doms, \sdoms, and \tdoms when restricted to specific graph classes, which was first surveyed by Galby et al.~\cite{Galby2020}.
We also added another column showing the known parameterized complexity. 
\cref{fig:bigpicture} visualizes the big picture and the inclusions among the graph classes.

A polynomial-time algorithm was shown for graphs of bounded \textit{mim}-width by Galby et al.~\cite{Galby2020}.
Because a branch decompositions of constant \textit{mim}-width can be found for \textit{permutation}, \textit{convex}, \textit{interval}, \textit{(circular k-)} \textit{trapezoid}, \textit{circular permutation}, \textit{Dilworth-$k$}, \textit{$k$-polygon}, \textit{circular arc}, \textit{complements of $d$-degenerate}, \textit{bipartite permutation} and \textit{convex bipartite} graphs in polynomial time~\cite{Belmonte2011}, this algorithm immediately implies the existence of a polynomial time algorithm for these graph classes as well. 
Additionally, Galby et al.~\cite{Galby2020} resolved the complexity for \textit{bipartite permutation} and \textit{convex bipartite} graphs.

Furthermore, new polynomial time algorithms have been devised for \textit{strongly chordal}~\cite{Tripathi2021}, \textit{AT-free}~\cite{Kloks2021} and \textit{block}~\cite{Henning2022} graphs and a new linear time algorithm has been found for interval graphs~\cite{Pradhan2021} beating the previous $\mathcal{O}(n^2)$-algorithm given by Henning and Pandey~\cite{Henning2019}.
On the negative side, \sdom on \textit{circle graphs}~\cite{Kloks2021} graphs, and \textit{undirected path graphs}~\cite{Henning2019} was shown to be \NPc.

There seems to be a symmetry between these three domination problems where the complexity of many graph classes mirrors each other, but it looks like \sdoms has more in common with \doms than with \tdoms.
For instance, on \textit{chordal bipartite} graphs, \sdoms and \doms are \NPc, while it is polynomial-time solvable for \tdoms. 
Contrary, \sdoms and \doms have a polynomial-time algorithm for \textit{strongly chordal} graphs, but \tdoms is \NPc.
In all currently known cases, \sdom follows the complexity of \dom in the classical and the parameterized setting.

This thesis goes the first step into approaching the problem from the perspective of \textit{parameterized complexity}.
If not mentioned otherwise, the problem is always parameterized by the size of a solution.
We have also added the parameterized complexity for those graph classes which we used as an orientation for the complexities of \sdom.
We were able to prove \WTWO-hardness for \textit{bipartite} and \textit{split} graphs by giving parameterized reducing from \doms by solution size. We provide an explicit construction of a kernel for the \textit{planar} case, which exists for \doms~\cite{Alber2004} and \tdoms~\cite{Garnero2019} as well.

\begin{center}
    \begin{table}[t]
    \begin{minipage}[th]{\linewidth}
    \setcounter{mpfootnote}{\value{footnote}}
    \renewcommand{\thempfootnote}{\arabic{mpfootnote}}

\resizebox{1.0\textwidth}{!}{

    \begin{tabularx}{1.5\textwidth}{lllllll}

        \arrayrulecolor{TUMBlue}\toprule

        \textbf{Graph Class}                  & \multicolumn{2}{c}{\textbf{\dom}}                       & \multicolumn{2}{c}{\textbf{\sdom}}           & \multicolumn{2}{c}{\textbf{\tdom}}                                                                                                                                \\
        \cmidrule(lr){2-3} \cmidrule(lr){4-5} \cmidrule(lr){6-7}
                                              & classical                                               & Parameterized                                & classical                                               & Parameterized              & classical                                    & Parameterized               \\
        bipartite                             & \NPcs~\cite{Bertossi1984}                               & \WTWOhs~\cite{Raman2008}                                      & \NPcs~\cite{Henning2019}                                & \WTWOhs (\cref{lemma:bipartite})             & \NPcs~\cite{Pfaff1983}                       & ?                             \\
        
        line graph of bipartite               & \NPcs~\cite{Korobitsin1992}                             & ? & \NPcs~\cite{Galby2020}                                  &     ?                       & \NPcs~\cite{McRae1995}                       &  ?                            \\
        
        circle                                & \NPcs~\cite{Keil1993}                                   & \WONEhs~\cite{Bousquet2012}                  & \NPcs~\cite{Kloks2021}                                  & ?  & \NPcs~\cite{McRae1995}                       & \WONEhs~\cite{Bousquet2012} \\
        
        chordal                               & \NPcs~\cite{Booth1982}                                  & \WTWOhs~\cite{Raman2008}                     & \NPcs~\cite{Henning2019}                                & \WTWOhs (\cref{lemma:splitgraph})               & \NPcs~\cite{Laskar1983}                      & \WONEhs~\cite{Chang1998} by \textit{split}                            \\
        
        $s$-chordal , $s > 3$                          & \NPcs~\cite{Liu2011}                                    & \WTWOhs~\cite{Liu2011}                       & ?                                                     & ?                         & \NPcs~\cite{Liu2011}                         & \WONEhs~\cite{Liu2011}      \\
        
        split                                 & \NPcs~\cite{Bertossi1984}                               & \WTWOhs~\cite{Raman2008}         & \NPcs~\cite{Henning2019}                                & \WTWOhs (\cref{lemma:splitgraph})             & \NPcs~\cite{Laskar1983}                      & \WONEhs~\cite{Chang1998}    \\
        
        3-claw-free                           & \NPcs~\cite{Cygan2011}                                  & \FPTt~\cite{Cygan2011}                        & ?                                               & ? & \NPcs~\cite{McRae1995}                       & ? \\
        
        $t$-claw-free, $t>3$                  & \NPcs~\cite{Cygan2011}                                  & \WTWOhs~\cite{Cygan2011}                     & ?                                           & ?                    & \NPcs~\cite{McRae1995}                       & ?               \\
        
        chordal bipartite                     & \NPcs~\cite{Mueller1987}                                & ?                                 & \NPcs~\cite{Henning2019}                                & ?                      & \multicolumn{2}{c}{\Ptt~\cite{Damaschke1990}}                               \\
        
        planar                                & \NPcs~\cite{Garey1979}                                        & \FPTt~\cite{Alber2004}                        & \NPcs                                                   & \FPT (\cref{thm:central})                       & \NPcs                                        & \FPTt~\cite{Garnero2018}     \\
        
        undirected path                                & \NPcs~\cite{Booth1982}                                   & \FPTt~\cite{Figueiredo2022} & \NPcs~\cite{Henning2022}  & ?                     & \NPcs~\cite{Lan2014}                         & ?                     \\

        dually chordal                        & \multicolumn{2}{c}{\Ptt~\cite{Brandstaedt1998} }         & \multicolumn{2}{c}{?\footnotemark} &                           \multicolumn{2}{c}{\Ptt~\cite{Kratsch1997}}                                                                            \\
        
        strongly chordal                      & \multicolumn{2}{c}{\Ptt~\cite{Farber1984} }            & \multicolumn{2}{c}{\Ptt~\cite{Tripathi2021}}  & \NPcs~\cite{Farber1984}                                 &                                                                                                         \\
        
        AT-free                               & \multicolumn{2}{c}{\Ptt~\cite{Kratsch2000}}              & \multicolumn{2}{c}{\Ptt~\cite{Kloks2021} }    & \multicolumn{2}{c}{\Ptt~\cite{Kratsch2000}}                                                                                                                        \\
        
        tolerance                             & \multicolumn{2}{c}{\Ptt~\cite{Giannopoulou2016}}                         & \multicolumn{2}{c}{?}                                                  & \multicolumn{2}{c}{?}                                                                    \\
        
       block                        &                                                      \multicolumn{2}{c}{\Ptt~\cite{Farber1984} }                                          & \multicolumn{2}{c}{\Ptt~\cite{Henning2022}}              & \multicolumn{2}{c}{\Ptt~\cite{Chang1989}}                                                                       \\
        
        interval                  & \multicolumn{2}{c}{\Ptt~\cite{Chang1998a}}                                          & \multicolumn{2}{c}{\Ptt~\cite{Pradhan2021}} &                                         \multicolumn{2}{c}{\Ptt~\cite{Bertossi1986}}                       \\

        \midrule
        bounded clique-width                  & \multicolumn{2}{c}{\Ptt~\cite{Courcelle1990}}            & \multicolumn{2}{c}{\Ptt~\cite{Courcelle1990}} & \multicolumn{2}{c}{\Ptt~\cite{Courcelle1990}}                                                                                                                      \\
        
        bounded mim-width                     & \multicolumn{2}{c}{\Ptt~\cite{Belmonte2011,BuiXuan2013}} & \multicolumn{2}{c}{\Ptt~\cite{Galby2020}}     & \multicolumn{2}{c}{\Ptt~\cite{Belmonte2011,BuiXuan2013}}                                                                                                           \\
        \midrule

        \midrule
        \bottomrule
    \end{tabularx}
    \footnotetext[1]{Galby et al. \cite{Galby2020} attempted it, but found a mistake in their reduction.}
}
\end{minipage}
    \caption{\textit{Comparison between the complexities of \dom, \sdom and \tdom in the classical and parameterized setting when parameterized by solution size. 
    Open problems are marked with an \textbf{?}.
    Note that \sdom follows more the complexities of \dom than \tdom which can be seen in the \underline{strongly chordal} and \underline{chordal bipartite} cases.}}\label{tab:complexities}
    \end{table}
\end{center}

We would like to mention that membership of \psdom in \FPT already follows from Alber et al.~\cite{Alber2002}.
Their algorithm \pdomp can be used to obtain an fpt algorithm with running time $\mathcal{O}(c^{\sqrt{k}}n)$ for some large constant $c$ and solution size $k$.
Although the kernel size we obtained is high, we think there is a lot of room for improvement by further adjusting the reduction rules and therefore shrinking the factor of the kernel size.
There are still many gaps open left for further research.
We will discuss them at the end of this thesis in \cref{ch:closing}.


%% file: pages/theory/pc/pc.tex
\section{Computational Complexity Theory}

Computational complexity investigates how many computational resources are required to solve a specific problem. 

%
%
%
%
%
%
%

We denote \Pt as the class of all problems that can be solved by a \textit{Deterministic Turing Machine} in polynomial time, whereas \NP contains all problems that can be solved by a \textit{Non-Deterministic} Turing Machine in polynomial time.
In other words \Pt contains problems that are \textit{efficiently solvable}, whereas \NP contains all problems whose solution can efficiently be verified. 
Note that $\Pt \subseteq \NP$, but the reverse direction is unknown.

\subsection{\NPcn}\label{ch:npc}
A significant advance in the early 1970s was the realization that some problems in \NP are \textit{at least as hard as} as any other problem in \NP. By spanning a whole ``web of reductions''~\cite{Arora2006} we now get strong evidence that none of these problems can be solved efficiently.
The first results in this new field were published independently by Cook~\cite{Cook1971} and Levin~\cite{Levin1973} after Karp~\cite{Karp1972} had introduced the notion of problem reductions.
The Cook-Levin-Theorem~\cite{Cook1971} states that the \SAT (\SATs) problem is \NPc and any problem in $NP$ can thus be reduced to \SATs. 
A single algorithm for one \NPc problem would instantly give a fast algorithm for the others as well. 
We refer the reader to~\cite{Arora2006} for a comprehensive introduction to classical complexity theory.

\begin{definition}[{Reductions, \NP-hardness and \NPcn~\cite{Arora2006}}]
We say that a language $A\subseteq \{0,1\}^*$ is \textit{polynomial-time karp reducible} to a language $B \subseteq \{0,1\}^*$ (denote $A \leq_p B$) if there is a poly-time computable function $f: \{0,1\}^* \rightarrow \{0,1\}^*$ such that for every $x \in \{0,1\}^*$, $x \in A$ if and only if $f(x) \in B$.

\noindent We say that a problem $B$ is \NPh if $A \leq_p B$ for every $A \in \NP$ and $B$ is \NPc if additionally $B \in NP$ holds.

\end{definition}
There are thousands of \NPc problems we do not expect to be solvable in polynomial time.
Whether $\Pt \overset{?}{=} \NP$ is still one of the biggest open questions in mathematics and bountied with one million dollars by the \textit{Clay Mathematical Institute}~\cite{Fortnow2021}. 
Most of the domination problems like \dom, \sdom, \tdom are \NPc.

We do not expect \NPc problems to have a polynomial-time algorithm, but strategies exist to cope with hardness. 
We can either give up the exactness of a solution to possibly find fast \textit{approximation algorithms} or abandon the search for a polynomial-time algorithm in favor of finding good \textit{Exact Exponential (EEA) Algorithms} instead.
A third technique is using additional structural parameters of a specific problem instance and therefore \textbf{restricting the input to special cases}. 
This idea led to the development of \textit{parameterized complexity}.

\section{Parameterized Complexity}\label{cha:param}

Introduced by Downey and Fellows~\cite{Downey1999a}, parameterized complexity extends the classical theory with a framework that allows a more dimensional analysis of computationally hard problems. 
The idea is to extract an arbitrary parameter $k$ and find an algorithm that is only exponential in a function $f(k)$ but only polynomial in the instance size.
$k$ denotes how complex the problem is: 
$k$ can be seen as a measure of the difficulty of a given instance.
If $k$ is small, the problem can still be considered tractable, although the underlying \NPh problem is generally intractable.
All definitions are taken from~\cite{Cygan2015} if not marked otherwise.

\begin{definition}[Parameterized Problem]
    A parameterized problem is a language $L\subseteq\Sigma^*\times \mathbb{N}$ where $\Sigma$ is a finite, fixed alphabet.
    For an instance $(x,k) \in \Sigma^*\times \mathbb{N}$, $k$ is called the \textit{parameter}.

    The \underline{size of an instance} of an instance $(x,k)$ of a parameterized problem is $\abs{(x,k)} = \abs{x} + k$ where the parameter $k$ is encoded in unary by convention.
\end{definition}

\subsection{Fixed-Parameter Tractability}
We say that a problem parameterized by some parameter $k$ is \textit{fixed-parameter tractable (fpt)} if problem instances of size $n$ can be solved in $f(k)n^{\mathcal{O}(1)}$ time for some function $f$ independent of $n$. 
Like the class \Pt can be seen as a notion of \textit{tractability} in classical complexity theory, there is an equivalent in parameterized complexity, which we denote as \FPTl (\FPT):

\begin{cc} [The Class \FPT]{cc:fpt}
    A parameterized problem $L\subseteq\Sigma^*\times\mathbb{N}$ is called \textit{fixed-parameter tractable} if there exists an algorithm A (called a \textit{fixed-parameter algorithm}), a computable function $f:\mathbb{N} \rightarrow \mathbb{N}$ and a constant c such that, given $(x,k) \in \Sigma^* \times \mathbb{N}$, the algorithm $\mathcal{A}$ correctly decides whether $(x,k) \in L$ in time bounded by $f(k) \cdot |(x,k)|^c$. The complexity class containing all fixed-parameter tractable problems is called \FPT.
\end{cc}
\subsection{Kernelization}

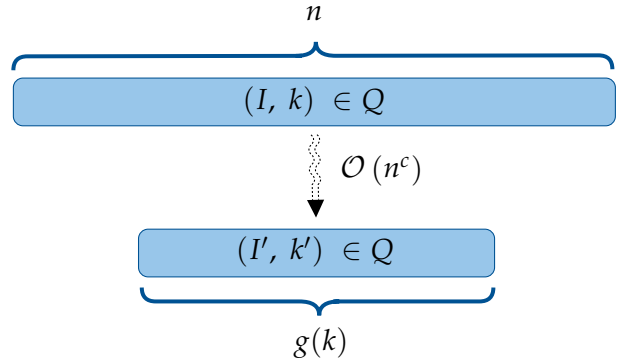
\begin{figure}
    \centering
    \input{fig/mathcha/kernelization.tikz}
    \caption[Idea of kernelization]{\textit{The Idea of Kernelization: Reducing an instance $(I,k) \in Q$ of size $n$ to a smaller instance $(I', k') \in Q$ in polynomial time. 
    The resulting size of the kernel is a function $g(k)$ only dependent on $k$.}
    }
    \label{fig:kernelization}
\end{figure}

A kernelization algorithm is a natural and intuitive way to approach problems and is a preprocessing procedure that simplifies parts of an instance before the actual solving algorithm runs. 
The \cref{fig:kernelization} visualizes this idea.
One can introduce different \textit{reduction rules} that iteratively reduce the instance until we are left with a small kernel. 
The size of this kernel must be merely dependent on the parameter $k$.

\begin{definition}[Kernelization and Reduction Rules]
A \textit{kernelization algorithm} or \textit{kernel} is an algorithm $\mathfrak{A}$ for a parameterized problem $Q$ that given an instance $(I,k)$ of $Q$ runs in polynomial time and returns an equivalent instance $(I', k')$ of $Q$. 
Moreover, we require that $size_{\mathfrak{A}}(k) \leq g(k)$ for some computable function $g:\mathbb{N} \rightarrow \mathbb{N}$.

A \underline{reduction rule} is a function $\phi:\Sigma^* \times \mathbb{N} \rightarrow \Sigma^* \times \mathbb{N}$ that maps an instance $(x,k)$ to an equivalent instance $(x',k')$ such that $\phi$ is computable in time polynomial in $\abs{x}$ and $k$.
A reduction rule is \underline{sound} (or \underline{safe}) if $(I, k) \in Q \Leftrightarrow (I',k') \in Q$.
\end{definition}

We can precisely define the kernel's size after executing a preprocessing algorithm $\mathfrak{A}$.

\begin{definition}[Output Size of a Preprocessing Algorithm] The output size of a preprocessing algorithms $\mathfrak{A}$ is defined as 

    \[\mathrm{size}_{\mathfrak{A}}(k) = \sup\{\abs{I'} + k': (I',k')= \mathfrak{A}(I,k), I \in \Sigma^* \} \]
\end{definition}

\noindent $size_{\mathfrak{A}}$ denotes the largest size of any instance $I$ after $\mathfrak{A}$ has been applied.
If we bound $\mathrm{size}_{\mathfrak{A}}$ by a polynomial in $k$, we say that the problem admits a \textbf{polynomial kernel} or a \textbf{linear} kernel analogously.

The following \cref{lemma:fptiskernel} shows the relation between the complexity class \FPT and a kernelization algorithm. 
If we find a kernelization algorithm $\mathfrak{A}$ for a (decidable) problem $P$, we immediately obtain an fpt algorithm.
First, we will run $\mathfrak{A}$ on the given instance in polynomial time and then solve the kernel with an exponential time algorithm.
The total running time is of order $\mathcal{O}(g(f(k)) \cdot \mathrm{poly}(n))$ and hence, fpt.
Surprisingly, also the converse is true:

\begin{lemma}\label{lemma:fptiskernel}
    A parameterized problem $Q$ is \FPT if and only if it admits a kernelization algorithm.
\end{lemma}

 We will use this property in \cref{ch:linkern} to construct a kernel for \psdom and showing membership in \FPT.

\subsection{Reductions and Parameterized Intractability}

It is natural to ask whether all (hard) problems are also fixed-parameter tractable.
The answer is no, and parameterized complexity has another tool to show that a problem is unlikely to be in \FPT.
The idea is to transfer the concepts of \NP-hardness from \Cref{ch:npc} and reductions from the classical setting to the parameterized world.
This raises the need for a new type of reduction that ensures that a reduced instance $(I', k')$ is not only created in fpt time, but the new parameter $k'$ depends only on the size of the parameter in the original instance.

There exists a whole hierarchy of classes $\FPT \subseteq \WONE \subseteq \WTWO \subseteq ... \subseteq \Wt \subseteq ...$, which is known as the \WHIERARCHY.
It is strongly believed that $\FPT \subsetneq \Wt$ and therefore, we do not expect the existence of an algorithm solving any \Wt-hard problem in fpt time.

\begin{definition}[Parameterized Reduction] Let $A,B\subseteq \Sigma^*\times\mathbb{N}$ two parameterized problems. A \textit{parameter preserving reduction} from $A$ to $B$ is an algorithm that, given an instance $(x,k)$ of $A$, outputs an instance $(x', k')$ of $B$ such that:
    \begin{itemize}
        \item $(x,k)$ is a \textcolor{darkgray}{\textbf{yes instance}} of A \textbf{iff} $(x',k')$ is a \textcolor{darkgray}{\textbf{yes instance}} of B,
        \item $k' \leq g(k)$ for some computable function $g$, and
        \item runs in fpt-time $f(k)\cdot |x|^{\mathcal{O}(1)}$ for some computable function f.
    \end{itemize}
\end{definition}

\cref{lem:cfptr,lem:trans} stated in~\cite{Cygan2015} show that this definition ensures that reductions are transitive and closed under fpt reductions.

\begin{lemma}[Closed Under fpt-reductions]\label{lem:cfptr}
    If there is a parameterized reduction from $A$ to $B$ and $B \in \FPT$, then $A \in \FPT$.
\end{lemma}

\begin{lemma}[Transitivity] \label{lem:trans}
    If there are parameterized reductions from $A$ to $B$ and from $B$ to $C$, then there is a parameterized reduction from $A$ to $C$.
\end{lemma}

If there exists a parameterized reduction transforming a \Wt-hard problem $A$ to another problem $B$, then $B$ is \Wt-hard as well.
We know that the \is problem is \WONE-complete and \dom is \WTWO-complete~\cite{Downey1995,Cygan2015}.
A more profound introduction is not required for this work, and we refer the interested reader to~\cite{Cygan2015, Fomin2019} for more details.


%% file: fig/mathcha/kernelization.tikz
\begin{tikzpicture}[x=0.75pt,y=0.75pt,yscale=-1,xscale=1]
\tikzset{every picture/.style={line width=0.75pt}} 

\draw  [color={rgb, 255:red, 0; green, 82; blue, 147 }  ,draw opacity=1 ][fill={rgb, 255:red, 152; green, 198; blue, 234 }  ,fill opacity=1 ] (129.8,74.8) .. controls (129.8,72.15) and (131.95,70) .. (134.6,70) -- (425.5,70) .. controls (428.15,70) and (430.3,72.15) .. (430.3,74.8) -- (430.3,89.2) .. controls (430.3,91.85) and (428.15,94) .. (425.5,94) -- (134.6,94) .. controls (131.95,94) and (129.8,91.85) .. (129.8,89.2) -- cycle ;
\draw  [dash pattern={on 0.75pt off 0.75pt}]  (281.37,98.46) .. controls (283.08,100.09) and (283.12,101.75) .. (281.49,103.46) .. controls (279.86,105.17) and (279.9,106.83) .. (281.61,108.46) .. controls (283.32,110.09) and (283.36,111.75) .. (281.73,113.46) .. controls (280.1,115.17) and (280.14,116.83) .. (281.85,118.46) .. controls (283.56,120.09) and (283.6,121.75) .. (281.97,123.46) -- (282.08,128.13) -- (282.15,131.13)(278.37,98.54) .. controls (280.08,100.16) and (280.12,101.82) .. (278.49,103.53) .. controls (276.86,105.24) and (276.9,106.9) .. (278.61,108.53) .. controls (280.32,110.16) and (280.36,111.82) .. (278.73,113.53) .. controls (277.1,115.24) and (277.14,116.9) .. (278.85,118.53) .. controls (280.56,120.16) and (280.6,121.82) .. (278.97,123.53) -- (279.08,128.21) -- (279.15,131.21) ;
\draw [shift={(280.87,140.17)}, rotate = 268.63] [fill={rgb, 255:red, 0; green, 0; blue, 0 }  ][line width=0.08]  [draw opacity=0] (8.93,-4.29) -- (0,0) -- (8.93,4.29) -- cycle    ;
\draw  [color={rgb, 255:red, 0; green, 82; blue, 147 }  ,draw opacity=1 ][fill={rgb, 255:red, 152; green, 198; blue, 234 }  ,fill opacity=1 ] (192.27,150.8) .. controls (192.27,148.15) and (194.42,146) .. (197.07,146) -- (365.2,146) .. controls (367.85,146) and (370,148.15) .. (370,150.8) -- (370,165.2) .. controls (370,167.85) and (367.85,170) .. (365.2,170) -- (197.07,170) .. controls (194.42,170) and (192.27,167.85) .. (192.27,165.2) -- cycle ;
\draw  [color={rgb, 255:red, 0; green, 82; blue, 147 }  ,draw opacity=1 ][line width=1.5]  (428.8,65.55) .. controls (428.79,60.88) and (426.46,58.55) .. (421.79,58.56) -- (290.81,58.79) .. controls (284.14,58.8) and (280.81,56.48) .. (280.8,51.81) .. controls (280.81,56.48) and (277.48,58.82) .. (270.81,58.83)(273.81,58.82) -- (136.79,59.06) .. controls (132.12,59.07) and (129.79,61.4) .. (129.8,66.07) ;
\draw  [color={rgb, 255:red, 0; green, 82; blue, 147 }  ,draw opacity=1 ][line width=1.5]  (193.9,175.35) .. controls (193.9,180.02) and (196.23,182.35) .. (200.9,182.35) -- (272.47,182.35) .. controls (279.14,182.35) and (282.47,184.68) .. (282.47,189.35) .. controls (282.47,184.68) and (285.8,182.35) .. (292.47,182.35)(289.47,182.35) -- (362.4,182.35) .. controls (367.07,182.35) and (369.4,180.02) .. (369.4,175.35) ;

\draw (242.9,72.9) node [anchor=north west][inner sep=0.75pt]    {$( I,\ k) \ \in Q$};
\draw (291.67,107.4) node [anchor=north west][inner sep=0.75pt]    {$\mathcal{O}\left( n^{c}\right)$};
\draw (239.5,148.4) node [anchor=north west][inner sep=0.75pt]    {$( I',\ k') \ \in Q$};
\draw (268.5,193.9) node [anchor=north west][inner sep=0.75pt]    {$g( k)$};
\draw (274.5,32.4) node [anchor=north west][inner sep=0.75pt]    {$n$};

\end{tikzpicture}

%% file: fig/tikz/ds-examples.tikz
\begin{tikzpicture}
	\begin{pgfonlayer}{nodelayer}
		\node [style=BLACK] (0) at (-14.5, 0) {};
		\node [style=DOM] (1) at (-12.5, 2) {};
		\node [style=BLACK] (2) at (-9.5, 2) {};
		\node [style=BLACK] (3) at (-7.5, 0) {};
		\node [style=DOM] (4) at (-8.5, -2) {};
		\node [style=BLACK] (5) at (-11.5, -3) {};
		\node [style=BLACK] (6) at (-4.5, 0) {};
		\node [style=DOM] (7) at (-2.5, 2) {};
		\node [style=BLACK] (8) at (0.5, 2) {};
		\node [style=DOM] (9) at (2.5, 0) {};
		\node [style=DOM] (10) at (1.5, -2) {};
		\node [style=BLACK] (11) at (-1.5, -3) {};
		\node [style=BLACK] (12) at (5.5, 0) {};
		\node [style=DOM] (13) at (7.5, 2) {};
		\node [style=DOM] (14) at (10.5, 2) {};
		\node [style=DOM] (15) at (12.5, 0) {};
		\node [style=DOM] (16) at (11.5, -2) {};
		\node [style=BLACK] (17) at (8.5, -3) {};
		\node [style=none] (18) at (-6.25, 4) {};
		\node [style=none] (19) at (-6.25, -4) {};
		\node [style=none] (20) at (4.25, 4) {};
		\node [style=none] (21) at (4.25, -4) {};
		\node [style=none] (22) at (-11.75, 5.25) {$\dom$};
		\node [style=none] (23) at (-0.75, 5.25) {$\sdom$};
		\node [style=none] (24) at (9.5, 5.25) {$\tdom$};
		\node [style=none] (26) at (-13.25, 3) {$d_1$};
		\node [style=none] (27) at (-3.25, 3) {$d_1$};
		\node [style=none] (28) at (6.75, 3) {$d_1$};
		\node [style=none] (29) at (-9.25, -1) {$d_2$};
		\node [style=none] (30) at (0.75, -1) {$d_2$};
		\node [style=none] (31) at (10.75, -1) {$d_2$};
		\node [style=none] (32) at (12.75, 1) {$d_3$};
		\node [style=none] (33) at (2.75, 1) {$d_3$};
		\node [style=none] (34) at (10.75, 3) {$d_4$};
	\end{pgfonlayer}
	\begin{pgfonlayer}{edgelayer}
		\draw (1) to (0);
		\draw (1) to (2);
		\draw (3) to (2);
		\draw (3) to (4);
		\draw (4) to (5);
		\draw (7) to (6);
		\draw (7) to (8);
		\draw (9) to (8);
		\draw (9) to (10);
		\draw (10) to (11);
		\draw (13) to (12);
		\draw (13) to (14);
		\draw (15) to (14);
		\draw (15) to (16);
		\draw (16) to (17);
		\draw [style=EDGEDASHED] (18.center) to (19.center);
		\draw [style=EDGEDASHED] (21.center) to (20.center);
	\end{pgfonlayer}
\end{tikzpicture}

%% file: fig/mathcha/classes.tikz
\tikzset{every picture/.style={line width=0.75pt}} 

\begin{tikzpicture}[x=0.75pt,y=0.75pt,yscale=-1,xscale=1]

\draw [fill={rgb, 255:red, 0; green, 101; blue, 189 }  ,fill opacity=1 ]   (-114.6,212.17) .. controls (-112.93,210.51) and (-111.26,210.52) .. (-109.6,212.19) .. controls (-107.93,213.86) and (-106.27,213.86) .. (-104.6,212.2) .. controls (-102.93,210.54) and (-101.27,210.54) .. (-99.6,212.21) .. controls (-97.94,213.88) and (-96.27,213.89) .. (-94.6,212.23) .. controls (-92.93,210.57) and (-91.27,210.57) .. (-89.6,212.24) .. controls (-87.93,213.91) and (-86.27,213.91) .. (-84.6,212.25) .. controls (-82.93,210.59) and (-81.26,210.6) .. (-79.6,212.27) .. controls (-77.93,213.94) and (-76.27,213.94) .. (-74.6,212.28) .. controls (-72.93,210.62) and (-71.27,210.62) .. (-69.6,212.29) .. controls (-67.93,213.96) and (-66.27,213.96) .. (-64.6,212.3) .. controls (-62.93,210.64) and (-61.26,210.65) .. (-59.6,212.32) .. controls (-57.93,213.99) and (-56.27,213.99) .. (-54.6,212.33) .. controls (-52.93,210.67) and (-51.27,210.67) .. (-49.6,212.34) .. controls (-47.94,214.01) and (-46.27,214.02) .. (-44.6,212.36) .. controls (-42.93,210.7) and (-41.27,210.7) .. (-39.6,212.37) .. controls (-37.93,214.04) and (-36.27,214.04) .. (-34.6,212.38) .. controls (-32.93,210.72) and (-31.27,210.72) .. (-29.6,212.39) .. controls (-27.94,214.06) and (-26.27,214.07) .. (-24.6,212.41) .. controls (-22.93,210.75) and (-21.27,210.75) .. (-19.6,212.42) .. controls (-17.93,214.09) and (-16.27,214.09) .. (-14.6,212.43) .. controls (-12.93,210.77) and (-11.26,210.78) .. (-9.6,212.45) .. controls (-7.93,214.12) and (-6.27,214.12) .. (-4.6,212.46) .. controls (-2.93,210.8) and (-1.27,210.8) .. (0.4,212.47) .. controls (2.07,214.14) and (3.73,214.14) .. (5.4,212.48) .. controls (7.07,210.82) and (8.74,210.83) .. (10.4,212.5) .. controls (12.07,214.17) and (13.73,214.17) .. (15.4,212.51) .. controls (17.07,210.85) and (18.73,210.85) .. (20.4,212.52) .. controls (22.06,214.19) and (23.73,214.2) .. (25.4,212.54) .. controls (27.07,210.88) and (28.73,210.88) .. (30.4,212.55) .. controls (32.07,214.22) and (33.73,214.22) .. (35.4,212.56) .. controls (37.07,210.9) and (38.73,210.9) .. (40.4,212.57) .. controls (42.06,214.24) and (43.73,214.25) .. (45.4,212.59) .. controls (47.07,210.93) and (48.73,210.93) .. (50.4,212.6) .. controls (52.07,214.27) and (53.73,214.27) .. (55.4,212.61) .. controls (57.07,210.95) and (58.74,210.96) .. (60.4,212.63) .. controls (62.07,214.3) and (63.73,214.3) .. (65.4,212.64) .. controls (67.07,210.98) and (68.73,210.98) .. (70.4,212.65) .. controls (72.07,214.32) and (73.73,214.32) .. (75.4,212.66) .. controls (77.07,211) and (78.74,211.01) .. (80.4,212.68) .. controls (82.07,214.35) and (83.73,214.35) .. (85.4,212.69) .. controls (87.07,211.03) and (88.73,211.03) .. (90.4,212.7) .. controls (92.06,214.37) and (93.73,214.38) .. (95.4,212.72) .. controls (97.07,211.06) and (98.73,211.06) .. (100.4,212.73) .. controls (102.07,214.4) and (103.73,214.4) .. (105.4,212.74) .. controls (107.07,211.08) and (108.73,211.08) .. (110.4,212.75) .. controls (112.06,214.42) and (113.73,214.43) .. (115.4,212.77) .. controls (117.07,211.11) and (118.73,211.11) .. (120.4,212.78) .. controls (122.07,214.45) and (123.73,214.45) .. (125.4,212.79) .. controls (127.07,211.13) and (128.74,211.14) .. (130.4,212.81) .. controls (132.07,214.48) and (133.73,214.48) .. (135.4,212.82) .. controls (137.07,211.16) and (138.73,211.16) .. (140.4,212.83) .. controls (142.06,214.5) and (143.73,214.51) .. (145.4,212.85) .. controls (147.07,211.19) and (148.73,211.19) .. (150.4,212.86) .. controls (152.07,214.53) and (153.73,214.53) .. (155.4,212.87) .. controls (157.07,211.21) and (158.73,211.21) .. (160.4,212.88) .. controls (162.06,214.55) and (163.73,214.56) .. (165.4,212.9) .. controls (167.07,211.24) and (168.73,211.24) .. (170.4,212.91) .. controls (172.07,214.58) and (173.73,214.58) .. (175.4,212.92) .. controls (177.07,211.26) and (178.74,211.27) .. (180.4,212.94) .. controls (182.07,214.61) and (183.73,214.61) .. (185.4,212.95) .. controls (187.07,211.29) and (188.73,211.29) .. (190.4,212.96) .. controls (192.07,214.63) and (193.73,214.63) .. (195.4,212.97) .. controls (197.07,211.31) and (198.74,211.32) .. (200.4,212.99) .. controls (202.07,214.66) and (203.73,214.66) .. (205.4,213) .. controls (207.07,211.34) and (208.73,211.34) .. (210.4,213.01) .. controls (212.06,214.68) and (213.73,214.69) .. (215.4,213.03) .. controls (217.07,211.37) and (218.73,211.37) .. (220.4,213.04) .. controls (222.07,214.71) and (223.73,214.71) .. (225.4,213.05) .. controls (227.07,211.39) and (228.73,211.39) .. (230.4,213.06) .. controls (232.06,214.73) and (233.73,214.74) .. (235.4,213.08) .. controls (237.07,211.42) and (238.73,211.42) .. (240.4,213.09) .. controls (242.07,214.76) and (243.73,214.76) .. (245.4,213.1) .. controls (247.07,211.44) and (248.74,211.45) .. (250.4,213.12) .. controls (252.07,214.79) and (253.73,214.79) .. (255.4,213.13) .. controls (257.07,211.47) and (258.73,211.47) .. (260.4,213.14) .. controls (262.07,214.81) and (263.73,214.81) .. (265.4,213.15) .. controls (267.07,211.49) and (268.74,211.5) .. (270.4,213.17) .. controls (272.07,214.84) and (273.73,214.84) .. (275.4,213.18) .. controls (277.07,211.52) and (278.73,211.52) .. (280.4,213.19) .. controls (282.06,214.86) and (283.73,214.87) .. (285.4,213.21) .. controls (287.07,211.55) and (288.73,211.55) .. (290.4,213.22) .. controls (292.07,214.89) and (293.73,214.89) .. (295.4,213.23) .. controls (297.07,211.57) and (298.73,211.57) .. (300.4,213.24) .. controls (302.06,214.91) and (303.73,214.92) .. (305.4,213.26) .. controls (307.07,211.6) and (308.73,211.6) .. (310.4,213.27) .. controls (312.07,214.94) and (313.73,214.94) .. (315.4,213.28) .. controls (317.07,211.62) and (318.74,211.63) .. (320.4,213.3) .. controls (322.07,214.97) and (323.73,214.97) .. (325.4,213.31) .. controls (327.07,211.65) and (328.73,211.65) .. (330.4,213.32) .. controls (332.07,214.99) and (333.73,214.99) .. (335.4,213.33) .. controls (337.07,211.67) and (338.74,211.68) .. (340.4,213.35) .. controls (342.07,215.02) and (343.73,215.02) .. (345.4,213.36) .. controls (347.07,211.7) and (348.73,211.7) .. (350.4,213.37) .. controls (352.06,215.04) and (353.73,215.05) .. (355.4,213.39) .. controls (357.07,211.73) and (358.73,211.73) .. (360.4,213.4) .. controls (362.07,215.07) and (363.73,215.07) .. (365.4,213.41) .. controls (367.07,211.75) and (368.73,211.75) .. (370.4,213.42) .. controls (372.06,215.09) and (373.73,215.1) .. (375.4,213.44) .. controls (377.07,211.78) and (378.73,211.78) .. (380.4,213.45) .. controls (382.07,215.12) and (383.73,215.12) .. (385.4,213.46) .. controls (387.07,211.8) and (388.74,211.81) .. (390.4,213.48) .. controls (392.07,215.15) and (393.73,215.15) .. (395.4,213.49) .. controls (397.07,211.83) and (398.73,211.83) .. (400.4,213.5) .. controls (402.06,215.17) and (403.73,215.18) .. (405.4,213.52) .. controls (407.07,211.86) and (408.73,211.86) .. (410.4,213.53) .. controls (412.07,215.2) and (413.73,215.2) .. (415.4,213.54) .. controls (417.07,211.88) and (418.73,211.88) .. (420.4,213.55) .. controls (422.06,215.22) and (423.73,215.23) .. (425.4,213.57) .. controls (427.07,211.91) and (428.73,211.91) .. (430.4,213.58) .. controls (432.07,215.25) and (433.73,215.25) .. (435.4,213.59) .. controls (437.07,211.93) and (438.74,211.94) .. (440.4,213.61) .. controls (442.07,215.28) and (443.73,215.28) .. (445.4,213.62) .. controls (447.07,211.96) and (448.73,211.96) .. (450.4,213.63) .. controls (452.07,215.3) and (453.73,215.3) .. (455.4,213.64) .. controls (457.07,211.98) and (458.74,211.99) .. (460.4,213.66) .. controls (462.07,215.33) and (463.73,215.33) .. (465.4,213.67) .. controls (467.07,212.01) and (468.73,212.01) .. (470.4,213.68) .. controls (472.06,215.35) and (473.73,215.36) .. (475.4,213.7) .. controls (477.07,212.04) and (478.73,212.04) .. (480.4,213.71) .. controls (482.07,215.38) and (483.73,215.38) .. (485.4,213.72) .. controls (487.07,212.06) and (488.73,212.06) .. (490.4,213.73) .. controls (492.06,215.4) and (493.73,215.41) .. (495.4,213.75) .. controls (497.07,212.09) and (498.73,212.09) .. (500.4,213.76) .. controls (502.07,215.43) and (503.73,215.43) .. (505.4,213.77) .. controls (507.07,212.11) and (508.74,212.12) .. (510.4,213.79) .. controls (512.07,215.46) and (513.73,215.46) .. (515.4,213.8) .. controls (517.07,212.14) and (518.73,212.14) .. (520.4,213.81) .. controls (522.07,215.48) and (523.73,215.48) .. (525.4,213.82) .. controls (527.07,212.16) and (528.74,212.17) .. (530.4,213.84) .. controls (532.07,215.51) and (533.73,215.51) .. (535.4,213.85) .. controls (537.07,212.19) and (538.73,212.19) .. (540.4,213.86) .. controls (542.06,215.53) and (543.73,215.54) .. (545.4,213.88) .. controls (547.07,212.22) and (548.73,212.22) .. (550.4,213.89) .. controls (552.07,215.56) and (553.73,215.56) .. (555.4,213.9) .. controls (557.07,212.24) and (558.73,212.24) .. (560.4,213.91) .. controls (562.06,215.58) and (563.73,215.59) .. (565.4,213.93) .. controls (567.07,212.27) and (568.73,212.27) .. (570.4,213.94) .. controls (572.07,215.61) and (573.73,215.61) .. (575.4,213.95) .. controls (577.07,212.29) and (578.74,212.3) .. (580.4,213.97) .. controls (582.07,215.64) and (583.73,215.64) .. (585.4,213.98) .. controls (587.07,212.32) and (588.73,212.32) .. (590.4,213.99) .. controls (592.07,215.66) and (593.73,215.66) .. (595.4,214) .. controls (597.07,212.34) and (598.74,212.35) .. (600.4,214.02) .. controls (602.07,215.69) and (603.73,215.69) .. (605.4,214.03) .. controls (607.07,212.37) and (608.73,212.37) .. (610.4,214.04) .. controls (612.06,215.71) and (613.73,215.72) .. (615.4,214.06) .. controls (617.07,212.4) and (618.73,212.4) .. (620.4,214.07) .. controls (622.07,215.74) and (623.73,215.74) .. (625.4,214.08) .. controls (627.07,212.42) and (628.74,212.43) .. (630.4,214.1) .. controls (632.07,215.77) and (633.73,215.77) .. (635.4,214.11) .. controls (637.07,212.45) and (638.73,212.45) .. (640.4,214.12) .. controls (642.07,215.79) and (643.73,215.79) .. (645.4,214.13) .. controls (647.07,212.47) and (648.74,212.48) .. (650.4,214.15) .. controls (652.07,215.82) and (653.73,215.82) .. (655.4,214.16) .. controls (657.07,212.5) and (658.73,212.5) .. (660.4,214.17) .. controls (662.06,215.84) and (663.73,215.85) .. (665.4,214.19) .. controls (667.07,212.53) and (668.73,212.53) .. (670.4,214.2) .. controls (672.07,215.87) and (673.73,215.87) .. (675.4,214.21) .. controls (677.07,212.55) and (678.73,212.55) .. (680.4,214.22) .. controls (682.06,215.89) and (683.73,215.9) .. (685.4,214.24) .. controls (687.07,212.58) and (688.73,212.58) .. (690.4,214.25) .. controls (692.07,215.92) and (693.73,215.92) .. (695.4,214.26) .. controls (697.07,212.6) and (698.74,212.61) .. (700.4,214.28) .. controls (702.07,215.95) and (703.73,215.95) .. (705.4,214.29) .. controls (707.07,212.63) and (708.73,212.63) .. (710.4,214.3) .. controls (712.07,215.97) and (713.73,215.97) .. (715.4,214.31) .. controls (717.07,212.65) and (718.74,212.66) .. (720.4,214.33) .. controls (722.07,216) and (723.73,216) .. (725.4,214.34) .. controls (727.07,212.68) and (728.73,212.68) .. (730.4,214.35) .. controls (732.06,216.02) and (733.73,216.03) .. (735.4,214.37) .. controls (737.07,212.71) and (738.73,212.71) .. (740.4,214.38) .. controls (742.07,216.05) and (743.73,216.05) .. (745.4,214.39) .. controls (747.07,212.73) and (748.73,212.73) .. (750.4,214.4) .. controls (752.06,216.07) and (753.73,216.08) .. (755.4,214.42) .. controls (757.07,212.76) and (758.73,212.76) .. (760.4,214.43) .. controls (762.07,216.1) and (763.73,216.1) .. (765.4,214.44) .. controls (767.07,212.78) and (768.74,212.79) .. (770.4,214.46) .. controls (772.07,216.13) and (773.73,216.13) .. (775.4,214.47) .. controls (777.07,212.81) and (778.73,212.81) .. (780.4,214.48) .. controls (782.07,216.15) and (783.73,216.15) .. (785.4,214.49) .. controls (787.07,212.83) and (788.74,212.84) .. (790.4,214.51) .. controls (792.07,216.18) and (793.73,216.18) .. (795.4,214.52) .. controls (797.07,212.86) and (798.73,212.86) .. (800.4,214.53) .. controls (802.06,216.2) and (803.73,216.21) .. (805.4,214.55) .. controls (807.07,212.89) and (808.73,212.89) .. (810.4,214.56) .. controls (812.07,216.23) and (813.73,216.23) .. (815.4,214.57) -- (816.6,214.57) -- (816.6,214.57) ;

\draw  [fill={rgb, 255:red, 233; green, 17; blue, 17 }  ,fill opacity=0.69 ]  (228.33,-33.4) -- (332.33,-33.4) -- (332.33,-4.4) -- (228.33,-4.4) -- cycle  ;
\draw (280.33,-18.9) node  [font=\small] [align=left] {\begin{minipage}[lt]{68pt}\setlength\topsep{0pt}
\begin{center}
\textbf{perfect}
\end{center}

\end{minipage}};
\draw    (90.67,-12.58) -- (194.67,-12.58) -- (194.67,36.42) -- (90.67,36.42) -- cycle  ;
\draw (142.67,11.92) node  [font=\small] [align=left] {\begin{minipage}[lt]{68pt}\setlength\topsep{0pt}
\begin{center}
Line Graph of Bipartite
\end{center}

\end{minipage}};
\draw  [fill={rgb, 255:red, 233; green, 17; blue, 17 }  ,fill opacity=0.69 ]  (103.67,59.27) -- (207.67,59.27) -- (207.67,88.27) -- (103.67,88.27) -- cycle  ;
\draw (155.67,73.77) node  [font=\small] [align=left] {\begin{minipage}[lt]{68pt}\setlength\topsep{0pt}
\begin{center}
\textbf{comparability}
\end{center}

\end{minipage}};
\draw    (195.2,159.27) -- (276.2,159.27) -- (276.2,188.27) -- (195.2,188.27) -- cycle  ;
\draw (235.7,173.77) node  [font=\small] [align=left] {\begin{minipage}[lt]{52.63pt}\setlength\topsep{0pt}
\begin{center}
tolerance ?
\end{center}

\end{minipage}};
\draw  [fill={rgb, 255:red, 184; green, 233; blue, 134 }  ,fill opacity=1 ]  (382.2,220.93) -- (445.2,220.93) -- (445.2,249.93) -- (382.2,249.93) -- cycle  ;
\draw (413.7,235.43) node  [font=\small] [align=left] {\begin{minipage}[lt]{40.39pt}\setlength\topsep{0pt}
\begin{center}
AT-free
\end{center}

\end{minipage}};
\draw  [fill={rgb, 255:red, 233; green, 17; blue, 17 }  ,fill opacity=0.69 ]  (293.53,45.93) -- (357.53,45.93) -- (357.53,74.93) -- (293.53,74.93) -- cycle  ;
\draw (325.53,60.43) node  [font=\small] [align=left] {\begin{minipage}[lt]{40.62pt}\setlength\topsep{0pt}
\begin{center}
\textbf{chordal}
\end{center}

\end{minipage}};
\draw  [fill={rgb, 255:red, 233; green, 17; blue, 17 }  ,fill opacity=0.69 ]  (296.67,144.93) -- (400.67,144.93) -- (400.67,173.93) -- (296.67,173.93) -- cycle  ;
\draw (348.67,159.43) node  [font=\small] [align=left] {\begin{minipage}[lt]{68pt}\setlength\topsep{0pt}
\begin{center}
\textbf{split}
\end{center}

\end{minipage}};
\draw  [fill={rgb, 255:red, 184; green, 233; blue, 134 }  ,fill opacity=1 ]  (608.67,549.93) -- (712.67,549.93) -- (712.67,578.93) -- (608.67,578.93) -- cycle  ;
\draw (660.67,564.43) node  [font=\small] [align=left] {\begin{minipage}[lt]{68pt}\setlength\topsep{0pt}
\begin{center}
$\displaystyle k$-tree, fixed $\displaystyle k$
\end{center}

\end{minipage}};
\draw  [color={rgb, 255:red, 0; green, 0; blue, 0 }  ,draw opacity=1 ][fill={rgb, 255:red, 184; green, 233; blue, 134 }  ,fill opacity=1 ]  (207.17,241.27) -- (331.17,241.27) -- (331.17,270.27) -- (207.17,270.27) -- cycle  ;
\draw (269.17,255.77) node  [font=\small] [align=left] {\begin{minipage}[lt]{81.83pt}\setlength\topsep{0pt}
\begin{center}
cocomparability
\end{center}

\end{minipage}};
\draw  [fill={rgb, 255:red, 184; green, 233; blue, 134 }  ,fill opacity=1 ]  (244.67,554.27) -- (348.67,554.27) -- (348.67,583.27) -- (244.67,583.27) -- cycle  ;
\draw (296.67,568.77) node  [font=\small] [align=left] {\begin{minipage}[lt]{68pt}\setlength\topsep{0pt}
\begin{center}
tree
\end{center}

\end{minipage}};
\draw    (-77.8,-29.73) -- (-21.8,-29.73) -- (-21.8,-0.73) -- (-77.8,-0.73) -- cycle  ;
\draw (-49.8,-15.23) node  [font=\small] [align=left] {\begin{minipage}[lt]{35.63pt}\setlength\topsep{0pt}
\begin{center}
circle
\end{center}

\end{minipage}};
\draw    (10.8,150.42) -- (83.8,150.42) -- (83.8,203.42) -- (10.8,203.42) -- cycle  ;
\draw (47.3,176.92) node  [font=\small] [align=left] {\begin{minipage}[lt]{47.1pt}\setlength\topsep{0pt}
\begin{center}
chordal Bipartite
\end{center}

\end{minipage}};
\draw  [fill={rgb, 255:red, 233; green, 17; blue, 17 }  ,fill opacity=0.69 ]  (28.3,106.47) -- (115.3,106.47) -- (115.3,135.47) -- (28.3,135.47) -- cycle  ;
\draw (71.8,120.97) node  [font=\small] [align=left] {\begin{minipage}[lt]{56.39pt}\setlength\topsep{0pt}
\begin{center}
\textbf{bipartite}
\end{center}

\end{minipage}};
\draw  [fill={rgb, 255:red, 184; green, 233; blue, 134 }  ,fill opacity=1 ]  (366.87,262.15) -- (494.87,262.15) -- (494.87,294.15) -- (366.87,294.15) -- cycle  ;
\draw (430.87,278.15) node  [font=\small] [align=left] {\begin{minipage}[lt]{84.59pt}\setlength\topsep{0pt}
\begin{center}
Circular $\displaystyle k$-trapezoid
\end{center}

\end{minipage}};
\draw  [fill={rgb, 255:red, 184; green, 233; blue, 134 }  ,fill opacity=1 ]  (211.6,294.24) -- (301.6,294.24) -- (301.6,324.24) -- (211.6,324.24) -- cycle  ;
\draw (256.6,309.24) node  [font=\small] [align=left] {\begin{minipage}[lt]{58.21pt}\setlength\topsep{0pt}
\begin{center}
$\displaystyle k$-trapezoid
\end{center}

\end{minipage}};
\draw  [fill={rgb, 255:red, 184; green, 233; blue, 134 }  ,fill opacity=1 ]  (354.53,310.45) -- (435.53,310.45) -- (435.53,334.45) -- (354.53,334.45) -- cycle  ;
\draw (395.03,322.45) node  [font=\small] [align=left] {\begin{minipage}[lt]{52.18pt}\setlength\topsep{0pt}
\begin{center}
trapezoid
\end{center}

\end{minipage}};
\draw  [fill={rgb, 255:red, 184; green, 233; blue, 134 }  ,fill opacity=1 ]  (228.2,356.05) -- (350.2,356.05) -- (350.2,393.05) -- (228.2,393.05) -- cycle  ;
\draw (289.2,374.55) node  [font=\small] [align=left] {\begin{minipage}[lt]{80.51pt}\setlength\topsep{0pt}
\begin{center}
bounded tolerance
\end{center}

\end{minipage}};
\draw  [fill={rgb, 255:red, 184; green, 233; blue, 134 }  ,fill opacity=1 ]  (84.37,401.11) -- (167.37,401.11) -- (167.37,431.11) -- (84.37,431.11) -- cycle  ;
\draw (125.87,416.11) node  [font=\small] [align=left] {\begin{minipage}[lt]{53.77pt}\setlength\topsep{0pt}
\begin{center}
permutation
\end{center}

\end{minipage}};
\draw  [fill={rgb, 255:red, 184; green, 233; blue, 134 }  ,fill opacity=1 ]  (31.37,486.49) -- (175.37,486.49) -- (175.37,514.49) -- (31.37,514.49) -- cycle  ;
\draw (103.37,500.49) node  [font=\small] [align=left] {\begin{minipage}[lt]{95.25pt}\setlength\topsep{0pt}
\begin{center}
bipartite permutation
\end{center}

\end{minipage}};
\draw  [fill={rgb, 255:red, 184; green, 233; blue, 134 }  ,fill opacity=1 ]  (-1.53,428.57) -- (62.47,428.57) -- (62.47,455.57) -- (-1.53,455.57) -- cycle  ;
\draw (30.47,442.07) node  [font=\small] [align=left] {\begin{minipage}[lt]{40.53pt}\setlength\topsep{0pt}
\begin{center}
convex
\end{center}

\end{minipage}};
\draw  [fill={rgb, 255:red, 184; green, 233; blue, 134 }  ,fill opacity=1 ]  (516.2,366.93) -- (583.2,366.93) -- (583.2,395.93) -- (516.2,395.93) -- cycle  ;
\draw (549.7,381.43) node  [font=\small] [align=left] {\begin{minipage}[lt]{43.11pt}\setlength\topsep{0pt}
\begin{center}
interval
\end{center}

\end{minipage}};
\draw  [fill={rgb, 255:red, 184; green, 233; blue, 134 }  ,fill opacity=1 ]  (417.53,513.27) -- (506.53,513.27) -- (506.53,542.27) -- (417.53,542.27) -- cycle  ;
\draw (462.03,527.77) node  [font=\small] [align=left] {\begin{minipage}[lt]{57.62pt}\setlength\topsep{0pt}
\begin{center}
threshold
\end{center}

\end{minipage}};
\draw  [fill={rgb, 255:red, 184; green, 233; blue, 134 }  ,fill opacity=1 ]  (468.53,416.6) -- (571.53,416.6) -- (571.53,445.6) -- (468.53,445.6) -- cycle  ;
\draw (520.03,431.1) node  [font=\small] [align=left] {\begin{minipage}[lt]{67.14pt}\setlength\topsep{0pt}
\begin{center}
Dilworth $\displaystyle k$
\end{center}

\end{minipage}};
\draw  [fill={rgb, 255:red, 184; green, 233; blue, 134 }  ,fill opacity=1 ]  (284.53,506.93) -- (369.53,506.93) -- (369.53,535.93) -- (284.53,535.93) -- cycle  ;
\draw (327.03,521.43) node  [font=\small] [align=left] {\begin{minipage}[lt]{55.35pt}\setlength\topsep{0pt}
\begin{center}
cograph
\end{center}

\end{minipage}};
\draw  [fill={rgb, 255:red, 184; green, 233; blue, 134 }  ,fill opacity=1 ]  (481.4,309.27) -- (564.4,309.27) -- (564.4,338.27) -- (481.4,338.27) -- cycle  ;
\draw (522.9,323.77) node  [font=\small] [align=left] {\begin{minipage}[lt]{53.72pt}\setlength\topsep{0pt}
\begin{center}
circular arc
\end{center}

\end{minipage}};
\draw  [fill={rgb, 255:red, 255; green, 255; blue, 255 }  ,fill opacity=1 ]  (516.37,24.22) -- (630.37,24.22) -- (630.37,58.22) -- (516.37,58.22) -- cycle  ;
\draw (573.37,41.22) node  [font=\small] [align=left] {\begin{minipage}[lt]{74.85pt}\setlength\topsep{0pt}
\begin{center}
dually chordal ? $ $
\end{center}

\end{minipage}};
\draw  [fill={rgb, 255:red, 184; green, 233; blue, 134 }  ,fill opacity=1 ]  (502.23,233.27) -- (618.23,233.27) -- (618.23,262.27) -- (502.23,262.27) -- cycle  ;
\draw (560.23,247.77) node  [font=\small] [align=left] {\begin{minipage}[lt]{76.39pt}\setlength\topsep{0pt}
\begin{center}
strongly chordal
\end{center}

\end{minipage}};
\draw  [fill={rgb, 255:red, 184; green, 233; blue, 134 }  ,fill opacity=1 ]  (35.9,299.3) -- (125.9,299.3) -- (125.9,344.3) -- (35.9,344.3) -- cycle  ;
\draw (80.9,321.8) node  [font=\small] [align=left] {\begin{minipage}[lt]{58.34pt}\setlength\topsep{0pt}
\begin{center}
circular permutation
\end{center}

\end{minipage}};
\draw  [fill={rgb, 255:red, 184; green, 233; blue, 134 }  ,fill opacity=1 ]  (-80.1,236.27) -- (-2.1,236.27) -- (-2.1,265.27) -- (-80.1,265.27) -- cycle  ;
\draw (-41.1,250.77) node  [font=\small] [align=left] {\begin{minipage}[lt]{50.18pt}\setlength\topsep{0pt}
\begin{center}
$\displaystyle k$-polygon
\end{center}

\end{minipage}};
\draw  [fill={rgb, 255:red, 184; green, 233; blue, 134 }  ,fill opacity=1 ]  (103.9,233.49) -- (176.9,233.49) -- (176.9,280.49) -- (103.9,280.49) -- cycle  ;
\draw (140.4,256.99) node  [font=\small] [align=left] {\begin{minipage}[lt]{46.78pt}\setlength\topsep{0pt}
\begin{center}
distance hereditary
\end{center}

\end{minipage}};
\draw  [fill={rgb, 255:red, 184; green, 233; blue, 134 }  ,fill opacity=1 ]  (261.83,405.74) -- (375.83,405.74) -- (375.83,464.74) -- (261.83,464.74) -- cycle  ;
\draw (318.83,435.24) node  [font=\small] [align=left] {\begin{minipage}[lt]{74.57pt}\setlength\topsep{0pt}
\begin{center}
$\displaystyle Free( P-4+tK_{1})$, fixed $\displaystyle k$
\end{center}

\end{minipage}};
\draw  [fill={rgb, 255:red, 230; green, 216; blue, 48 }  ,fill opacity=0.69 ]  (677.87,-20.4) -- (741.87,-20.4) -- (741.87,8.6) -- (677.87,8.6) -- cycle  ;
\draw (709.87,-5.9) node  [font=\small] [align=left] {\begin{minipage}[lt]{40.62pt}\setlength\topsep{0pt}
\begin{center}
\textbf{planar}
\end{center}

\end{minipage}};
\draw (663,191) node [anchor=north west][inner sep=0.75pt]   [align=left] {$\displaystyle NPc$};
\draw (682,226) node [anchor=north west][inner sep=0.75pt]   [align=left] {\textbf{P}};
\draw  [fill={rgb, 255:red, 233; green, 17; blue, 17 }  ,fill opacity=0.69 ]  (311.33,-99.5) -- (415.33,-99.5) -- (415.33,-70.5) -- (311.33,-70.5) -- cycle  ;
\draw (363.33,-85) node  [font=\small] [align=left] {\begin{minipage}[lt]{68pt}\setlength\topsep{0pt}
\begin{center}
\textbf{general}
\end{center}

\end{minipage}};
\draw  [fill={rgb, 255:red, 184; green, 233; blue, 134 }  ,fill opacity=1 ]  (614.23,409.17) -- (714.23,409.17) -- (714.23,437.17) -- (614.23,437.17) -- cycle  ;
\draw (664.23,423.17) node  [font=\small] [align=left] {\begin{minipage}[lt]{65.51pt}\setlength\topsep{0pt}
\begin{center}
block
\end{center}

\end{minipage}};
\draw  [fill={rgb, 255:red, 255; green, 255; blue, 255 }  ,fill opacity=1 ]  (619.5,114.42) -- (724.5,114.42) -- (724.5,145.42) -- (619.5,145.42) -- cycle  ;
\draw (672,129.92) node  [font=\small] [align=left] {\begin{minipage}[lt]{68.82pt}\setlength\topsep{0pt}
\begin{center}
undirected path
\end{center}

\end{minipage}};
\draw  [fill={rgb, 255:red, 233; green, 17; blue, 17 }  ,fill opacity=0.69 ]  (-28.67,24.26) -- (78.33,24.26) -- (78.33,57.26) -- (-28.67,57.26) -- cycle  ;
\draw (24.83,40.76) node  [font=\small] [align=left] {\begin{minipage}[lt]{69.95pt}\setlength\topsep{0pt}
\begin{center}
\textbf{triangle-free}
\end{center}

\end{minipage}};
\draw    (260.83,-4.4) -- (176.78,58.07) ;
\draw [shift={(175.17,59.27)}, rotate = 323.38] [color={rgb, 255:red, 0; green, 0; blue, 0 }  ][line width=0.75]    (10.93,-3.29) .. controls (6.95,-1.4) and (3.31,-0.3) .. (0,0) .. controls (3.31,0.3) and (6.95,1.4) .. (10.93,3.29)   ;
\draw    (276.97,-4.4) -- (239.51,157.32) ;
\draw [shift={(239.06,159.27)}, rotate = 283.04] [color={rgb, 255:red, 0; green, 0; blue, 0 }  ][line width=0.75]    (10.93,-3.29) .. controls (6.95,-1.4) and (3.31,-0.3) .. (0,0) .. controls (3.31,0.3) and (6.95,1.4) .. (10.93,3.29)   ;
\draw    (288.59,-4.4) -- (316.28,44.2) ;
\draw [shift={(317.27,45.93)}, rotate = 240.33] [color={rgb, 255:red, 0; green, 0; blue, 0 }  ][line width=0.75]    (10.93,-3.29) .. controls (6.95,-1.4) and (3.31,-0.3) .. (0,0) .. controls (3.31,0.3) and (6.95,1.4) .. (10.93,3.29)   ;
\draw    (328.92,74.93) -- (344.82,142.99) ;
\draw [shift={(345.28,144.93)}, rotate = 256.85] [color={rgb, 255:red, 0; green, 0; blue, 0 }  ][line width=0.75]    (10.93,-3.29) .. controls (6.95,-1.4) and (3.31,-0.3) .. (0,0) .. controls (3.31,0.3) and (6.95,1.4) .. (10.93,3.29)   ;
\draw    (357.53,70.7) .. controls (735.51,188.21) and (840.44,347.96) .. (672.32,549.93) ;
\draw [shift={(672.32,549.93)}, rotate = 309.77] [color={rgb, 255:red, 0; green, 0; blue, 0 }  ][line width=0.75]    (10.93,-3.29) .. controls (6.95,-1.4) and (3.31,-0.3) .. (0,0) .. controls (3.31,0.3) and (6.95,1.4) .. (10.93,3.29)   ;
\draw    (382.2,239.86) -- (333.15,246.77) ;
\draw [shift={(331.17,247.04)}, rotate = 351.99] [color={rgb, 255:red, 0; green, 0; blue, 0 }  ][line width=0.75]    (10.93,-3.29) .. controls (6.95,-1.4) and (3.31,-0.3) .. (0,0) .. controls (3.31,0.3) and (6.95,1.4) .. (10.93,3.29)   ;
\draw    (608.67,573.9) .. controls (556.14,588.12) and (470.09,588.26) .. (350.47,574.31) ;
\draw [shift={(348.67,574.1)}, rotate = 6.7] [color={rgb, 255:red, 0; green, 0; blue, 0 }  ][line width=0.75]    (10.93,-3.29) .. controls (6.95,-1.4) and (3.31,-0.3) .. (0,0) .. controls (3.31,0.3) and (6.95,1.4) .. (10.93,3.29)   ;
\draw    (129.91,88.27) -- (99.3,105.49) ;
\draw [shift={(97.56,106.47)}, rotate = 330.62] [color={rgb, 255:red, 0; green, 0; blue, 0 }  ][line width=0.75]    (10.93,-3.29) .. controls (6.95,-1.4) and (3.31,-0.3) .. (0,0) .. controls (3.31,0.3) and (6.95,1.4) .. (10.93,3.29)   ;
\draw    (65.45,135.47) -- (59.71,148.59) ;
\draw [shift={(58.9,150.42)}, rotate = 293.65] [color={rgb, 255:red, 0; green, 0; blue, 0 }  ][line width=0.75]    (10.93,-3.29) .. controls (6.95,-1.4) and (3.31,-0.3) .. (0,0) .. controls (3.31,0.3) and (6.95,1.4) .. (10.93,3.29)   ;
\draw    (228.33,-7.26) -- (196.62,-0.16) ;
\draw [shift={(194.67,0.28)}, rotate = 347.38] [color={rgb, 255:red, 0; green, 0; blue, 0 }  ][line width=0.75]    (10.93,-3.29) .. controls (6.95,-1.4) and (3.31,-0.3) .. (0,0) .. controls (3.31,0.3) and (6.95,1.4) .. (10.93,3.29)   ;
\draw    (366.87,289.57) -- (303.57,300.86) ;
\draw [shift={(301.6,301.21)}, rotate = 349.89] [color={rgb, 255:red, 0; green, 0; blue, 0 }  ][line width=0.75]    (10.93,-3.29) .. controls (6.95,-1.4) and (3.31,-0.3) .. (0,0) .. controls (3.31,0.3) and (6.95,1.4) .. (10.93,3.29)   ;
\draw    (266.84,270.27) .. controls (266.74,274.5) and (265.01,281.87) .. (261.65,292.38) ;
\draw [shift={(261.04,294.24)}, rotate = 288.08] [color={rgb, 255:red, 0; green, 0; blue, 0 }  ][line width=0.75]    (10.93,-3.29) .. controls (6.95,-1.4) and (3.31,-0.3) .. (0,0) .. controls (3.31,0.3) and (6.95,1.4) .. (10.93,3.29)   ;
\draw    (301.6,313.54) -- (352.54,318.4) ;
\draw [shift={(354.53,318.59)}, rotate = 185.45] [color={rgb, 255:red, 0; green, 0; blue, 0 }  ][line width=0.75]    (10.93,-3.29) .. controls (6.95,-1.4) and (3.31,-0.3) .. (0,0) .. controls (3.31,0.3) and (6.95,1.4) .. (10.93,3.29)   ;
\draw    (370.66,334.45) -- (328.57,355.17) ;
\draw [shift={(326.78,356.05)}, rotate = 333.79] [color={rgb, 255:red, 0; green, 0; blue, 0 }  ][line width=0.75]    (10.93,-3.29) .. controls (6.95,-1.4) and (3.31,-0.3) .. (0,0) .. controls (3.31,0.3) and (6.95,1.4) .. (10.93,3.29)   ;
\draw    (45.62,203.42) -- (31.45,426.58) ;
\draw [shift={(31.32,428.57)}, rotate = 273.63] [color={rgb, 255:red, 0; green, 0; blue, 0 }  ][line width=0.75]    (10.93,-3.29) .. controls (6.95,-1.4) and (3.31,-0.3) .. (0,0) .. controls (3.31,0.3) and (6.95,1.4) .. (10.93,3.29)   ;
\draw [color={rgb, 255:red, 0; green, 0; blue, 0 }  ,draw opacity=1 ]   (47.31,455.57) -- (84.33,485.24) ;
\draw [shift={(85.89,486.49)}, rotate = 218.7] [color={rgb, 255:red, 0; green, 0; blue, 0 }  ,draw opacity=1 ][line width=0.75]    (10.93,-3.29) .. controls (6.95,-1.4) and (3.31,-0.3) .. (0,0) .. controls (3.31,0.3) and (6.95,1.4) .. (10.93,3.29)   ;
\draw    (121.87,431.11) -- (107.62,484.56) ;
\draw [shift={(107.1,486.49)}, rotate = 284.93] [color={rgb, 255:red, 0; green, 0; blue, 0 }  ][line width=0.75]    (10.93,-3.29) .. controls (6.95,-1.4) and (3.31,-0.3) .. (0,0) .. controls (3.31,0.3) and (6.95,1.4) .. (10.93,3.29)   ;
\draw    (228.2,390.07) -- (169.3,405.06) ;
\draw [shift={(167.37,405.55)}, rotate = 345.73] [color={rgb, 255:red, 0; green, 0; blue, 0 }  ][line width=0.75]    (10.93,-3.29) .. controls (6.95,-1.4) and (3.31,-0.3) .. (0,0) .. controls (3.31,0.3) and (6.95,1.4) .. (10.93,3.29)   ;
\draw    (426.5,334.45) -- (514.33,367.95) ;
\draw [shift={(516.2,368.66)}, rotate = 200.87] [color={rgb, 255:red, 0; green, 0; blue, 0 }  ][line width=0.75]    (10.93,-3.29) .. controls (6.95,-1.4) and (3.31,-0.3) .. (0,0) .. controls (3.31,0.3) and (6.95,1.4) .. (10.93,3.29)   ;
\draw    (563.94,395.93) .. controls (630.81,458.31) and (612.21,498.75) .. (508.11,517.28) ;
\draw [shift={(506.53,517.55)}, rotate = 350.11] [color={rgb, 255:red, 0; green, 0; blue, 0 }  ][line width=0.75]    (10.93,-3.29) .. controls (6.95,-1.4) and (3.31,-0.3) .. (0,0) .. controls (3.31,0.3) and (6.95,1.4) .. (10.93,3.29)   ;
\draw    (369.53,523.43) -- (415.54,525.59) ;
\draw [shift={(417.53,525.68)}, rotate = 182.69] [color={rgb, 255:red, 0; green, 0; blue, 0 }  ][line width=0.75]    (10.93,-3.29) .. controls (6.95,-1.4) and (3.31,-0.3) .. (0,0) .. controls (3.31,0.3) and (6.95,1.4) .. (10.93,3.29)   ;
\draw    (511.33,445.6) -- (471.76,511.55) ;
\draw [shift={(470.73,513.27)}, rotate = 300.96] [color={rgb, 255:red, 0; green, 0; blue, 0 }  ][line width=0.75]    (10.93,-3.29) .. controls (6.95,-1.4) and (3.31,-0.3) .. (0,0) .. controls (3.31,0.3) and (6.95,1.4) .. (10.93,3.29)   ;
\draw    (345.09,173.93) .. controls (318.03,270.62) and (353.43,383.73) .. (451.29,513.27) ;
\draw [shift={(451.29,513.27)}, rotate = 232.93] [color={rgb, 255:red, 0; green, 0; blue, 0 }  ][line width=0.75]    (10.93,-3.29) .. controls (6.95,-1.4) and (3.31,-0.3) .. (0,0) .. controls (3.31,0.3) and (6.95,1.4) .. (10.93,3.29)   ;
\draw    (529.64,338.27) -- (542.12,365.12) ;
\draw [shift={(542.96,366.93)}, rotate = 245.07] [color={rgb, 255:red, 0; green, 0; blue, 0 }  ][line width=0.75]    (10.93,-3.29) .. controls (6.95,-1.4) and (3.31,-0.3) .. (0,0) .. controls (3.31,0.3) and (6.95,1.4) .. (10.93,3.29)   ;
\draw    (463.15,294.15) -- (491.85,308.38) ;
\draw [shift={(493.64,309.27)}, rotate = 206.36] [color={rgb, 255:red, 0; green, 0; blue, 0 }  ][line width=0.75]    (10.93,-3.29) .. controls (6.95,-1.4) and (3.31,-0.3) .. (0,0) .. controls (3.31,0.3) and (6.95,1.4) .. (10.93,3.29)   ;
\draw    (572.29,58.22) -- (561.28,231.27) ;
\draw [shift={(561.16,233.27)}, rotate = 273.64] [color={rgb, 255:red, 0; green, 0; blue, 0 }  ][line width=0.75]    (10.93,-3.29) .. controls (6.95,-1.4) and (3.31,-0.3) .. (0,0) .. controls (3.31,0.3) and (6.95,1.4) .. (10.93,3.29)   ;
\draw    (148.2,88.27) .. controls (113.41,153.53) and (91.96,223.51) .. (83.83,298.17) ;
\draw [shift={(83.71,299.3)}, rotate = 276.11] [color={rgb, 255:red, 0; green, 0; blue, 0 }  ][line width=0.75]    (10.93,-3.29) .. controls (6.95,-1.4) and (3.31,-0.3) .. (0,0) .. controls (3.31,0.3) and (6.95,1.4) .. (10.93,3.29)   ;
\draw    (91.63,344.3) -- (117.85,399.3) ;
\draw [shift={(118.71,401.11)}, rotate = 244.51] [color={rgb, 255:red, 0; green, 0; blue, 0 }  ][line width=0.75]    (10.93,-3.29) .. controls (6.95,-1.4) and (3.31,-0.3) .. (0,0) .. controls (3.31,0.3) and (6.95,1.4) .. (10.93,3.29)   ;
\draw    (-40.72,265.27) .. controls (-42.54,333.81) and (-1.28,379.41) .. (83.09,402.08) ;
\draw [shift={(84.37,402.42)}, rotate = 194.82] [color={rgb, 255:red, 0; green, 0; blue, 0 }  ][line width=0.75]    (10.93,-3.29) .. controls (6.95,-1.4) and (3.31,-0.3) .. (0,0) .. controls (3.31,0.3) and (6.95,1.4) .. (10.93,3.29)   ;
\draw    (272.04,-4.4) .. controls (229.96,68.55) and (189.83,147.43) .. (151.66,232.21) ;
\draw [shift={(151.08,233.49)}, rotate = 294.22] [color={rgb, 255:red, 0; green, 0; blue, 0 }  ][line width=0.75]    (10.93,-3.29) .. controls (6.95,-1.4) and (3.31,-0.3) .. (0,0) .. controls (3.31,0.3) and (6.95,1.4) .. (10.93,3.29)   ;
\draw    (150.74,280.49) .. controls (206.77,410.89) and (257.05,486.17) .. (301.55,506.34) ;
\draw [shift={(302.9,506.93)}, rotate = 203.23] [color={rgb, 255:red, 0; green, 0; blue, 0 }  ][line width=0.75]    (10.93,-3.29) .. controls (6.95,-1.4) and (3.31,-0.3) .. (0,0) .. controls (3.31,0.3) and (6.95,1.4) .. (10.93,3.29)   ;
\draw    (321.64,464.74) -- (325.46,504.94) ;
\draw [shift={(325.65,506.93)}, rotate = 264.57] [color={rgb, 255:red, 0; green, 0; blue, 0 }  ][line width=0.75]    (10.93,-3.29) .. controls (6.95,-1.4) and (3.31,-0.3) .. (0,0) .. controls (3.31,0.3) and (6.95,1.4) .. (10.93,3.29)   ;
\draw    (34.11,203.42) .. controls (-103.76,470.65) and (-33.58,593.28) .. (244.67,571.32) ;
\draw [shift={(244.67,571.32)}, rotate = 175.49] [color={rgb, 255:red, 0; green, 0; blue, 0 }  ][line width=0.75]    (10.93,-3.29) .. controls (6.95,-1.4) and (3.31,-0.3) .. (0,0) .. controls (3.31,0.3) and (6.95,1.4) .. (10.93,3.29)   ;
\draw    (-49.33,-0.73) -- (-41.64,234.27) ;
\draw [shift={(-41.57,236.27)}, rotate = 268.13] [color={rgb, 255:red, 0; green, 0; blue, 0 }  ][line width=0.75]    (10.93,-3.29) .. controls (6.95,-1.4) and (3.31,-0.3) .. (0,0) .. controls (3.31,0.3) and (6.95,1.4) .. (10.93,3.29)   ;
\draw    (148.08,280.49) .. controls (183.8,393.42) and (229.28,484.34) .. (284.54,553.23) ;
\draw [shift={(285.37,554.27)}, rotate = 231.13] [color={rgb, 255:red, 0; green, 0; blue, 0 }  ][line width=0.75]    (10.93,-3.29) .. controls (6.95,-1.4) and (3.31,-0.3) .. (0,0) .. controls (3.31,0.3) and (6.95,1.4) .. (10.93,3.29)   ;
\draw    (336.17,74.93) .. controls (381.52,139.81) and (446.21,192.39) .. (530.24,232.66) ;
\draw [shift={(531.51,233.27)}, rotate = 205.48] [color={rgb, 255:red, 0; green, 0; blue, 0 }  ][line width=0.75]    (10.93,-3.29) .. controls (6.95,-1.4) and (3.31,-0.3) .. (0,0) .. controls (3.31,0.3) and (6.95,1.4) .. (10.93,3.29)   ;
\draw    (715.12,8.6) .. controls (908.15,536.57) and (784.26,728.12) .. (343.44,583.27) ;
\draw [shift={(343.44,583.27)}, rotate = 18.19] [color={rgb, 255:red, 0; green, 0; blue, 0 }  ][line width=0.75]    (10.93,-3.29) .. controls (6.95,-1.4) and (3.31,-0.3) .. (0,0) .. controls (3.31,0.3) and (6.95,1.4) .. (10.93,3.29)   ;
\draw    (227.78,188.27) .. controls (173.53,282.67) and (179.53,338.51) .. (245.77,355.8) ;
\draw [shift={(246.77,356.05)}, rotate = 194.19] [color={rgb, 255:red, 0; green, 0; blue, 0 }  ][line width=0.75]    (10.93,-3.29) .. controls (6.95,-1.4) and (3.31,-0.3) .. (0,0) .. controls (3.31,0.3) and (6.95,1.4) .. (10.93,3.29)   ;
\draw    (345.13,-70.5) -- (300.11,-34.65) ;
\draw [shift={(298.54,-33.4)}, rotate = 321.47] [color={rgb, 255:red, 0; green, 0; blue, 0 }  ][line width=0.75]    (10.93,-3.29) .. controls (6.95,-1.4) and (3.31,-0.3) .. (0,0) .. controls (3.31,0.3) and (6.95,1.4) .. (10.93,3.29)   ;
\draw    (311.33,-85.41) .. controls (107.69,-88.94) and (-7.81,-70.79) .. (-35.14,-30.95) ;
\draw [shift={(-35.95,-29.73)}, rotate = 302.4] [color={rgb, 255:red, 0; green, 0; blue, 0 }  ][line width=0.75]    (10.93,-3.29) .. controls (6.95,-1.4) and (3.31,-0.3) .. (0,0) .. controls (3.31,0.3) and (6.95,1.4) .. (10.93,3.29)   ;
\draw    (415.33,-84.36) .. controls (521.94,-86.6) and (611.22,-65.5) .. (683.19,-21.07) ;
\draw [shift={(684.27,-20.4)}, rotate = 211.89] [color={rgb, 255:red, 0; green, 0; blue, 0 }  ][line width=0.75]    (10.93,-3.29) .. controls (6.95,-1.4) and (3.31,-0.3) .. (0,0) .. controls (3.31,0.3) and (6.95,1.4) .. (10.93,3.29)   ;
\draw    (402.65,-70.5) .. controls (459.27,-52.71) and (509.76,-21.59) .. (554.11,22.87) ;
\draw [shift={(555.45,24.22)}, rotate = 225.33] [color={rgb, 255:red, 0; green, 0; blue, 0 }  ][line width=0.75]    (10.93,-3.29) .. controls (6.95,-1.4) and (3.31,-0.3) .. (0,0) .. controls (3.31,0.3) and (6.95,1.4) .. (10.93,3.29)   ;
\draw    (279.74,-4.4) -- (269.84,239.27) ;
\draw [shift={(269.76,241.27)}, rotate = 272.33] [color={rgb, 255:red, 0; green, 0; blue, 0 }  ][line width=0.75]    (10.93,-3.29) .. controls (6.95,-1.4) and (3.31,-0.3) .. (0,0) .. controls (3.31,0.3) and (6.95,1.4) .. (10.93,3.29)   ;
\draw    (365.61,-70.5) -- (411.11,218.96) ;
\draw [shift={(411.42,220.93)}, rotate = 261.07] [color={rgb, 255:red, 0; green, 0; blue, 0 }  ][line width=0.75]    (10.93,-3.29) .. controls (6.95,-1.4) and (3.31,-0.3) .. (0,0) .. controls (3.31,0.3) and (6.95,1.4) .. (10.93,3.29)   ;
\draw    (664.24,437.17) .. controls (670.52,526.09) and (565.33,569.79) .. (348.67,568.26) ;
\draw [shift={(348.67,568.26)}, rotate = 0.41] [color={rgb, 255:red, 0; green, 0; blue, 0 }  ][line width=0.75]    (10.93,-3.29) .. controls (6.95,-1.4) and (3.31,-0.3) .. (0,0) .. controls (3.31,0.3) and (6.95,1.4) .. (10.93,3.29)   ;
\draw    (669.71,145.42) .. controls (657.56,237.73) and (622.66,311.28) .. (565.03,366.11) ;
\draw [shift={(564.16,366.93)}, rotate = 316.64] [color={rgb, 255:red, 0; green, 0; blue, 0 }  ][line width=0.75]    (10.93,-3.29) .. controls (6.95,-1.4) and (3.31,-0.3) .. (0,0) .. controls (3.31,0.3) and (6.95,1.4) .. (10.93,3.29)   ;
\draw    (671.59,145.42) -- (664.66,407.18) ;
\draw [shift={(664.6,409.17)}, rotate = 271.52] [color={rgb, 255:red, 0; green, 0; blue, 0 }  ][line width=0.75]    (10.93,-3.29) .. controls (6.95,-1.4) and (3.31,-0.3) .. (0,0) .. controls (3.31,0.3) and (6.95,1.4) .. (10.93,3.29)   ;
\draw    (370.37,-70.5) .. controls (460.67,113.74) and (485.57,224.25) .. (445.09,261.06) ;
\draw [shift={(443.84,262.15)}, rotate = 319.81] [color={rgb, 255:red, 0; green, 0; blue, 0 }  ][line width=0.75]    (10.93,-3.29) .. controls (6.95,-1.4) and (3.31,-0.3) .. (0,0) .. controls (3.31,0.3) and (6.95,1.4) .. (10.93,3.29)   ;
\draw [color={rgb, 255:red, 155; green, 155; blue, 155 }  ,draw opacity=1 ]   (367.74,-70.5) -- (515.05,414.69) ;
\draw [shift={(515.63,416.6)}, rotate = 253.11] [color={rgb, 255:red, 155; green, 155; blue, 155 }  ,draw opacity=1 ][line width=0.75]    (10.93,-3.29) .. controls (6.95,-1.4) and (3.31,-0.3) .. (0,0) .. controls (3.31,0.3) and (6.95,1.4) .. (10.93,3.29)   ;
\draw    (357.53,66.85) -- (617.54,119) ;
\draw [shift={(619.5,119.39)}, rotate = 191.34] [color={rgb, 255:red, 0; green, 0; blue, 0 }  ][line width=0.75]    (10.93,-3.29) .. controls (6.95,-1.4) and (3.31,-0.3) .. (0,0) .. controls (3.31,0.3) and (6.95,1.4) .. (10.93,3.29)   ;
\draw    (311.33,-77.19) .. controls (138.43,-53.56) and (45.74,-20.18) .. (33.3,22.95) ;
\draw [shift={(32.95,24.26)}, rotate = 284.03] [color={rgb, 255:red, 0; green, 0; blue, 0 }  ][line width=0.75]    (10.93,-3.29) .. controls (6.95,-1.4) and (3.31,-0.3) .. (0,0) .. controls (3.31,0.3) and (6.95,1.4) .. (10.93,3.29)   ;
\draw    (34.49,57.26) -- (62.3,104.75) ;
\draw [shift={(63.31,106.47)}, rotate = 239.65] [color={rgb, 255:red, 0; green, 0; blue, 0 }  ][line width=0.75]    (10.93,-3.29) .. controls (6.95,-1.4) and (3.31,-0.3) .. (0,0) .. controls (3.31,0.3) and (6.95,1.4) .. (10.93,3.29)   ;

\end{tikzpicture}

%% file: pages/content/whardness.tex
\section{Fixed-Parameter Intractability}

We will start by giving some intractability results and show that \sdom parameterized by the natural parameterization is \WTWOhs-hard on \textit{bipartite graphs} and \textit{triangle-free graphs} by fpt-reducing from \dom on bipartite graphs
\dom on bipartite graphs is known to be \WTWOhs when parameterized with natural parameterization \cite{Raman2008}.

\begin{figure}[ht]
    \label{fig:bipartiteConstruction}
    \begin{equation*}
        \input{fig/tikz/wbipartite.tikz}
    \end{equation*}
\caption[Construction bipartite]{\textit{Reducing to a bipartite $G'$ from the bipartite graph $K_{m,n}$ by duplicating all vertices and adding exactly two forced witnesses.}}
\end{figure}
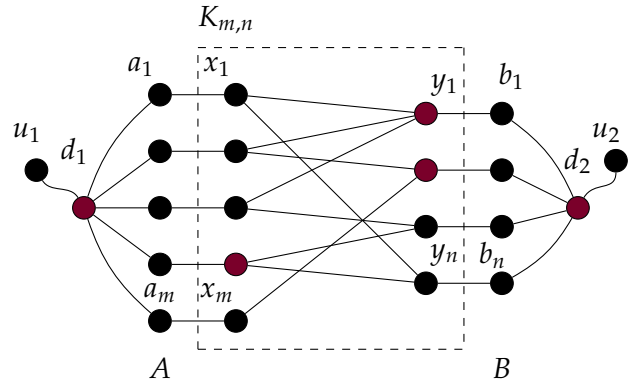

\begin{theorem}\label{lemma:bipartite}
    \sdom parameterized by solution size is \WTWOhs-hard when restricted to bipartite graphs.
\end{theorem}

\begin{proof}
    We reduce from \dom and consider only $X, Y \neq \emptyset$.
    Given a bipartite $G = ( X \cup Y, E)$, we construct a bipartite graph $G' = (\{X' \cup Y'\},E')$:
    \begin{enumerate}[topsep=0pt,itemsep=0ex,partopsep=1ex,parsep=1ex]
        \item For each $x_i \in X$, we add a new vertex $a_i \in A$  and an edge $\{x_i, a_i\}$ in between. 
        \item For each $y_j \in Y$, we add a new vertex $b_j \in B$ and an edge $\{y_j, b_j\}$ in between.
        \item We add two $P_2$'s with the vertices $u_j,d_j$ and connect their ends to all $a_i$ (resp. $b_i$).
    \end{enumerate}

    $G'$ is bipartite because $A$ and $B$ form an independent set on $G'$ that can be cross-wise attached to $X$ and $Y$. 
    $X' = X \cup \{u_2,d_1\} \cup B$ and $Y' = X \cup \{u_1,d_2\} \cup A$ form the partitions of the new bipartite $G'$.
    It is left to show that $G$ has a ds $D$ of size $k$ if and only if $G'$ has an sds $D'$ of size $k' = k + 2$.
 
    First, assume a ds $D$ in $G$ of size $k$. 
    We know that $D' = D\cup \{d_1,d_2\}$ is an sds in $G'$ of size $k' = k + 2$, because $d_1$ dominates $u_1$ and all $a_i \in A$; $d_2$ dominates $u_2$ and all $b_i \in B$. 
    The same vertices dominate the rest as they were in $G$, but now they all have either $d_1$ or $d_2$ as a witness.
    Therefore,  $\forall v \in (D \cap X) \cup (D \cap Y): (d(v, d_1) = 2 \vee d(v, d_2)=2)$.

    Contrary, assume an sds $D'$ in $G'$ with size $k'$. 
    Wlog, assume that $d_1, d_2 \in D$ and $u_1,u_2 \notin D$, because choosing $d_i$ is preferred to $u_i$.
    By technical assumption $X, Y \neq \emptyset$, $u_i$ can not be a witness for $d_i$.
    If $a_i,b_i \in D'$ we replace it with $x_i$ and $y_i$ preserving the size $D$.
    A $a_i \in A$ can only be used to dominate their neighboring $x_i$ ($b_i \in B$ for $y_i$), because $\abs{N(a_i)} = 2$ and $d_1,d_2\in D'$.

    As $d_1$ and $d_2$ suffice to provide a witness for every vertex in the graph, this operation is sound.
    Hence, $D = D' \setminus \{ d_1,d_2\}$ gives a ds in $G$ of size $ k = k' + 2$.

    As $G'$ can be constructed in linear time and $k$ does not depend on the input size, this reduction is an fpt reduction.  
    Because \dom is \WTWOhs-hard on bipartite graphs~\cite[Th. 1]{Raman2008}, we conclude that \sdom is \WTWOhs-hard when parameterized by solution size as well.
\end{proof}

We will prove intractability for \textit{split} graphs. 
We use the fact that every dominating set in a split graph can directly be mapped to a corresponding semitotal dominating set.

\begin{theorem}\label{lemma:splitgraph}
    \sdom is \WONEhs-hard when restricted to \textit{split} and \textit{chordal} graphs.
\end{theorem}

\begin{figure}[ht]
    \begin{equation*}
        \input{fig/tikz/split.tikz}
    \end{equation*}
\caption[Constructing split graph]{\textit{Obtaining an sds $D'$ form a given ds $D$ in a split graph by flipping all vertices $d_i$ to their corresponding neighbor in the clique $K$.
The clique $K$ is highlighted in {\setulcolor{TUMBlue}\ul{blue}}
Note that in $D$, no witnesses are available as $d(d_1,d_5) = 3$.
After the flip operation, this is fixed.}} \label{fig:splitgraph}
\end{figure}
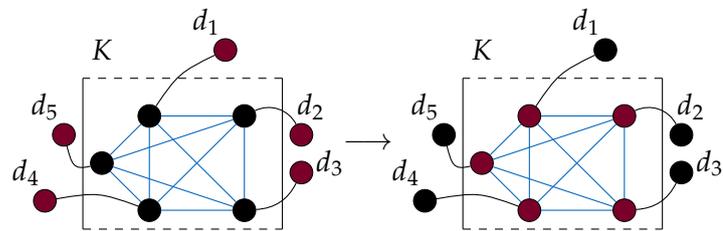

\begin{proof}

    We reduce from \dom on \textit{split} graphs by showing that any ds in a split graph can be mapped to a sds on the same graph. 
    Given a split graph ${G = \{V = (K \cup I), E\}}$ with $\abs{V} \geq 2$ and a ds $D$ of size $k$, we can immediately obtain an sds $D'$ by flipping a few vertices:
    If $I \cap D \neq \emptyset$ we replace them respectively with one arbitrary neighbor in $K$.
    All vertices in $I$ are still being dominated and $D \cap K \neq \emptyset$ is sufficient to preserve domination $K$. 
    Note that it is necessary to catch the case, where $D \subseteq I$ and witnesses are missing, which will only be guaranteed after the flip operation. (see \cref{fig:splitgraph}).
    We now assume $D \subseteq K$ and set $D = D'$.
    If $|D'| > 1$, we immediately obtain an sds as $D' \subseteq K$ and $k_1,k_2\in K$ witness each other.
    If $D' = \{d\}$, we add any neighbor of $d$ to $D'$.
    
    For the sake of completeness, if $|V| \leq 1$, we instantly reject it as there is no witness available.
    In all cases $k' \leq k + 1$ and because \dom is \WTWOhs-hard on \textit{split} graphs~\cite{Raman2008} the claim follows.
\end{proof}

%% file: fig/tikz/wbipartite.tikz
\begin{tikzpicture}
	\begin{pgfonlayer}{nodelayer}
		\node [style=BLACK] (14) at (-7, -3.25) {};
		\node [style=DOM] (15) at (-7, -1.75) {};
		\node [style=BLACK] (16) at (-7, -0.25) {};
		\node [style=BLACK] (17) at (-7, 1.25) {};
		\node [style=BLACK] (18) at (-7, 2.75) {};
		\node [style=BLACK] (19) at (-2, -2.25) {};
		\node [style=BLACK] (20) at (-2, -0.75) {};
		\node [style=DOM] (21) at (-2, 0.75) {};
		\node [style=DOM] (22) at (-2, 2.25) {};
		\node [style=none] (23) at (-7.5, 3.5) {$x_1$};
		\node [style=none] (24) at (-7.5, -2.5) {$x_m$};
		\node [style=none] (25) at (-1.5, 3) {$y_1$};
		\node [style=none] (26) at (-1.5, -1.5) {$y_n$};
		\node [style=BLACK] (27) at (-9, 2.75) {};
		\node [style=BLACK] (28) at (-9, 1.25) {};
		\node [style=BLACK] (29) at (-9, -0.25) {};
		\node [style=BLACK] (30) at (-9, -1.75) {};
		\node [style=BLACK] (31) at (-9, -3.25) {};
		\node [style=BLACK] (34) at (0, -2.25) {};
		\node [style=BLACK] (35) at (0, -0.75) {};
		\node [style=BLACK] (36) at (0, 0.75) {};
		\node [style=BLACK] (37) at (0, 2.25) {};
		\node [style=DOM] (38) at (-11, -0.25) {};
		\node [style=DOM] (39) at (2, -0.25) {};
		\node [style=none] (40) at (0.25, 3.25) {$b_1$};
		\node [style=none] (41) at (-0.25, -1.5) {$b_n$};
		\node [style=none] (42) at (-9.5, 3.5) {$a_1$};
		\node [style=none] (43) at (-9, -2.5) {$a_m$};
		\node [style=BLACK] (44) at (3, 1) {};
		\node [style=BLACK] (45) at (-12.25, 0.75) {};
		\node [style=none] (49) at (-8, 4) {};
		\node [style=none] (50) at (-1, 4) {};
		\node [style=none] (51) at (-1, -4) {};
		\node [style=none] (52) at (-8, -4) {};
		\node [style=none] (53) at (-7.25, 4.75) {$K_{m,n}$};
		\node [style=none] (54) at (2, 1) {$d_2$};
		\node [style=none] (55) at (-11.25, 1.25) {$d_1$};
		\node [style=none] (57) at (-12.5, 1.75) {$u_1$};
		\node [style=none] (58) at (2.75, 1.75) {$u_2$};
		\node [style=none] (60) at (-9, -4.5) {$A$};
		\node [style=none] (61) at (0, -4.5) {$B$};
	\end{pgfonlayer}
	\begin{pgfonlayer}{edgelayer}
		\draw (18) to (22);
		\draw (22) to (17);
		\draw (17) to (21);
		\draw (20) to (16);
		\draw (15) to (20);
		\draw (16) to (22);
		\draw (21) to (14);
		\draw (19) to (15);
		\draw (19) to (18);
		\draw (27) to (18);
		\draw (22) to (37);
		\draw (36) to (21);
		\draw (17) to (28);
		\draw (29) to (16);
		\draw (30) to (15);
		\draw (31) to (14);
		\draw (19) to (34);
		\draw (20) to (35);
		\draw [bend right=15] (27) to (38);
		\draw (38) to (28);
		\draw (29) to (38);
		\draw (30) to (38);
		\draw [bend left=15] (31) to (38);
		\draw [bend right=15] (34) to (39);
		\draw (35) to (39);
		\draw [bend left=15] (37) to (39);
		\draw (36) to (39);
		\draw [in=120, out=-60] (45) to (38);
		\draw [in=60, out=-90, looseness=1.25] (44) to (39);
		\draw [style=EDGEDASHED] (50.center) to (51.center);
		\draw [style=EDGEDASHED] (52.center) to (51.center);
		\draw [style=EDGEDASHED] (52.center) to (49.center);
		\draw [style=EDGEDASHED] (49.center) to (50.center);
	\end{pgfonlayer}
\end{tikzpicture}

%% file: fig/tikz/split.tikz
\begin{tikzpicture}
	\begin{pgfonlayer}{nodelayer}
		\node [style=BLACK] (0) at (-3.25, 1.5) {};
		\node [style=BLACK] (1) at (-5.75, 1.5) {};
		\node [style=BLACK] (2) at (-5.75, -1) {};
		\node [style=BLACK] (3) at (-3.25, -1) {};
		\node [style=BLACK] (9) at (-7, 0.25) {};
		\node [style=none] (11) at (-7.5, 2.5) {};
		\node [style=none] (12) at (-2.25, 2.5) {};
		\node [style=none] (13) at (-2.25, -1.5) {};
		\node [style=none] (14) at (-7.5, -1.5) {};
		\node [style=none] (18) at (-7, 3.25) {$K$};
		\node [style=none] (33) at (-4.25, 4) {$d_1$};
		\node [style=DOM] (34) at (-1.75, 1) {};
		\node [style=DOM] (36) at (-1.75, 0) {};
		\node [style=DOM] (37) at (-8, 1) {};
		\node [style=DOM] (38) at (-3.75, 3.25) {};
		\node [style=DOM] (39) at (-8.5, -0.75) {};
		\node [style=none] (40) at (-1.5, 1.75) {$d_2$};
		\node [style=none] (41) at (-1, 0.25) {$d_3$};
		\node [style=none] (42) at (-9, 0) {$d_4$};
		\node [style=none] (43) at (-8.5, 1.75) {$d_5$};
		\node [style=DOM] (44) at (6.75, 1.5) {};
		\node [style=DOM] (45) at (4.25, 1.5) {};
		\node [style=DOM] (46) at (4.25, -1) {};
		\node [style=DOM] (47) at (6.75, -1) {};
		\node [style=DOM] (48) at (3, 0.25) {};
		\node [style=none] (49) at (2.5, 2.5) {};
		\node [style=none] (50) at (7.75, 2.5) {};
		\node [style=none] (51) at (7.75, -1.5) {};
		\node [style=none] (52) at (2.5, -1.5) {};
		\node [style=none] (53) at (3, 3.25) {$K$};
		\node [style=none] (54) at (5.75, 4) {$d_1$};
		\node [style=BLACK] (55) at (8.25, 1) {};
		\node [style=BLACK] (56) at (8.25, 0) {};
		\node [style=BLACK] (57) at (2, 1) {};
		\node [style=BLACK] (58) at (6.25, 3.25) {};
		\node [style=BLACK] (59) at (1.5, -0.75) {};
		\node [style=none] (60) at (8.5, 1.75) {$d_2$};
		\node [style=none] (61) at (9, 0.25) {$d_3$};
		\node [style=none] (62) at (1, 0) {$d_4$};
		\node [style=none] (63) at (1.5, 1.75) {$d_5$};
		\node [style=none] (64) at (0, 0.75) {$\longrightarrow$};
	\end{pgfonlayer}
	\begin{pgfonlayer}{edgelayer}
		\draw [style=BLUE] (0) to (3);
		\draw [style=BLUE] (3) to (2);
		\draw [style=BLUE] (1) to (2);
		\draw [style=BLUE] (0) to (1);
		\draw [style=BLUE] (1) to (3);
		\draw [style=BLUE] (0) to (2);
		\draw [style=BLUE] (9) to (1);
		\draw [style=BLUE] (2) to (9);
		\draw [style=BLUE] (9) to (3);
		\draw [style=BLUE] (0) to (9);
		\draw [style=EDGEDASHED] (14.center) to (13.center);
		\draw (12.center) to (13.center);
		\draw [style=EDGEDASHED] (12.center) to (11.center);
		\draw [style=EDGE] (11.center) to (14.center);
		\draw [style=EDGE, bend left=45] (0) to (34);
		\draw [style=EDGE, bend right] (3) to (36);
		\draw [style=EDGE, bend left=15, looseness=1.25] (39) to (2);
		\draw [style=EDGE, in=645, out=-165, looseness=1.50] (9) to (37);
		\draw [style=EDGE, bend right=15, looseness=1.25] (38) to (1);
		\draw [style=BLUE] (44) to (47);
		\draw [style=BLUE] (47) to (46);
		\draw [style=BLUE] (45) to (46);
		\draw [style=BLUE] (44) to (45);
		\draw [style=BLUE] (45) to (47);
		\draw [style=BLUE] (44) to (46);
		\draw [style=BLUE] (48) to (45);
		\draw [style=BLUE] (46) to (48);
		\draw [style=BLUE] (48) to (47);
		\draw [style=BLUE] (44) to (48);
		\draw [style=EDGEDASHED] (52.center) to (51.center);
		\draw (50.center) to (51.center);
		\draw [style=EDGEDASHED] (50.center) to (49.center);
		\draw [style=EDGE] (49.center) to (52.center);
		\draw [style=EDGE, bend left=45] (44) to (55);
		\draw [style=EDGE, bend right] (47) to (56);
		\draw [style=EDGE, bend left=15, looseness=1.25] (59) to (46);
		\draw [style=EDGE, in=645, out=-165, looseness=1.50] (48) to (57);
		\draw [style=EDGE, bend right=15, looseness=1.25] (58) to (45);
	\end{pgfonlayer}
\end{tikzpicture}

%% file: pages/content/planarsds.tex
\chapter{A Linear Kernel for Planar Semitotal Domination}\label{ch:linkern}

\vspace*{-50pt}
\begin{figure}[ht]
\includegraphics[width=.3\textwidth, right]{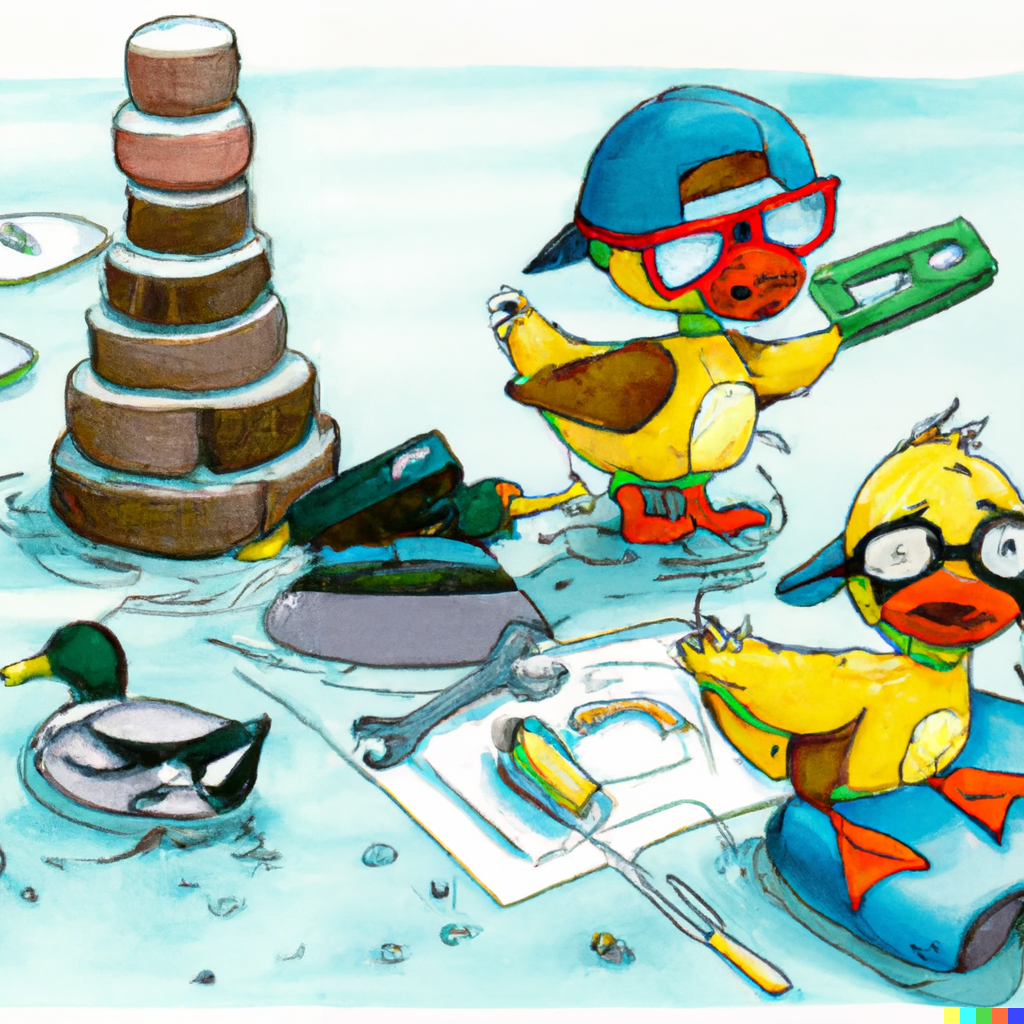}
        \captionsetup{textformat=empty,labelformat=blank}
        \caption[Generated with Dalle-E. Knowledge Cutoff 09-2022]{Generated with Dall-E. \url{https://labs.openai.com/}. ``more ducks building and working on  a high stack in the water using an excavator and tools and machines''}
\end{figure}

\epigraph{\itshape The best way to explain it is to do it.}{Lewis Caroll, \textit{Alice in Wonderland}}

We will present a polynomial-time preprocessing procedure giving a linear kernel for \psdom parameterized by solution size. Based on the technique first introduced by Alber, Fellows, and Niedermeier~\cite{Alber2004} in 2004, an abundance of similar results to other domination problems emerged which gave us the belief we can transfer these results to \sdom. 
\cref{tbl:kernels} gives an overview of the status of various kernels for the planar case on various domination problems.
All of these results introduce reduction rules bounding the number of vertices inside so-called ``regions'' which can be obtained by a decomposition of the planar graph. 

\begin{table}[h]
\begin{minipage}[th]{\linewidth}
\setcounter{mpfootnote}{\value{footnote}}
\renewcommand{\thempfootnote}{\arabic{mpfootnote}}

\begin{tabularx}{\textwidth}{lcX}
\textbf{Problem} & \textbf{Best Kernel} & \textbf{Source} \\
\pdom &  $67k$ &~\cite{Diekert2005}\footnotemark\\
\ptdom &  $410k$ &~\cite{Garnero2018}\footnotemark \\
\psdom & $\kernelsize k$ & \cref{thm:central} \\
& & \\
\peddom & $14k$  &~\cite[Th. 2]{Guo2007} \\
\pefdom &  $84k$ &~\cite[Th. 4]{Guo2007} \\
\prbdom &  $43k$ &~\cite{Garnero2017} \\
\pcdom & $130k$  &~\cite{Luo2013} \\
\pdirdom & Linear  &~\cite{Alber2006}  \\
\end{tabularx}

\footnotetext[1]{Halseth's master thesis~\cite{Halseth2016} claims a bound of $43k$, but no conference or journal version was found.}
\footnotetext[2]{Improved their own results from first $694k$~\cite[arXiv v2]{Garnero2018}}
\setcounter{footnote}{\value{mpfootnote}}
\end{minipage}
\caption{An overview about existing kernels for planar dominating problems.}
\label{tbl:kernels}
\end{table}
In the following years, this approach bore fruits for other planar problems as well:
a ${}^{11}{\mskip -5mu/\mskip -3mu}_3$ kernel for \name{Connected Vertex Cover}\xspace given in~\cite{Kowalik2013},
$624k$ for \name{Maximum Triangle Packing}\xspace,
$40k$ for \name{Induced Matching}\xspace,
$13k$ for \name{Feedback Vertex Set}\xspace and further linear kernels for
\name{Full-Degree Spanning Tree}\xspace and
\name{Cycle Packing}\xspace~\cite{Wang2011, Kanj2011, Bonamy2016, Guo2006, Garnero2019}.

In the upcoming years, this approach was generalized to larger graph classes. 
Fomin and Thilikos~\cite{Fomin2004} started by proving that the initial reduction rules in~\cite{Alber2004} can be extended to obtain a linear kernel on graphs with bounded genus $g$ for \dom.
Gutner~\cite{Gutner2009} advanced in 2008 by giving a linear kernel for $K_{3,h}$-topological-minor-free graph classes and a polynomial kernel for $K_h$-topological-minor-free graph classes. 
In 2012 Philip, Raman, and Sikdar~\cite{Philip2012} showed that $K_{i,j}$-free graph classes admit a polynomial kernel. 
In an attempt to further expand these ideas to other problems, Bodlaender et al.~\cite{Bodlaender2016} proved that all problems expressible in counting monadic second-order logic satisfying a coverability property admit a polynomial kernel on graphs of bounded genus $g$. 
Interestingly from a theoretical point of view, the constants in these meta-theorems for the kernels obtained are too large to be of practical interest. 
The question of how an efficient kernel for the \psdom problem can be constructed remains. 

In this chapter, we will transfer the linear kernel with ``reasonable'' small constants for \ptdom described by Garnero and Sau~\cite[arXiv v2]{Garnero2018} to \psdom. 
We modified the original reduction rules to preserve the witness properties of an sds. 

 \paragraph{The Main Idea} A planar graph \G with a given vertex set $D \subseteq V$ can be decomposed into at most $(3 \cdot \abs{D}-6)$ so-called ``regions'' (\cref{def:region}). 
 If $D$ is a given sds of size $\abs{D}$, the total number of regions in this decomposition depends \underline{linearly} on the size of $D$. 
 We define \textit{reduction rules} (\cref{rgl:rone,rgl:rtwo,rgl:rthree}) to iteratively reduce the number of vertices around a region.
 After, we bound the size of a resulting graph by proving that for a fixed $k$ each region has only a constant number of vertices nearby. 
Such a reduction gives us a \textit{kernel} for \psdom.

Interestingly, the reduction rules do not rely on the decomposition itself, but rather consider the neighborhood of every pair of vertices in the graph. 
The decomposition has just been used as a tool to analyze the kernel size after the reduction.

\section{Definitions}

Before stating the reduction rules, we need definitions to capture the ``nice'' properties we are to exploit. 
They are equal to those given by Garnero and Sau for \ptdom in~\cite[arXiv v2]{Garnero2018} and for \prbdom in~\cite{Garnero2017} which in turn reused ideas introduced by Alber, Fellows and Niedermeier~\cite{Alber2004} for \pdom.
The main idea is to partition the neighborhoods of both a single vertex and a pair of vertices respectively into three distinct subsets which intuitively classify how much these vertices are confined and how closely they are related to the rest of the graph.
Recall for the following definition that ${N(v) = \{u \in V : \{u,v\} \in E \}}$ and ${N[v] = N(v) \cup \{v\}}$ the closed neighborhood of a vertex $v$.

\begin{definition}
    \label{def:nv}
    Let \G be a graph and let $v \in V$. We split $N(v)$ into three subsets:
    \begin{align}
        N_1(v) & = \{u \in N(v) : N(u) \setminus N[v] \neq \emptyset \}              \\
        N_2(v) & = \{u \in N(v)\setminus N_1(v) : N(u) \cap N_1(v) \neq \emptyset \} \\
        N_3(v) & = N(v) \setminus (N_1(v) \cup N_2(v))
    \end{align}
    For $i,j \in [1,3]$, we denote $N_{i,j} (v) := N_i(v) \cup N_j(v)$. Furthermore, we call a vertex $v'$ \textit{confined} by a vertex $v$, if $N(v') \subseteq N[v]$.
\end{definition}

\begin{figure}[h]
    \label{fig:neighborhoodSingle}
    \begin{equation*}
        \input{fig/tikz/neighborhoods-single-vertex.tikz}
    \end{equation*}
   \caption[The neighbordhood of a single Vertex $v$]{\textit{The neighborhood of a single vertex v split to $N_1(v)$ ({\setulcolor{NONE}\ul{purple}}), $N_2(v)$ ({\setulcolor{NTWO}\ul{blue}}), and $N_3(v)$ ({\setulcolor{NTHREE}\ul{olive}}).
    $N_1(v)$'s are those having neighbors outside $N(v)$, $N_2(v)$'s are a buffer between $N_1(v)$ and $N_3(v)$, and $N_3(v)$-vertices are confined in $N(v)$}.}
\end{figure}
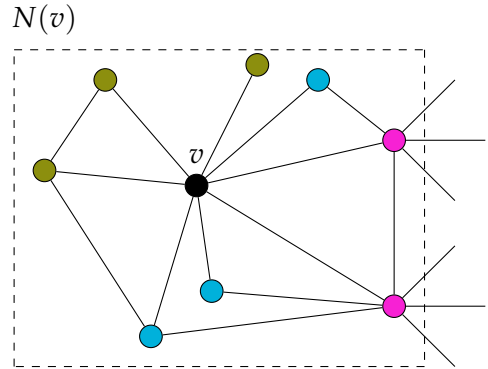

We will shortly discuss the definition of these sets:

\begin{xltabular}{\textwidth}{lX}
\textbf{$\mathbf{N_1(v)}$} & are all the neighbors of $v$ having at least one neighbor outside of $N(v)$ and connect $v$ with the rest of the graph. 
They are the only vertices with the power to dominate vertices outside the neighborhood of $v$. \\

\textbf{$\mathbf{N_2(v)}$} & contains all neighbors of $v$ not from $N_1(v)$ with at least one neighbor in $N_1(v)$. 
These vertices do not have any function as dominators and are placed in between a vertex from $N_1(v)$ and those from  $N_3(v) \cup \{ v \}$. 
They are useless as witnesses because either we can replace them with $v$ (sharing the same neighborhood) or we replace them with a $z \in N_1(v)$ if they function as a witness for $v$. \\

\textbf{$\mathbf{N_3(v)}$} & vertices are sealed off from the rest of the graph. 
They are useless as dominators: For all $z \in N_3(v)$, $N(z) \subseteq N(v)$ by definition and thus, we would always prefer $v$ as a dominating vertex instead of $z$. 
They can still be necessary as a witness for $v$ if $N_1(v) \cup N_2(v) =\emptyset$ but this can only happen if $v$ forms a connected component with only $N_3(v)$ vertices as neighbors. 
We will be using this observation in \cref{rgl:rone} to shrink $\abs{N_3(v)} \leq 1$.
\end{xltabular}

Next, we will extend this notation to a pair of vertices which we will later use in  \cref{rgl:rtwo} to reduce the neighborhood of two vertices. We will classify how strongly the joined neighborhood $N(v) \cup N(w)$ of two vertices is connected to the rest of the graph.

\begin{definition}
    Let \G be a graph and $v,w \in V$. We denote by ${N(v,w) := N(v) \cup N(w)}$ the joined neighborhood $N(v) \cup N(w)$ of the pair $v,w$ and split $N(v,w)$ into three distinct subsets:
    \begin{align}
        N_1(v,w) & = \{u \in N(v,w) \mid N(u) \setminus (N(v,w)\cup \{v,w\}) \neq \emptyset \}  \\
        N_2(v,w) & = \{u \in N(v,w)\setminus N_1(v,w) \mid N(u) \cap N_1(v,w) \neq \emptyset \} \\
        N_3(v,w) & =  N(v,w) \setminus (N_1(v,w) \cup N_2(v,w))
    \end{align}
    For $i,j \in [1,3]$, we denote $N_{i,j}(v,w) = N_i(v,w) \cup N_j(v,w)$.
\end{definition}

Similiar as before, \textbf{$\mathbf{N_1(v,w)}$} contains vertices with at least one neighbor outside $N[v] \cup N[w]$, \textbf{$\mathbf{N_2(v, w)}$}-vertices are in between those from $N_3(v,w) \cup \{v, w\}$ and $N_1(v,w)$, and \textbf{$\mathbf{N_3(v,w)}$} contains vertices isolated from the rest of the graph. 

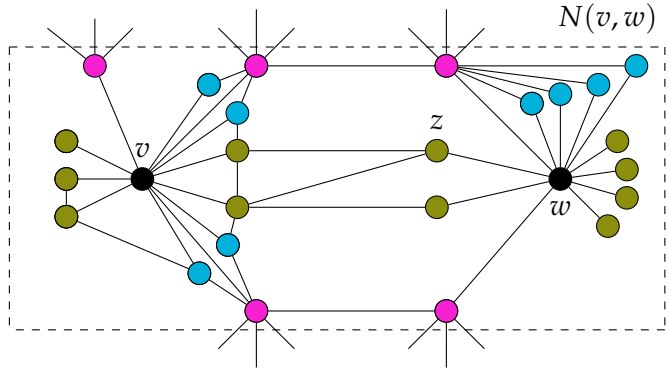
\begin{figure}[]
    \begin{equation*}
        \input{fig/tikz/neighborhoods-two-vertices.tikz}
    \end{equation*}
    \caption[The neighborhood of a pair of vertices]{\textit{The neighborhood of a pair of vertices. Vertices from $N_3(v,w)$ are colored {\setulcolor{NTHREE}\ul{olive}}, $N_2(v,w)$'s {\setulcolor{NTWO}\ul{blue}} and $N_1(v,w)$'s {\setulcolor{NONE}\ul{purple}}.
    Note that $z \in N_1(w)$, because there is an edge to a neighbor of $v$, but $z \notin N_1(v,w)$ (and rather $z \in N_3(v,w)$)}.}
    \label{fig:neighborhoodDouble}
\end{figure}

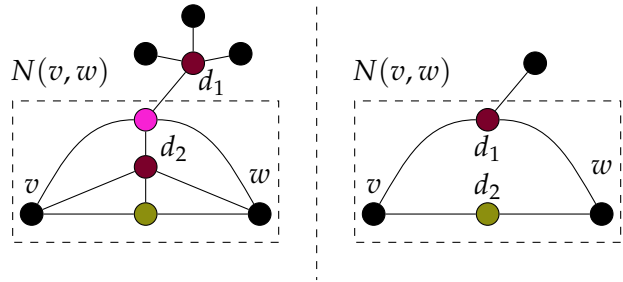
\begin{figure}[]
        \begin{equation*}
            \input{fig/tikz/ntwoimportant.tikz}
    \end{equation*}
    \caption[Example for $N_2(v,w)$ dominating]{Contrary to $N_i(v)$, vertices from all $N_i(v,w)$ can be dominators. 
    Left: $\{d_1, d_2\}$ with $d_2 \in N_2(v,w)$ form the only minimum sds. Right: $d_1 \in N_1(v,w)$ and $d_2 \in N_3(v,w)$ optimal.}
    \label{fig:alldominating}
\end{figure}

A vertex $v \in N_i(v)$ is not necessarily also in $N_i(v,w)$! Observe the vertex $z$ in \cref{fig:neighborhoodDouble}. 
Unlike the sets $N_1(v), N_2(v)$ and $N_3(v)$, in every one of the distinct sets $N_i(v,w)$ ($i \in [3]$) there can be vertices that belong to an sds. See \cref{fig:alldominating} for examples.

\subsection{Reduced Graph}

Before stating the reduction rules, we want to clarify when we consider a graph to be \textit{reduced}. 

\begin{definition}[\cite{Garnero2017}]\label{def:reduced}
    A graph \G is \underline{reduced under} a set of rules if either none of them can be applied to $G$ or the application of any of them creates a graph isomorphic to $G$.
\end{definition}

\cref{def:reduced} differs from the definition usually used where a graph $G$ is \textit{reduced} under a set of reduction rules if none of them can be applied to G anymore (compare e.g.~\cite{Fomin2019}). Some of our reduction rules (\cref{rgl:rone} or \cref{rgl:rtwo}) could be applied \textit{ad infinitum} creating an endless loop that does not change $G$ anymore. Our definition guarantees termination in that case. 
All of the given reduction rules are local and only need the neighborhood of at most two vertices.
They are replaced partially with gadgets of constant size.  
Checking whether the application of one of the rules created an isomorphic graph can therefore be accomplished in constant time because only the neighbors of these vertices might have changed and recognizing this can already be done while each rule itself is active. 
If all rules created an isomorphic graph, we exit the reduction procedure.

\subsection{Regions in Planar Graphs}

Alber, Fellows and Niedermeier~\cite{Alber2004} gave a novel approach to look at planar graphs. In their analysis, they stated a constructive algorithm that decomposes a planar graph into local ``regions''. Intuitively, assume that we have a fixed plane embedding of a planar graph \G. If we pick two distinct vertices $v$ and $w$ from a given \sdom $D \subseteq V$ that are at most of distance two apart, we can try to find two distinct paths from $v$ to $w$ that span up the boundaries of a face and enclose as many other vertices as possible. 

The following definitions are based on those given by Garnero and Sau in~\cite[arXiv v2]{Garnero2018} and will lead towards a clean definition of a \textit{region} and what we understand as a \dreg. More detailed explanations and concrete examples can be found in their paper.

\begin{definition}
    Two simple paths $P_1, P_2$ in a plane graph G are \underline{confluent} if at least one of the following statements holds:
    
    \begin{enumerate}
        \item they are vertex-disjoint;
        \item they are edge-disjoint and for every common vertex $u$, if $v_i, w_i$ are the neighbors of $u$ in $p_i$, for $i \in [1,2]$, it holds that $[v_1, w_1, v_2, w_2]$;
        \item they are confluent after contracting common edges.
    \end{enumerate}
\end{definition}


\begin{definition}
    Let \G be a plane graph and let $v,w \in V$ be two distinct vertices. A \underline{region $R(v,w)$} (also denoted as $vw$-region $R$) is a closed subset of the plane, such that:
    \begin{enumerate}
        \item the boundary of $R(v,w)$ is formed by two confluent simple $vw$-paths with length at most 3
        \item every vertex in $R(v,w)$ belongs to $N(v,w)$, and
        \item the complement of $R(v,w)$ in the plane is connected.
    \end{enumerate}
    
    The \underline{poles} of R are the vertices $v$ and $w$. 
    The boundary paths are the two $vw$-paths that form $\partial R$.
    We denote with $\partial R$ the set of vertices on the boundary of $R$ (including the poles) and by $V(R)$ the set of vertices laying (on the plane embedding) in $R$. 
    Furthermore, we call $\abs{V(R)}$ the \underline{size} of the region of the region.
    
\end{definition}

\begin{definition}
    Two regions $R_1$ and $R_2$ are non-crossing, if:
    \begin{enumerate}
        \item $(R_1 \setminus \partial R_1) \cap R_2 = (R_2 \setminus \partial R_2) \cap R_1 = \emptyset$, and
        \item the boundary paths of $R_1$ are pairwise confluent with the ones in $R_2$.
    \end{enumerate}
\end{definition}

We now have all the definitions ready to formally define a maximal \dreg on planar graphs:

\begin{definition}\label{def:region}
    Given a plane graph \G and $D\subseteq V$, a \underline{\dreg} of G is a set $\mathfrak{R}$ of regions with poles in D such that: 
    \begin{enumerate}
        \item for any $vw$-region $R \in \mathfrak{R} $, it holds that $D \cap V(R) = \{v, w\}$, and
        \item all regions are pairwise non-crossing.
    \end{enumerate}
    We define $V(\mathfrak{R}) = \bigcup\limits_{R \in \mathfrak{R}} V(R)$ to be all vertices enclosed in the region. 
    
    \noindent A \dreg is \underline{maximal} if there is no region $R \notin \mathfrak{R}$ such that $\mathfrak{R}' = \mathfrak{R} \cup \{R\}$ is a \dreg~with $V(\mathfrak{R}) \subsetneq V(\mathfrak{R}')$.
\end{definition}

Intuitively, the first condition ensures that only a boundary path can be shared and the second one is that these boundary paths are indeed the frontier between these regions.
\cref{fig:maxRegionDecompose} gives an example of how to decompose a graph into a maximal \dreg for a given sds $D$ of size $3$.

\begin{figure}[!ht]
    \begin{equation*}
        \input{fig/tikz/region-example.tikz}
    \end{equation*}
     \caption[Region Decomposition]{\textit{Left: A maximal \dreg $\mathfrak{R}$, where $D = \{d_1,d_2,d_3\}$ form am sds. 
   There are two regions between $d_2$ and $d_1$ ({\setulcolor{MATHARED}\ul{red}} and {\setulcolor{MATHABLUE}\ul{purple}}), one region between $d_1$ and $d_3$ ({\setulcolor{MATHABLUE}\ul{purple}}) and one region between $d_2$ and $d_3$ ({\setulcolor{MATHAGREEN}\ul{green}}). 
    Observe that this \dreg, some neighbors of $d_1$ are not covered by any $vw$-region for any $v,w \in D$. 
    Our reduction rules are going to take care of them and bound this number of vertices to obtain the kernel. 
    Right: The corresponding underlying multigraph $G_{\mathfrak{R}}$. Every edge denotes a region between $d_i$ and $d_j$.}}\label{fig:maxRegionDecompose}
\end{figure}

We introduce a special subset of a region, namely a \textit{simple region} where every vertex is a common neighbor of $v$ and $w$. 
They will appear in many unexpected astonishing places and are an important tool to operate on small parts of a plane graph.
We will use the upcoming \cref{rgl:rthree} to bound the size of \textit{simple regions}. 

\begin{definition}
    A simple $vw$-region is a $vw$-region such that:
    \begin{enumerate}
        \item its boundary paths have length at most 2, and
        \item $V(R) \setminus \{v,w\} \subseteq N(v) \cap N(w)$.
    \end{enumerate}

\end{definition}

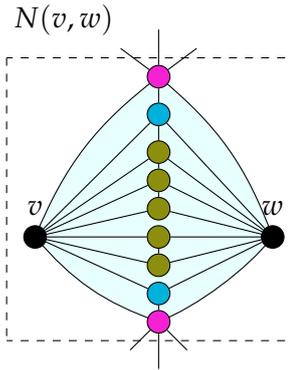
\begin{figure}[!ht]
    \begin{equation*}
        \input{fig/tikz/simple-region-example.tikz}
    \end{equation*}
   \caption[A simple region]{\textit{A simple region with two vertices from $N_1(v,w)$ ({\setulcolor{NONE}\ul{purple}}) setting the boundary, two vertices from $N_2(v,w)$ ({\setulcolor{NTWO}\ul{blue}}) and some vertices from $N_3(v,w)$ ({\setulcolor{NTHREE}\ul{olive}}) in between.}}
   \label{fig:simpleRegionExample}
\end{figure}

\cref{fig:simpleRegionExample} shows an example of a simple region containing 9 distinct vertices.

In the analysis, we will also use properties of the \textit{underlying multigraph} of a \dreg $\mathfrak{R}$. Refer to \cref{fig:maxRegionDecompose} for an example.

\begin{definition}\label{def:unterlyingMG}
    Let \G be a plane graph, let $D \subseteq V$ and let $\mathfrak{R}$ be a \dreg of G. The underlying multigraph $G_\mathfrak{R} = (V_\mathfrak{R}, E _\mathfrak{R})$ of $\mathfrak{R}$ is such that  $V_\mathfrak{R} = D$ and there is an edge $\{v,w\} \in E_\mathfrak{R}$ for each vw-region $R(v,w) \in \mathfrak{R}$.
\end{definition}

\section{The Big Picture}
The following \Cref{fig:overview} gives an overview on how to obtain a linear kernel for \psdom.
 We will first define three polynomial-time reduction rules \cref{rgl:rone,rgl:rtwo,rgl:rthree}  and prove that they preserve the solution size $k$. 
We will use the existence of a maximal \dreg~$\mathfrak{R}$ on planar graphs to bound the number of vertices that fly around a given region $R \in \mathfrak{R}$ after the reduction rules have been exhaustively applied. 
Additionally, \cref{rgl:rone} helps us to bound on the number of vertices that are not enclosed by any $R$. 

We will often encounter hidden simple regions which are reduced by \cref{rgl:rthree} and therefore of constant size by \cref{lemma:simpleregionbound}. As we know that the total number of regions R in the \dreg is linear in $k$, we obtained a linear kernel for the \psdom as well.

\begin{figure}[!ht]
    \begin{equation*}
    \scalebox{0.78}
    {
        \input{fig/tikz/obtainingKernel.tikz}
    }
    \end{equation*}
    \caption[Structure of the Proofs]{\textit{The plan for obtaining a linear kernel for \psdom.
    After \cref{rgl:rone,rgl:rtwo,rgl:rthree} have been applied, we analyze their impact and we prove the following bounds for any sds $D$.
    In \circled{1}, we try to bound the maximal number of vertices \underline{inside} a region $R$ by analyzing the disjoint sets $N_1(v,w), N_2(v,w)$ and $N_3(v,w)$ independently for two given poles $v,w \in D$. 
    Then, in \circled{2}, we bound the number of vertices that lay \underline{outside} and are not enclosed into any region of a maximal \dreg $\mathfrak{R}$. 
    The arrows point to these regions.
    Finally, in \circled{3}, we use the fact that a $\mathfrak{R}$ exists with at most $3 \cdot \abs{D} -6$ regions.}} \label{fig:overview}
\end{figure}

\section{The Reduction Rules}

Following the ideas proposed by Garnero and Sau~\cite[arXiv v2]{Garnero2018}, we state adjusted reduction rules leading to a linear kernel after exhaustive application. 
Especially in \cref{rgl:rtwo}, we relied on the second revision of the paper electronically submitted to \textit{arXiv}.
In the later version, the authors improved their results by giving a better kernel size, but it turned out that these rules just work for \tdom, but not for \sdom anymore.
After looking deeper into the structure of a simple region, we were able to give a slightly more complex reduction \cref{rgl:rthree} having the same bounds as in~\cite{Garnero2018}.  
Our main challenge was to preserve the witness properties of an sds.
This is because a vertex inside a region can be important as a witness for vertices in another region. 
A tds has fewer side effects as the witnesses are direct neighbors of the dominators and they do not have too much influence on the rest of the graph as the witnesses have in an sds.
Therefore, we had to make sure that the effect of vertices to more distanced vertices is preserved by the reduction.

\subsection{Reduction Rule I: Shrinking $N_3(v)$}

\begin{figure}[!ht]
    \begin{equation*}
        \input{fig/tikz/ruleOne.tikz}
    \end{equation*}
    \caption[Application of \cref{rgl:rone}]{\textit{Simplifying $N_{23}(v)$: As $N_3(v) \geq 1$, we remove $N_{23}(v)$ and add a new witness $v'$. $N_1(v)$ remains untouched.}}
    \label{fig:ruleOne}
\end{figure}
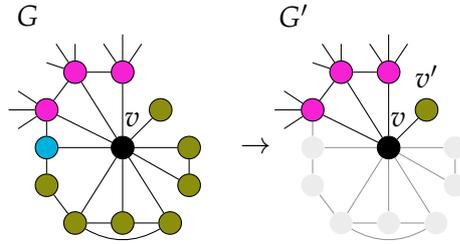

The idea of the first rule is the observation that a vertex $v' \in N_{2,3}(v)$ dominates $v$ and possibly vertices from $N_2(v)$ and $N_3(v)$. As $N(v') \subseteq N(v)$ and the \cref{fact:witnessTwin} that a witness for $v'$ is also a witness for $v$, we can use $v$ instead of $v'$ as a dominating vertex. Therefore, we can remove $N_{2,3}$ from the graph. Nevertheless, $v'$ can be a witness for $v$ itself and might be required in a solution. Our rule ensures that at least one $N_3(v)$-vertex is preserved. An example of this rule is shown in \cref{fig:ruleOne}.

\begin{fact}\label{fact:witnessTwin}
Let \G, $v \in V$ and $v' \in N_{2,3}(v)$. Any witness $w \neq v$ for $v'$ is also a witness for $v$. 
\end{fact}
\begin{proof}

Assume $v' \in N_{2,3}(v)$ and a witness $w \neq v$ for $v$ with $d(w,v') \leq 2$. 
By definition of $N_{2,3}(v)$, $N(v') \subseteq N[v]$ and $v'$ is \textit{confined} inside the neighborhood of $v$. 
Therefore every path from $v'$ to the witness $w$ within two steps must pass at least one other vertex $p \in N(v) \cup \{v\}$ and $\{w\} \in N(p)$.
If $p = v$ then $w$ is a direct witness for $v$ and if $p \in N(v)$, there exists a path of length 2 from $v$ to $w$.
\end{proof}

\begin{minipage}{\textwidth}
\begin{rgl}\label{rgl:rone}
    Let \G be a graph and let $v \in V$. If $\abs{\Nthreev} \geq 1$:
    \begin{itemize}
        \item remove $N_{2,3}(v)$ from G,
        \item add a vertex $v'$ and an edge $\{v, v'\}$.
    \end{itemize}
\end{rgl}
\end{minipage}

~\\
\noindent We will now prove the correctness of this rule.

\begin{lemma}\label{lemma:correctnessone}
    Let \G be a graph and let $v \in V$. If $G'$ is the graph obtained by applying \cref{rgl:rone} on $G$, then $G$ has an sds of size $k$ if and only if $G'$ has an sds of size $k$.
\end{lemma}
\begin{proof}
        Assume $D$ to be an sds set in $G$ of size $k$. 
        Because \cref{rgl:rone} was applied, we know that $N_{3}(v) \neq \emptyset$ in $G$.
        $N_3(v)$ must be dominated, so $\abs{D \cap (N_{2,3}(v) \cup \{v\})} \neq \emptyset$ in $G$. 
        Denote one of those as $d$.
        If $v \notin D$, we know by \cref{fact:witnessTwin} that a witness for $d \neq v$ is also a witness for $v$ and therefore, we replace $d$ with $v$ in $D$.
            Now assume $v \in D$ and $N_{2,3}(v)$ is already dominated by $v$.
        If $\abs{D \cap (N_{2,3}(v))} \geq 1$, set $D' = D \setminus N_{2,3}(v) \cup \{v'\}$, else $D' = D$. 
        A $z \in D \cap N_{2,3}(v)$ could have been a witness for $v$ and therefore we choose $v' \in D'$ in the first case preserving witnesses. 
        In both cases, $v'$ is dominated by $v$ and $\abs{D} \leq \abs{D'}$.

        Let $D'$ be an sds in $G'$. We assume that $v \in D'$ because $v'$ has to be dominated and $v$ is a better choice than $v'$.
        If $v' \in D'$ we have to preserve a witness for $v$ in $G$. We know $N_3(v) \neq \emptyset$ and therefore replace it with an arbitrary vertex $d \in N_3(v)$ in $G$. 
        If a witness for $v$ came from outside of $N_{2,3}(v,w)$, they have not been touched by the reduction
        Therefore, if $v' \in D'$, we set $D = D' \cup \{d\} \setminus \{v'\}$ for a $d \in N_3(v)$ and otherwise $D = D'$. 
        In both cases, $N_{2,3}(v)$ is dominated by $v$ and $\abs{D} = \abs{D'}$.
\end{proof}

\begin{lemma}\label{complex:rone}
    A plane graph $G$ of $n$ vertices is reduced under \cref{rgl:rone} in time $\mathcal{O}(n)$.
\end{lemma}
\begin{proof}
    As \cref{rgl:rone} stayed the same, the proof directly follows from the two-phase algorithm proposed in~\cite[Lemma 2]{Alber2004}. 
\end{proof}

Note that we need our definition of a reduced instance given in \cref{def:reduced}. 
If \cref{rgl:rthree} is being applied, it will still leave us with a vertex $z\in N_3(v)$ allowing this rule to be applied over and over again.
\subsection{Reduction Rule II: Shrinking the Size of a Region}

The second rule is the heart of the whole reduction and minimizes the neighborhood of two distinct vertices. The rule follows Garnero and Sau's approach~\cite{Garnero2018} for \ptdom. Especially \cref{rgl:rtwo} given in~\cite[arXiv v2]{Garnero2018} was not transferable to \psdom, because it heavily relies on the property of a total dominating set that a witness $w$ for $v$ \textbf{must} be a direct neighbor of $w$. 
In an sds, the witness is allowed to be further away which must be taken into account when reducing the graph.


It can be observed that in the worst case, four vertices are required to semitotally dominate $N(v,w)$ for $v,w \in V$: $v$, $w$ and two witnesses for them. 
For instance, observe the graph consisting of two distinct $K_{1,m}$ with $m \in \mathbb{N}$ with centers $v$ and $w$.

Before we give the concrete reduction rule, we need to define three sets. Intuitively, we first try to find a set $\tilde D \subseteq N_{2,3}(v,w)$ of size at most three dominating $N_3(v,w)$ without using $v$ or $w$. If no such set exists, we allow $v$ (resp. $w$) and try to find one again. 
If we now find such a set, we can conclude that $v$ ($w$) must be part of a solution.
If such a set does not exist, we set $v,w$ and two neighbors to be in $D$, which is guaranteed to be a solution.

\begin{definition}\label{def:dvv}
Let \G be a graph and let $v,w \in V$. We now consider all the sets that can dominate $N_3(v,w)$:
\begin{align}
    \Dvw & = \{ \tilde D \subseteq N_{2,3}(v,w)            \mid N_3(v,w) \subseteq \bigcup_{v \in \tilde D} N(v),\ |\tilde D| \leq 3                  \} \\
    \Dv  & = \{ \tilde D \subseteq N_{2,3}(v,w) \cup \{v\} \mid N_3(v,w) \subseteq \bigcup_{v \in \tilde D} N(v),\ |\tilde D| \leq 3,\ v \in \tilde D \} \\
    \Dw  & = \{ \tilde D \subseteq N_{2,3}(v,w) \cup \{w\} \mid N_3(v,w) \subseteq \bigcup_{v \in \tilde D} N(v),\ |\tilde D| \leq 3,\ w \in \tilde D \}
\end{align}

Furthermore, we shortly denote $\bigcup \Dv = \bigcup\limits_{D \in \Dv}D $ and $\bigcup \Dw = \bigcup\limits_{D \in \Dw}D$.
\end{definition}

$\Dvw$ contains subsets of $N_{2,3}(v,w)$ of size at most three that dominate $N_3(v,w)$.
If $\Dvw \neq \emptyset$, we cannot reduce much because all subsets could be part of a minimum solution that does not use $v$ and $w$ at all.
On the other hand, the sets $\Dv, \Dw$ try to activate $v$ (resp. $w$) and two more vertices in $N_{2,3}(v, w)$ to dominate $N_3(v,w)$ with less than four vertices.
If at least one of $\Dv,\Dw \neq \emptyset$, we know that there are better solutions than just selecting both $v,w$ and two neighbors with $v$ or $w$ and then we can reduce $N(v,w)$ respectively.

Note: Assuming that $v$ and $w$ are closely connected with $d(v,w) \leq 2$, it might suffice to consider only sets of size at most three, because an intermediate vertex could witness $v$ and $w$ at the same time. 
In the later analysis, the \dreg exactly creates regions around $N(v,w)$ requiring at least one path from $v$ to $w$ of length two. 
As the following rule is only used to locally investigate such regions, we could add the requirement of a distance of two to it and work with sets of size at most three. 
We believe that this could further improve the kernel and a deeper discussion can be found at the end of this thesis, in  \cref{ch:closing}.

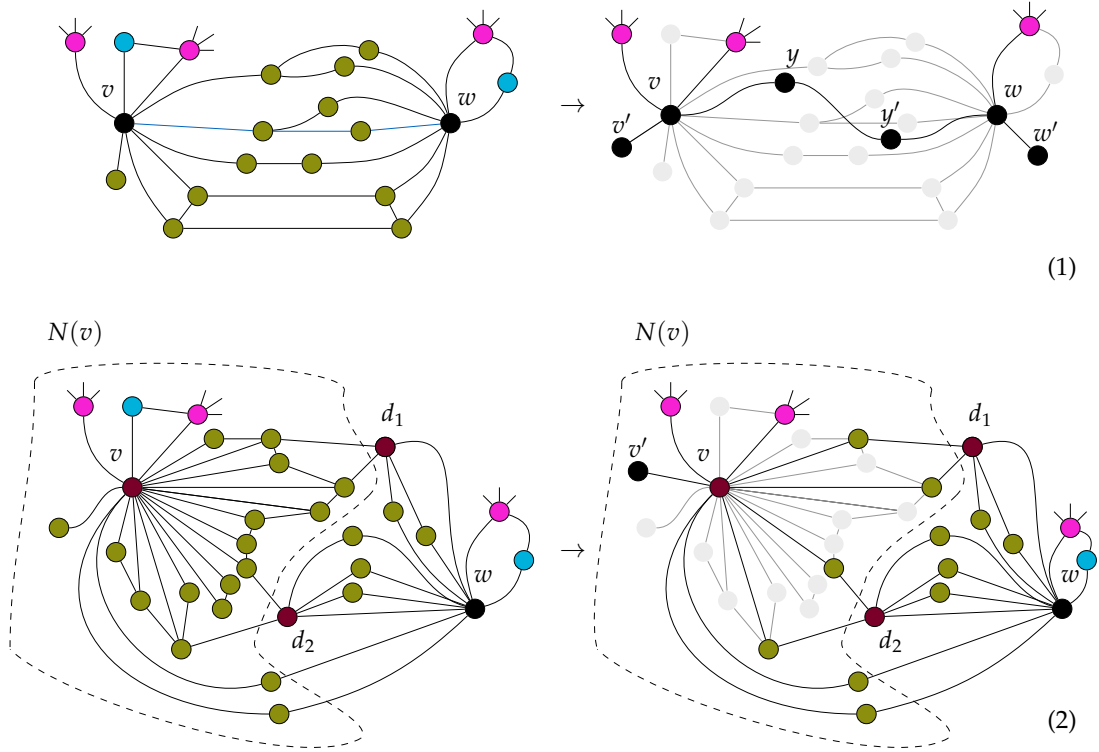
\begin{figure}[ht]
    \begin{equation*}
    \resizebox{\textwidth}{!}{
        \input{fig/tikz/ruletwo.tikz}
    }
    \end{equation*}
    \caption[Application of \cref{rgl:rtwo}]{\textit{An application of \cref{rgl:rtwo}: (1) $\Dv = \Dw = \emptyset$ and \cref{case:c1} applies.
    Both $v$ and $w$ must be in the sds and we can remove $N_{2,3}(v,w)$ and add $\{v',w'\}$. Furthermore, we need to preserve a path of length 3 from $v$ to $w$ by adding $\{y,y'\}$ as well. 
    (2) \cref{case:c2} has been applied and the $N_{2,3}(v)$ removed. Observe that those vertices that cross the dotted line are vertices from $N_1(v)$ and not removed. 
    ~\cref{case:c3} is symmetric.}
    }
    \label{fig:ruleTwo}
\end{figure}

We are now ready to state \cref{rgl:rtwo}. An exemplary application is shown in \cref{fig:ruleTwo}.

\begin{rgl}\label{rgl:rtwo}
    Let \G be a graph and $v, w$ be two distinct vertices from $V$. If $\mathbf{\Dvw = \emptyset}$ (\cref{def:dvv}) we apply the following:
    \begin{caseof}
        \case{if $\Dv =  \emptyset$ and $D_w = \emptyset$\label{case:c1}}

        \vspace{-5mm}
        \begin{itemize}
            \item Remove $N_{2,3}(v,w)$
            \item Add vertices $v'$ and $w'$ and two edges $\{v, v'\}$ and $\{w, w'\}$
            \item If there was a common neighbor of $v$ and $w$ in $N_{2,3}(v,w)$, add another vertex $y$ and two connecting edges  $\{v, y\}$ and $\{y, w\}$
            \item If there was no common neighbor of $v$ and $w$ in $N_{2,3}(v,w)$, but at least one path of length three from $v$ to $w$ via only vertices from $N_{2,3}(v,w)$, add two vertices $y$ and $y'$ and connecting edges $\{v,y\}$, $\{y, y'\}$ and $\{y', w\}$
        \end{itemize}
        \case{if $\Dv \neq  \emptyset$ and $D_w = \emptyset$\label{case:c2}}

        \vspace{-5mm}
        \begin{itemize}
            \item Remove $N_{2,3}(v)$
            \item Add $\{v, v'\}$
        \end{itemize}
        
        \case{if $\Dv =  \emptyset$ and $D_w \neq \emptyset$\label{case:c3}} This case is symmetrical to \cref{case:c2}.
    \end{caseof}
\end{rgl}

In \cref{case:c1}, we know by \cref{fact:f2} that $v$ and $w$ must be in $D$. 
Therefore, we introduce two forcing vertices $v'$ and $w'$ in $G'$ and remove $N_{2,3}(v,w)$ as these vertices are dominated by $v$ and $w$.
Removing $N_{2,3}(v,w)$ entirely, could lose paths of length less than three and destroy solutions: If $d(v,w) \leq 2$ then $v$ can directly witness $w$ and if $d(v,w) = 3$, one vertex on this path could be a witness for both.


Again by \cref{fact:f2} we know for \cref{case:c2,case:c3} that $v \in D$ (resp. $w \in D$) and similar to \cref{rgl:rone} we can simplify the neighborhood $N_{2,3}(v)$ ($N_{2,3}(w)$). 
\cref{fact:f1} states, that these vertices are only useful for witnessing $v$, but do not go beyond what $v$ already witnesses. 
Observe that removing $N_{2,3}$ cannot break any connectivity as all vertices in  $N_{2,3}(v)$ are confined in $v$. 

The case where $\Dv \neq \emptyset \wedge \Dw \neq \emptyset$ is not necessary for reasons clarified later.
Before proving \cref{rgl:rtwo} we will deduce some \textit{facts} which are implied by the definitions above. These facts justify the definition of the sets $\Dvw$, $\Dv$ and $\Dw$.

\begin{fact}\label{fact:f1}
    Let \G be a graph, let $v,w \in V$, and let $G'$ be the graph obtained by the application of \cref{rgl:rtwo} on $v,w$. If $\Dvw = \emptyset$, then $G$ has a solution of size at most $k$ if and only if it has a solution of size at most $k$ containing at least one of the two vertices $\{v,w \}$.
\end{fact}
\begin{proof}
Because $\Dvw = \emptyset$, an sds of $G$ must contain at least one of $\{v, w \}$ or at least four vertices from $N_{2,3}(v,w)$. 
In the second case, these four vertices can be replaced with $v$, $w$ and two neighbors of $v$ and $w$ as witnesses.
In all cases $k$ becomes either smaller or stays unchanged.
\end{proof}

The second fact states that if  $\Dv$ (resp. $\Dw$) is empty, too, we only need to consider solutions containing $w$ (resp. $v$):

\begin{fact}\label{fact:f2}
    Let \G be a graph, let $v,w \in V$, and let $G'$ be the graph obtained by the application of \cref{rgl:rtwo} on $v, w$. If $\Dvw = \emptyset$ and $\Dw = \emptyset$ (resp. $\Dv = \emptyset$) then $G'$ has a solution of size at most $k$ if and only if it has a solution containing $v$ of size at most $k$ (resp. $w$).
\end{fact}
\begin{proof}
As $\Dv = \emptyset$, no set of the form $\{v\}$, $\{v, u\}$ or $\{v, u, u'\}$ with $u, u' \in N_{2,3}(v,w)$ can dominate $N_3(v,w)$. 
Since also $\Dvw = \emptyset$ any sds of $G$ must contain $v$ or at least four other vertices by \cref{fact:f1}.
In the latter case, we replace these four vertices with $v$, $w$ and two additional neighbors as witnesses.
Again, $k$ becomes either smaller or stays unchanged.
\end{proof}

Using these two facts, we can now prove the correctness of \cref{rgl:rtwo}.

\begin{lemma}\label{lemma:correctnesstwo}
    Let \G be a plane graph, $v, w \in V$ and \GB be the graph obtained after application of \cref{rgl:rtwo} on the pair $\{v, w\}$. Then $G$ has an sds of size at most $k$ if and only if $G'$ has an sds of size at most $k$.
\end{lemma}
\begin{proof}

We will prove the claim by analyzing the different cases independently.     

Assume an sds $D$ in $G$ and by assumption $\Dvw = \emptyset$. 
We show that $G'$ has an sds with $\abs{D'} \leq \abs{D}$. 
    \begin{enumerate}
        \item Assume $ \Dv = \emptyset  \wedge \Dw = \emptyset $. By \cref{fact:f2}, both $v, w \in D$.
        Therefore, $v'$, $w'$, and potentially $y$ and $y'$ are dominated by $v$ or $w$ in $G'$.
        
        We have three cases: Either $v$ and $w$ have their own witnesses ($d(v,w) > 2$), or they share one witness on a path from $v$ to $w$ ($d(v,w)= 2$) is required, or they witness each other directly. ($d(v,w) <2$).
        Not that the rule preserves these distances.

        We will now build an sds $D'$ in $G'$ depending on which vertices from $D \cap N_{2,3}(v,w)$ have been removed. 

        \begin{itemize}
            \item If the rule has not removed any $d \in D$, we simply set $D' = D$. 
            If $v$ was a witness for $w$ (and vice versa), \cref{rgl:rtwo} will preserve the distance by introducing the vertex $y$. 
            Furthermore, a direct edge $\{v,w\}$ will be preserved and therefore, no witness relations being destroyed.

            \item If $d(v,w) > 3$, then $v$ and $w$ are not sharing any common witnesses. 
            If the rule has removed a vertex from $D \cap N(v)$, we set $D' = D \setminus N_{2,3}(v,w) \cup \{v'\}$.
            If the rule has removed a vertex from $D \cap N(w)$, we  set $D' = D \setminus N_{2,3}(v,w) \cup \{w'\}$.
            If the rule has removed a vertex from $(D \cap N(v))$ and a vertex from $(D \cap N(w))$, we set $D' = D \setminus N_{2,3}(v,w) \cup \{v', w'\}$.
            \item If $d(v,w) = 3$ and the vertices $y$ and $y'$ get introduced preserving one path from $v$ to $w$, because there has been a path via $N_{2,3}(v,w)$-vertices containing a single witness for both $v$ and $w$.
            If the rule removed a dominating vertex from $D \cap N_{2,3}(v, w)$, we set $D' = D \setminus N_{2,3}(v,w) \cup \{y\}$. Note that we could also choose $y' \in D'$, because $y$'s only function is to be a single witness for $v$ and $w$ and every other vertex it could be a witness for, will also be witnessed by $v,w \in D'$ (\cref{fact:f1}).
            \item If $d(v,w) \leq 2$, then $v$ and $w$ directly witness each other and the reduction must preserve this relation, which is accomplished by introducing the single bridging vertex $y$. 
            Even if the rule has removed a vertex $z \in D \cap N_{2,3}(v,w)$, we can ignore that, because \cref{fact:f1} states that $v$ and $w$ will witness the same vertices as $z$ did. Hence, we set $D' = D \setminus N_{2,3}(v,w)$.
        \end{itemize}
        
        In all of the cases, it follows that $D'$ is an sds of $G'$ with $\abs{D'} \leq \abs{D}$.
        
        \item  Now, assume $ \Dv \neq \emptyset  \wedge \Dw = \emptyset$: As $\Dw = \emptyset$ and \cref{fact:f2}, we know that $v \in D$ and $v$ dominates $N_{2,3}(v)$. 
        If a vertex $d \in D \cap N_{2,3}(v)$ was removed, we set $D' = D \setminus N_{2,3}(v) \cup \{v'\}$, and $D' = D$ otherwise.
        Deleting $d$ does not destroy the witness properties of the graph, because by \cref{fact:f1} $v$ already witnesses everything $d$ could. 
        If $d$ was a witness for $v$, we have replaced it with $v'$ in $G'$ as the new witness.
        All witnesses outside of $N_{2,3}(v)$  for $v$ have not been modified and clearly, $\abs{D'} \leq \abs{D}$ holds.
        \item  $ \Dv = \emptyset  \wedge \Dw \neq  \emptyset $: The proof is symmetrical to the previous case.
    \end{enumerate}

    Let $D'$ be an sds in $G'$ and $\Dvw =  \emptyset$. 
    We show that $G$ has an sds $D$ with $\abs{D} \leq \abs{D'}$ by case distinction. 
    \begin{enumerate}
        \item  $\Dv = \emptyset \wedge \Dw = \emptyset$: In any case we know that $v,w \in D$ to dominate $v'$ and $w'$ and therefore dominating $N_{2,3}(v,w)$ in $G$. 
        To preserve the distance $d(v,w)$ the rule might have introduced additional vertices $y$ and $y'$ in the following two cases:            
        \begin{itemize}
            \item If only $y$ was introduced we know that there was a common neighbor $n \in N(v) \cap N(w)$ of $v$ and $w$. $y$ allows $v$ to witness $w$ (and vice versa) and is not part of a solution itself. (assuming $y \notin D'$). Hence, we set $D = D'$.
            \item If $y$ and $y'$ were added, a solution could use one of them to provide a single witness for $v$ and $w$. There exists a path $p = (v, n_1, n_2, w)$ from $v$ to $w$ in $G$ only using vertices from $N_{2,3}(v,w)$. As $n_1$ and $n_2$ both witness $v$ and $w$, we put one of them in $D$ if at least one of $y$ or $y$ are dominating vertices in $G'$.
            Hence, if $y \in D'$ or $y \in D'$, we set $D = D' \setminus \{y,y'\} \cup \{n_1\}$.
        \end{itemize}
        \item  $\Dv \neq \emptyset \wedge \Dw = \emptyset$: Clearly, $v \in D'$ to dominate $v'$. If $v \in D'$, we set $D =  D' \setminus \{v'\} \cup d$ for some vertex $d \in N_{2,3}(v,w)$ and otherwise $D = D'$. If $v'$ was the witness of $v$, it is now replaced by $d$ and $D$ is an SDS with $\abs{D} \leq \abs{D'}$.
        \item  $\Dv = \emptyset \wedge \Dw \neq \emptyset$: Symmetrical to previous case.
    \end{enumerate} 
    In all cases, we have shown that $\abs{D} \leq \abs{D'}$ and $D$ is an sds of $G$.

\end{proof}

We will now prove that the reduction takes polynomial time:

\begin{corollary}\label{complex:rtwo}
\cref{rgl:rtwo} can be applied in $\mathcal{O}(d(v) + d(w))$ time on two vertices $v$ and $w$.
\end{corollary}
\begin{proof}
As mentioned in~\cite{Garnero2018}, we can construct $\Dvw, \Dv$ and $\Dw$ in $\mathcal{O}(2^{\sqrt{3}})(d(v) + d(w))$ fpt time thanks to the algorithn \pdomp~\cite{Alber2002}.
The transformation can be done again in order $\mathcal{O}(d(v) + d(w))$.
\end{proof}

\subsection{Reduction Rule III: Shrinking Simple Regions}

We will now introduce a rule that simplifies \textit{simple regions}.
This reduction rule will be our \href{https://en.wikipedia.org/wiki/Swiss_Army_knife}{swiss-army-knife} used in many places a \textit{simple region} can be found.
Interestingly, the idea of having such a rule separately was introduced in a later version of Garnero and Sau's~\cite{Garnero2018} paper for \ptdom.
In their first revision, this rule was circuitously included in the definition of \cref{rgl:rtwo}.
Later they decided to decouple it into a separate rule, which makes the arguments easier to follow and improves the kernel size by allowing a more sophisticated analysis.

Recall that in a \textit{simple region} $R$ the entire neighborhood of the poles $v$ and $w$ is shared.  
By planarity, a \textit{simple region} has at most two vertices from $N_1(v,w)$ (namely the border $\partial$ R), two vertices from $N_2(v,w)$ connected to the border and unlimited $N_3(v,w)$ vertices squeezed in the middle.
Unlike in a normal $vw$-region, every \textit{simple region} can be semitotally dominated by at most two vertices: $v$ and $w$, because $v$ instantly witnesses $w$ and $V(R)$ is dominated by them as well and we can assume $V(R) \cap D = \emptyset$.
This does not hold for a tds, because $v$ and $w$ do not witness each other and probably a vertex from inside the region must be used as a witness for a pole.
But on the other side, the witness for a vertex $d \in D \cap V(R \setminus \partial R)$ in an sds can have a witness outside the region and a solution without $v$ or $w$ can exist. 
In a tds where we know that $v$ or $w$ must be part of a solution, we can replace all inner vertices with one vertex $y$ and simulate an $\mathnormal{OR}$ gadget.
If $y \in D$ in a tds $D$, then immediately either $v \in D$ or $w \in D$ to witness $y$ as well. But if $y \in D$ then $v \in D$ or $w \in D$ to dominate $y$.
However, for an sds the situation is different because the witness for $y$ is not necessarily $v$ or $w$.

A tds is easier to handle because of the strict witness property and therefore, we had to come up with a novel reduction rule for \textit{simple regions}.

\begin{rgl}\label{rgl:rthree}
    Let \G be a plane graph, $v, w \in V$ and $R$ be a simple region between $v$ and $w$. If $\abs{V(R) \setminus \{v, w\}} \geq 5$ apply the following:

    \begin{caseof}
        \case{If $G[R\setminus\partial R] \cong P_3$, then:\label{case:rto}}

            \vspace{-5mm}
            \begin{itemize}
                    \item remove $V(R\setminus\partial R)$
                    \item add vertex $y$ with edges $\{v, y\}$ and $\{y, w\}$
            \end{itemize}
        \case{If $G[R\setminus\partial R] \ncong P_3$, then\label{case:rtt}}

            \vspace{-5mm}
            \begin{itemize}
                    \item remove $V(R\setminus\partial R)$
                    \item add vertices $y$, $y'$ and four edges $\{v,y\}$, $\{v, y'\}$, $\{y, w\}$ and $\{y', w\}$
            \end{itemize}
        \end{caseof}
Recap that we denoted $\partial R$ as the set of boundary vertices of the  \textit{simple region} $R$, which includes $v$ and $w$ and possibly up to two vertices on the border of $R$. 
We defined $G[V]$ to be the induced subgraph on the vertices in $V$.

\end{rgl}
Before proving the correctness of this rule, we will quickly discuss the idea behind it.

In \cref{case:rtt} we need at most two vertices to dominate $V(R \setminus \partial R)$ because only an induced path with three vertices could be dominated by one vertex.
The best way to do that is using $v$, $w$ or both and adding two vertices $y$ and $y'$ to simulate an $\mathrm{OR}$ gadget.
Either $\{y,y'\} \subseteq D$, $v \in D$ or $w \in D$ for any sds $D$ in $G$.
If we would add only one vertex $y$ (see \cref{case:rto}), then $y \in D$ dominates $v$ and $w$, and could potentially be witnessed by another neighbor outside $V(R \setminus \partial R)$.
This would lead to $v,w \notin D$ in $G'$ and a smaller solution, although $G$ requires at least one of them.
To preserve this property, our construction needs to add two vertices.

\begin{figure}[!ht]
    \begin{equation*}
        \input{fig/tikz/rulethree.tikz}
    \end{equation*}
    \caption[Application of \cref{rgl:rthree}]{\textit{Both cases of the application of \cref{rgl:rthree}. Left: the vertices inside the region are not isomorphic to a $P_3$, which means that \cref{case:rtt} will be applied and two new vertices being added. Right: They are isomorphic to a $P_3$ and we can replace the whole inner region with one single vertex by \cref{case:rto}.}}
    \label{fig:rulethree}
\end{figure}
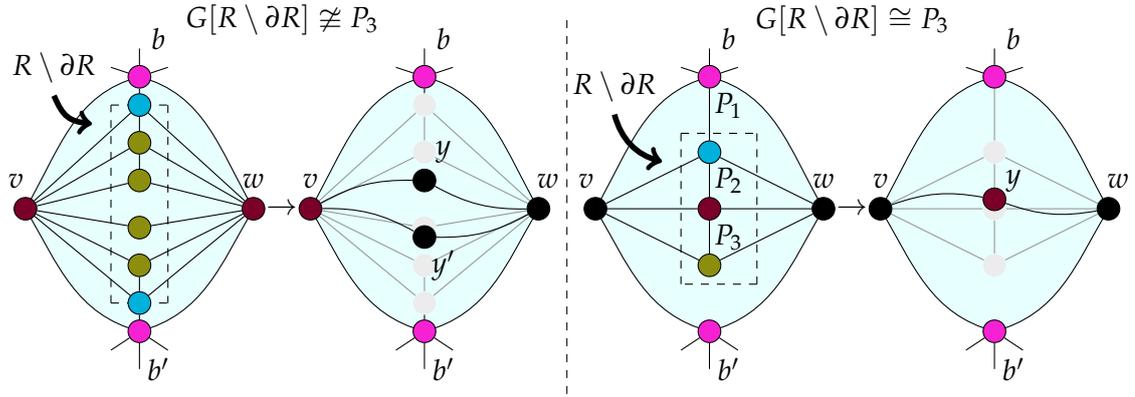

In \cref{case:rto} we dominate $V(R \setminus \partial R) \cong P_3$ with one single vertex $p_2$ in the middle without using $v$ or $w$.
This vertex is represented by the new vertex $y$ in $G'$ which witnesses and dominates the same vertices as $p_2$ in $G$.
If we would use the same gadget as in \cref{case:rtt}, we deteriorate our solution because  $v$ or $w$ would be forced into an sds $D$ in $G'$, but only $p_2 \in D$ might be enough in $G$.
See \cref{fig:rulethree} for an example of both cases.

\begin{lemma}[Correctness of \cref{rgl:rthree}]\label{lemma:correctnessthree}
    Let \G be a plane graph, $v, w \in V$ and \GB be the graph obtained after application of \cref{rgl:rthree} on the pair $\{v, w\}$. 
    Then $G$ has an sds of size $k$ if and only if $G'$ has an sds of size $k$.
\end{lemma}
\begin{proof}
        Consider an sds $D$ in $G$. We show that $G'$ also has an sds with $\abs{D'} \leq \abs{D}$. 
        By assumption, we have $\abs{V(R) \setminus \{v, w\}} \geq 5$, $R$ is a simple region and therefore $d(v,w) \leq 2$.
        We can assume that the border $\partial R$ consists of exactly two vertices $b, b'$, because if $\abs{\partial R \setminus \{v,w\}}  < 2$, the region's boundary path would not enclose an area.
        Consequently, $\abs{V(R \setminus \partial R)} \geq 3$.

        Observe that if a vertex $v' \in V(R \setminus \partial R) \cap D$ together with $v \in D$ or $w\in D$, we replace them with both $v$ and $w$ and set $D' = D \setminus \{v'\} \cup \{v,w\}$.  
        If $v'$ was used as a witness for $v$ (or $w$), we know that $v$ and $w$ witness each other in a simple region. 
        If $V(R \setminus \partial R) \cap D = \emptyset$, we just set $D' = D$. 
        It will be sufficient to only analyze cases where $\abs{D \cap V(R \setminus \partial R)} \leq 1$, because otherwise, we replace them again with $v$ and $w$. 

        It will be enough to only consider cases where no border vertex is dominating. 
        If $\abs{\partial R \setminus \{v,w\} \cap D} \geq 1$ then there is at least one other $v' \in V(R) \cap D$ to dominate all at least three vertices inside the region.
        If a pole $v \in D$ or $w \in D$, we set $D' = D \setminus V(R \setminus \partial R)$, otherwise there is at least one vertex $d \in (V(R) \setminus \partial R) \cap D$ and we replace it with one of the poles arbitrarily: $D' = D \setminus V(R \setminus \partial R) \cup \{v\}$.
        This works, because at least one $b \in \partial R \setminus \{v, w\}$ and $b,v$ together witness the same vertices as $d$ did: $b$ witnesses all vertices in $N(w) \cup N(v)$ and $v$ those neighbors from the opposite border vertices to $b$.

        \vspace*{1em}
        
        In summary, we only need to consider only $v,w \notin D$, $\partial R \cap D = \emptyset$, $\abs{V(R \setminus \partial R)} \geq 3$ and $\abs{(V(R) \setminus \partial R) \cap D} \leq 1$.

        First assume \cref{case:rto} has applied on $G$ and $V(R \setminus \partial R) \cong P_3$. 
        We denote this induced path as $(p_1,p_2,p_3)$.         
         If neither $\{p_1,p_2,p_3\} \cap D = \emptyset$, we set $D' = D$, trivially preserving an sds. 
         Otherwise, $p_2 \in D$ is forced because this is the only way to dominate $V(R \setminus \partial R)$ with exactly one vertex inside. 
         We set $D' = D \setminus \{p2\} \cup \{y\}$ and dominate $\{y,y'\}$ and witnesse $N(v) \cup N(w)$ which are the same as $p_2$ did. 
         This case is depicted in \cref{fig:rulethree} on the \textit{right}.

         Now assume \cref{case:rtt} has applied and $V(R \setminus \partial R) \ncong P_3$. 
         Let us denote this induced path as $(p_1,p_2,p_3)$. 
         We observe that without contradicting the planarity of $G$ the induced subgraph of the vertices $G[V(R \setminus \partial R)]$ is a subgraph $P_{\min({3, \abs{V(R \setminus \partial R)}})}$. 

          By assumption, we either have a $P_3$ with at least one missing edge or a longer path. 
          In both cases, it is impossible to dominate this path with only one single vertex from the path.
         Hence, we set $D' = D$ and the path is dominated by $v$ or $w$
         In both cases $\abs{D} = \abs{D'}$
         
         Consider an sds $D$ in $G'$ of size $\abs{D}$. 
         We show that $G$ has an sds with $\abs{D} \leq \abs{D'}$. 
         We analyze both cases separately.

         Assume $V(R \setminus \partial R) \cong P_3$ and \cref{case:rto} has applied and $V(R \setminus R)$ replaced by one single vertex $y$. 
         We denote the induced path in $G$ as $(p_1,p_2,p_3)$.
         If $y \in D'$, we set $D = D' \setminus \{y\} \cup \{p_2\}$. 
         We know that $p_2$ dominates $p_1$,$p_3$ in $G$ and witnesses the same vertices as in $G'$, namely $N(v) \cup N(w) \cup \{b, b'\}$.
         Otherwise, we just set $D = D'$. 

         Contrary assume $V(R \setminus \partial R) \ncong P_3$ and \cref{case:rtt} has been applied.
         $V(R \setminus R)$ was replaced by two vertices $y$ and $y'$.
         Observe that neither $y$ nor $y'$ are useful in $D'$ in $G'$: If both $y, y' \in D'$, we replace them by $v$ and $w$.
         If only $y \in D$ (resp. $y' \in D$), we need either $v \in D'$ or $w \in D'$ to dominate $y$ ($y'$) and again, we replace them by $v$ and $w$. 

         $y$ or $y'$ are unimporant as witnesses, $v$ and $w$ witness each other and everything that could be witnesed by $y$ or $y'$ is witnessed by $\{v,w\}$ as well.
         This simulates an $\mathnormal{OR}$ gadget in $\{v, w\}$.
        Hence $D = D'$ and $\abs{D} = \abs{D'}$.

        In all cases, the solution sizes only changes by a constant.
\end{proof}
    
The application of \cref{rgl:rthree} gives us a bound on the number of vertices inside a \sr. 
\begin{corollary}\label{lemma:simpleregionbound}
    Let \G be a graph, $v, w\in V$ and R a \sr~ between $v$ and $w$. If \cref{rgl:rthree} has been applied, this simple region has a size of at most 4.
\end{corollary}

\begin{proof}
    If $\abs{V(R) \setminus \{v, w\}} < 5$ then the rule would not have changed G and the size of the region would already be smaller than 5.
    Assuming $\abs{V(R) \setminus \{v, w\}} \geq 5$ in both cases $V(R \setminus \partial R)$ gets removed and at most two new vertices added. As the boundary in a simple region contains at most two vertices distinct from $v$ and $w$, the size of the simple region is bounded by at most four.
\end{proof}

The runtime of \cref{rgl:rthree} is polynomial. 
Note that there could be more than one $vw$-region, but every execution of \cref{rgl:rthree} reduces the graph and therefore \cref{rgl:rthree} might simply be applied again on $v$ and $w$.
In this claim, it is to suffice to observe one fixed $vw$-region between $v$ and $w$.

\begin{corollary}\label{complex:rthree}
        \cref{rgl:rthree} on two vertices $v$ and $w$ is applied in time $\mathcal{O}(d(v) + d(w))$.
\end{corollary}

\begin{proof}
Constructing one simple $vw$-region can be done by the algorithm proposed in~\cite{Alber2004} in time $\mathcal{O}(d(v) +(dw))$.
Check whether an $P_3$ is induced inside $V(R \setminus \partial R)$ can be done in constant time and the reduction itself is constant in $\max(d(v), d(w))$.
\end{proof}
    
\section{Bounding the Size of the Kernel}

We now put all the pieces together and prove the main result: a kernel whose size is bounded by a linear function dependent only on the solution size $k$. 
For that purpose, we distinguish between those vertices that are covered inside a region in a maximal \dreg and those that are not. 
In both cases, our reduction rules bound the number of vertices to a constant size for a fixed region.
\cref{lemma:numRegions} states that for any given dominating set $D$, we can partition the whole graph into a linear number of regions and we know that we have linearly many vertices left in the whole graph.
In particular, we show in the next sections that given an sds $D$ of size $k$, there exists a maximal \dreg $\mathfrak{R}$ such that:

\begin{enumerate}[topsep=0pt,itemsep=-1ex,partopsep=1ex,parsep=1ex]
    \item Each region of $\mathfrak{R}$ contains at most $89$ vertices (\cref{chp:sizeRegion});
    \item $V(\mathfrak{R})$ covers most vertices of $V$. There are at most $97 \cdot \abs{D}$ vertices outside of any region (\cref{cpt:outside});
    \item $\mathfrak{R}$ has only at most $3 \abs{D} - 6$ regions (\cref{cpt:numRegions}).
\end{enumerate}

The combination of these three statements will give us a linear kernel. 
\cref{fig:overview} gives a visualization of this plan.

\subsection{Bounding the Size of a Region}\label{chp:sizeRegion}

We start with a more fine-grained analysis of the impact of the different cases of \cref{rgl:rtwo} on a $vw$-region. 
The main idea is to count the number of simple regions in the $vw$-region and then use the bound on the size of a simple region after \cref{rgl:rthree} was applied exhaustively. 
The bound was obtained in \cref{lemma:simpleregionbound}.   

\begin{lemma}\label{lemma:nonecover}
    Given a plane Graph \G and a $vw$-region $R$, let $D$ be a semitotal dominating set and let $\mathfrak{R}$ be a maximal \dreg of G. 
    For any $vw$-region $R \in \mathfrak{R}$ it holds that $\abs{N_1(v,w) \cap V(R)} \leq 4$ and these vertices lay exactly on the boundary $\partial R$ of $R$. 
\end{lemma}
\begin{proof}
The same argument as proposed by Alber, Fellows and Niedermeier~\cite{Alber2004}, and again used by Garnero and Sau~\cite[Proposition 2]{Garnero2019} applies here as well:
Let $P_1 = (v, u_1, u_2,w)$ and $P_2 = (v, u_3, u_4,w)$ be the two boundary paths enclosing the $vw$-region $R$. By the definition of a region, they have a length of at most 3. Because every vertex in $R$ belongs to $N(v,w)$, but a vertex from $N_1(v,w)$ also has neighbors outside $N(v,w)$, it \emph{must} lie on one of the boundary paths $P_1, P_2$.
Therefore, $R$ has at most four boundary vertices and $\abs{N_1(v,w) \cap V(R)} \leq 4$.

The worst case occurs when the two confluent paths $P_1$ and $P_2$ are vertex-disjoint. 
\end{proof}

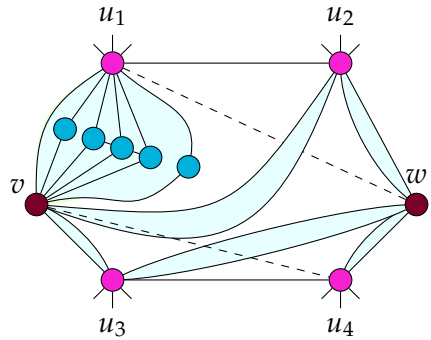
\begin{figure}[!ht]
    \begin{equation*}
        \input{fig/tikz/maxntwo.tikz}
    \end{equation*}
    \caption[Bounding number of simple regions with $N_2(v,w)$ inside a $vw$-region R]{\textit{Bounding the maximum number of simple regions inside a region $R(v,w)$. $N_2(v,w)$ is covered by 6 {\setulcolor{NTWO}\ul{blue}} regions. The two dashed edges indicate that they are among the 8 possible pairs of vertices, but a simple region between them would contradict the planarity.}}
    \label{fig:maxntwoinside}
\end{figure}

\begin{lemma}\cite[See Fact 5, arXiv]{Garnero2018}\label{lemma:ntwocover}
    Given a reduced plane graph \G and a $vw$-region $R$, $N_2(v,w) \cap V(R)$ can be covered by at most 6 simple regions.
\end{lemma}
\begin{proof}
    Let $P_1 = (v,u_1, u_2,w)$ and $P_2 = (v, u_3, u_4, w)$ be the two boundary paths of $R$.
    As in the previous \cref{lemma:nonecover}, the worst case is achieved if they are vertex-disjoint. Otherwise, a smaller bound would be obtained.

    By definition of $N_2(v,w)$, vertices from $N_2(v,w) \cap V(R)$ are common neighbors of $v$ or $w$ and one of $\{u_1,u_2,u_3,u_4\}$.
    By planarity, we can cover $N_2(v,w) \cap V(R)$ with at most 6 simple regions among 8 pairs of vertices (See \cref{fig:maxntwoinside}).

\end{proof}

We continue by giving a constant bound on the number of simple regions that cover all  $N_3(v,w)$ vertices in a given region.

\begin{lemma}\label{lemma:rtwosr}
    Given a plane Graph $G = (V,E)$ reduced under \cref{rgl:rtwo}  and a region R(v, w), if $\Dv \neq \emptyset $ (resp. $\Dw \neq \emptyset$), $N_3(v,w) \cap V(R)$ can be covered by: 
    \begin{enumerate}
        \item $11$ \sr s~if $\Dw \neq \emptyset$ ($\Dw \neq \emptyset$),
        \item $14$ \sr s~if $N_{2,3}(v) \cap N_3(v,w) = \emptyset$.
    \end{enumerate}
\end{lemma}

Observe that in the first case, we can assume that no case of \cref{rgl:rtwo} has been applied, but the claim is a direct consequence of the assumption $\Dv \neq \emptyset$ and $\Dw \neq \emptyset$. If \cref{case:c2} or \cref{case:c3} have been applied, $N_{2,3}(v,w)$ gets reduced and the second case can be applied. For the sake of completeness, we will restate (a slightly adjusted version of) the proof from Garnero and Sau~\cite[Fact 6, arXiv v2]{Garnero2018}. 

Note that this analysis provides a not necessarily tight upper bound and analyzing it more sophisticated will likely yield a better bound.
This could possibly be improved, since taking both $\Dv \neq \emptyset$ and $\Dw \neq \emptyset$, our regions might get even more restricted.

\begin{proof}
        We partition $N_3(v,w)$ into the distinct $N_3(v,w) \setminus N(w)$, $N_3(v,w) \setminus N(v)$ and $N_3(v,w) \cap N(v) \cap N(w)$ and then analyze how many simple regions can there be in the worst case.

    \begin{enumerate}
        \item Because $\Dv \neq \emptyset$ there exists $D = \{v,u,u'\} \in \Dv$ (a smaller set will give a better bound). 
        By definition we know that $D$ dominates $N_3(v,w)$ and also $N_3(v,w) \setminus N(v)$. 
        Therefore, all vertices in $N_3(v,w) \setminus N(v)$ must be neighbors of $w$ and either $u$ or $u'$ and in the worst case at most three simple regions are required. 
        By assumption, $\Dw \neq \emptyset$ as well, and therefore $N_3(v,w) \setminus N(w)$ is bounded by at most three simple regions, too.
        By planarity, we can cover the remaining common neighbors in $N_3(v,w) \cap N(v) \cap N(w)$ with at most 5 vertices and in total, we can cover \textbf{$\mathbf{N_3(v,w) \cap R(v,w)}$ by at most $\mathbf{3 + 3 +5 = 11}$} simple regions.

        \item The proof in~\cite{Garnero2018} holds in our case as well.

    
    \end{enumerate}

    Cases 2 to 4 of \cref{fig:nthreeinside} visualize these simple regions around $N_3(v,w) \cap V(R)$ with simple regions in the relevant cases.\footnote{In revision 2018 of~\cite{Garnero2018}, Garnero and Sau removed this proof, because they changed \cref{rgl:rtwo} and the overall proof was tuned.}
    
\end{proof}

\begin{lemma}[\#Vertices inside a Region after \cref{rgl:rone,rgl:rtwo,rgl:rthree}]\label{lemma:inside}
    Let \G be a plane graph reduced under \cref{rgl:rone,rgl:rtwo,rgl:rthree}. Furthermore, let $D$ be an sds of G and let $v,w \in D$. Any $vw$-region R contains at most 87 vertices distinct from its poles.
\end{lemma}
\begin{proof} 
    By \cref{lemma:nonecover,lemma:ntwocover} and \cref{lemma:simpleregionbound} to bound the number of vertices inside a simple region, we know that $\abs{N_1(v,w) \cap V(R)} \leq 4$. 
    Furthermore, $\abs{N_2(v,w) \cap V(R)} \leq 6 \cdot 4 = 24$, because after the reduction a simple region has at most 4 vertices distinct from its poles and has at most 6 simpler regions covering all $N_2(v, w)$.
    
    It is remaining to bound for \nthi, but gladly, \cref{rgl:rtwo} reduced them! We will distinguish between the cases of \cref{rgl:rtwo}. 
    \Cref{fig:nthreeinside} shows the worst-case amount of simple regions the individual cases can have.
    
    \begin{caseofz}
        \casez{\cref{rgl:rtwo} has \textbf{not} been applied in the following two cases: Either $\Dvw \neq \emptyset$ or $\Dvw = \emptyset \wedge \Dv \neq \emptyset \wedge \Dw \neq \emptyset$:}

        \begin{enumerate}
            \item If $\Dvw \neq \emptyset$, there exists a set $\tilde D = \{d_1,d_2,d_3\} \in \Dvw$ of at most three vertices dominating $N_3(v,w)$. We observe that vertices from \nthi~ are common neighbors of either v or w (by the definition of a $vw$-region) and at least one vertex from $\tilde D$, because someone has to dominate them and we know that only the poles or vertices in $\tilde D$ come into question.
            Without violating planarity, we can span at most 6 distinct simple regions. Using the bound of simple regions (worst case shown in \cref{lemma:simpleregionbound}) and including $\abs{\tilde D} = 3$, we can conclude $\nthi \leq 6 \cdot 4 + 3 = 27$.
            \item If $\Dvw = \emptyset$, $\Dv \neq \emptyset$ and $\Dw \neq \emptyset$, we can apply \cref{lemma:rtwosr} and although \cref{rgl:rtwo} has not changed the graph $G$, we can cover $R$ with at most $11$ simple regions giving us $\nthi \leq 11 \cdot 4 = 44$ vertices.
        \end{enumerate}
        
        \casez{If \cref{rgl:rtwo} \Cref{case:c1} has been applied,} \ntwi was entirely removed and \nthi replaced by at most four new vertices $v', w'$ and $y$ and $y'$. Hence $\nthi \leq 4$.

        
        \casez{If \cref{rgl:rtwo} \Cref{case:c2,case:c3} have been applied, } we know that $N_{2,3}(v) \cap N_3(v,w) \subseteq N_{2,3}(v)$ was removed and replaced by one single vertex. Applying \cref{lemma:rtwosr}, we can cover $N_3(v,w) \setminus \{v'\} \cap V(R)$ with at most $14$ simple regions giving us $\abs{\nthi} \leq 14 \cdot 4 + 1 = 57$.
        We other case is symmetrical.
    \end{caseofz}
    
    \begin{figure}[!ht]
        \begin{equation*}
            \input{fig/tikz/nthreeinside.tikz}
        \end{equation*}
        \caption[Bounding number of simple regions with inside a $vw$-region R]{\textit{Showing the worst case scenarios for the different cases in \cref{lemma:inside}: \textbf{Case 0.1}: $\Dvw$ is nonempty and we have three vertices that can dominate $N_{2,3}$ alone. They can span simple regions with the $N_3(v,w)$ vertices.
         \textbf{Case 1}: $N_{2,3}$ was removed and four vertices introduced.
        \textbf{Case 2 and 3}: At most $14$ simple regions after $N_{2,3}$ has been replaced by a single $v'$. 
        \textbf{Case 0.2}: $\Dvw$, $\Dv$ and $\Dw$ are all empty, so the rule has not changed anything and we can cover $N_3(v, w)\cap V(R)$ with at most 11 simple regions.}}
        \label{fig:nthreeinside}
    \end{figure}
    
    All in all, as $V(R) = \{v, w\} \cup (N_1(v,w) \cup N_2(v,w) \cup N_3(v,w)) $ we get 
    
    \[V(R) \leq 2 + 4 + 24 + \max(27, 44, 4, 57) = 87 \]
\end{proof}

We have proved the first step and bounded the number of vertices that lay inside a  single region to be at most $87$.

\subsection{Number of Vertices Outside the Decomposition}\label{cpt:outside}
We continue to bound the number of vertices that do not lay inside any region of a maximal \dreg $\mathfrak{R}$, that is, we bound the size of $V \setminus \VR$. \Cref{rgl:rone} ensures that we only have a small amount of $N_3(v)$-pendants. We then try to cover the rest with as few simple regions as possible, because, by application of \cref{rgl:rthree}, these simple regions are of constant size.

The following \cref{lemma:noneinside} states that no vertex from $N_1(v)$ will be outside of a maximal \dreg which was already proven by {Alber, Fellows and Niedermeier~\cite[Lemma 6]{Alber2004}}.

\begin{lemma}\label{lemma:noneinside}
    Let \G be a plane graph and $\mathfrak{R}$ be a maximal \dreg of a dominating set $D$. If $u \in N_1(v)$ for some vertex $v \in D$ then $u \in V(\mathfrak{R})$.
\end{lemma}

In the following, we define $\dGR(v) = \abs{\{R(v,w) \in \mathfrak{R}, w \in D\}}$ to be the number of regions in $\mathfrak{R}$ having $v$ as a pole. 

\begin{corollary}
    Let \G be a graph and D be a set. For any maximal \dreg~$\mathfrak{R}$ on $G$ it holds that $\sum_{v \in D} \dGR(v) = 2 \cdot \abs{\mathfrak{R}}$.
\end{corollary}
\begin{proof}\label{lemma:polesBound}
    
    The proof follows directly from the handshake lemma applied to the underlying multigraph~\GR~ where every edge between $v,w \in D$ represents a region between $v$ and $w$ in $\mathfrak{R}$.
\end{proof}

\begin{proposition}\label{lemma:outside}
    Let \G be a plane graph reduced under \cref*{rgl:rone,rgl:rtwo} and let D be a semitotal dominating set of $G$. For a maximal \dreg $\mathfrak{R}$,  $\abs{V \setminus (V(\mathfrak{R}) \cup D)} \leq 97 \abs{D}$.
\end{proposition}

With slight modifications, the proof given by Garnero and Sau~\cite[arXiv v2]{Garnero2018} will also apply for \sdom. Although we assume $G$ to be entirely reduced, the following proof only relies on \cref{rgl:rone,rgl:rthree}. The proof uses the observation that vertices from $N_2(v)$ that lie outside of a region must span simple regions between those from $\{v\} \cup N_1(v)$.

\begin{figure}[!ht]
    \begin{equation*}
        \input{fig/tikz/kernelSize.tikz}
    \end{equation*}
    \caption[Vertices from $N_2(v)$ laying outside]{\textit{Bounding the number of $N_2(v)$-vertices around a dominating vertex $v$ given a maximal \dreg $\mathfrak{R}$. $v$ is a pole of $R_1, R_2,...R_j$ and can span simple regions with the help of $N_2(v)$-vertices to at most two $N_1(v)$-vertices per $R_i$. Each region has at most four vertices in $N_1(v,w) \subseteq N_1(v)$ on the boundary of $R_j$, but only at most two can be used for a simple region: For Example trying to construct a simple region between $v$ and $n_2$ would contradict the maximality of $\mathfrak{R}$. Furthermore, because rule \cref{rgl:rone} has removed all but \textbf{one} vertex from $N_3(v)$, we intuitively can span two regions to each of the $N_1(v)$-vertices. Furthermore, the size of these simple regions is bounded after the application of \cref{rgl:rthree}.}}
    \label{fig:kernelSize}
\end{figure}
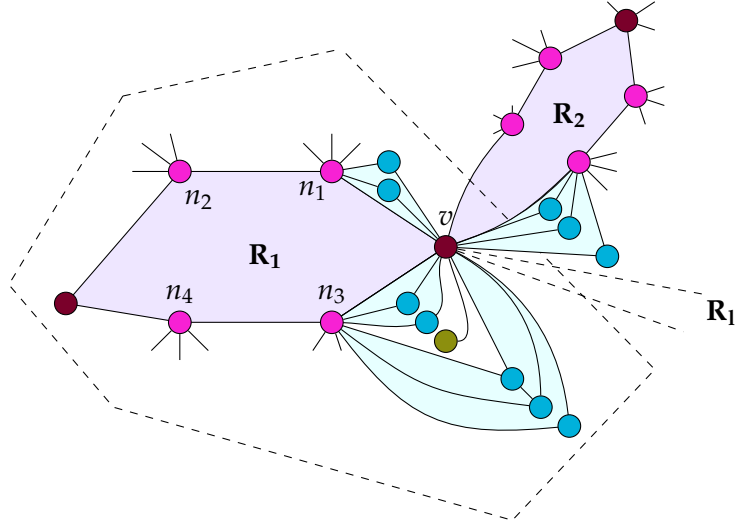

\begin{proof}
    We will follow the proof proposed by Alber, Fellows, Niedermeier~\cite[Proposition 2]{Alber2004} and use the size bound of a simple region we have proven in \cref{lemma:simpleregionbound}. 
    In particular, we are going to show that $V \setminus V(\mathfrak{R}) \leq 48 \cdot \abs{\mathfrak{R}} + 2 \cdot \abs{D}$. \Cref{lemma:numRegions} will then give the desired bound.
    
    Let $D$ be an sds, $\mathfrak{R}$ be a maximal \dreg and $v \in D$. Since $D$ dominates all vertices in the graph, we can consider $V$ as $\bigcup_{v \in D}N(v)$ and thus, we only need to bound the sizes of $N_1(v)\setminus \VR$, $N_2(v) \setminus \VR$ and $N_3(v) \setminus \VR$ separately.
    
    \noindent$\mathbf{N_3(v)}$: As we know that \cref{rgl:rone} has been exhaustively applied, we trivially see that $\abs{N_3(v)} \leq 1$ and hence, 
    
    \[\abs{\bigcup_{v \in D}N_3(v) \setminus V(\mathfrak{R})} \leq \abs{D}\]
    
    \noindent$\mathbf{N_2(v)}$: According to Garnero and Sau~\cite[Proposition 2]{Garnero2018}, we know that $N_2(v) \setminus V(\mathfrak{R})$ in a reduced graph can be covered by at most $4 \dGR(v)$ simple regions between $v$ and some vertices from $N_1(v)$ on the boundary of a region in $\mathfrak{R}$. 
    \cref{fig:kernelSize} gives some intuition, but intuitively, we can span two simple regions to each of the vertices from $N_1(v)$ on the two border vertices for each $R \in \mathfrak{R}$.
    
    Because we assume $G$ to be reduced, by \cref{lemma:simpleregionbound} a simple region can have at least $4$ vertices distinct from its poles and hence,
    
    \begin{equation}
        \begin{split}
            \abs{\bigcup_{v \in D} N_2(v) \setminus V(\mathfrak{R})} &\leq 4 \sum_{v \in D}4 \cdot \dGR(v) \\
            &= 16 \cdot \sum_{v \in D}\dGR(v) \\
            &\overset{\text{\cref{lemma:polesBound}}}{\leq} 32 \abs{\mathfrak{R}}
        \end{split}
    \end{equation}
    
    \noindent$\mathbf{N_1(v)}$: Because every sds is a ds, we can apply \cref{lemma:noneinside} and conclude that $N_1(v) \subseteq V(\mathfrak{R})$. 
    Hence,
    
    \[\abs{\bigcup_{v \in D}N_1(v) \setminus V(\mathfrak{R})} = 0\]
    
    \noindent Summing up these three upper bounds for each $v \in D$ we obtain the result using the equation from \cref{lemma:numRegions}:
    
    \begin{equation}
        \begin{alignedat}[b]{2}
            \abs{V \setminus V(\mathfrak{R}) \cup D)} &\leq 32 \cdot \abs{\mathfrak{R}} + \abs{D} & \zerotext[4em]{(\cref{lemma:numRegions})}\\ 
            &\leq 32 \cdot (3 \abs{D} - 6) + \abs{D} &\\
            &\leq  96 \cdot \abs{D} + \abs{D} &\\
            &= 97 \cdot \abs{D}
        \end{alignedat}
    \end{equation}
    
\end{proof}

\subsection{Bounding the Number of Regions}\label{cpt:numRegions}

We are now utilizing the final tool in our toolbox. Alber, Fellows and Niedermeier~\cite[Proposition 1]{Alber2004} gave an explicit greedy algorithm to construct a maximal \dreg for a \dom. 
The existence of this algorithm is the core for all following works because they always involve region decompositions. For example, Garnero and Sau used it for \rbdom~\cite{Garnero2017} and \tdom~\cite{Garnero2018}. This missing puzzle piece will now assemble everything we have set up so far giving us the linear kernelization we are looking for.

For the following lemma, Alber, Fellows and Niedermeie~\cite{Alber2004} required a reduced instance and their reduction rules for \pdom differed from ours. Luckily, they do not rely on any specific properties following from a reduced graph and therefore, we can just use it for our kernelization algorithm as well.

This was already observed by Garnero and Sau~\cite{Garnero2018} and a more formal proof along with the description of the algorithm was provided.

Because every semitotal dominating set is indeed also a dominating set, we can safely apply it to \psdom as well. 

\begin{lemma}[{\cite[Proposition 1 and Lemma 5]{Alber2004}}]\label{lemma:numRegions}
    Let G be a reduced plane graph and let $D$ be a sds with $\abs{D}\geq 3$. There is a maximal D-region decomposition of G such that $\abs{R} \leq 3 \cdot \abs{D} - 6$.
\end{lemma}

Furthermore, the overall reduction procedure runs in polynomial time:

\begin{lemma}\label{lemma:runtime}
    A plane graph G can be reduced by \cref{rgl:rone,rgl:rtwo,rgl:rthree} in time $\mathcal{O}(\abs{(G)}^3)$
\end{lemma}
\begin{proof} 
Testing whether \cref{rgl:rone} can be applied on every vertex takes $\sum_{v\in V} \mathcal{O}(d(v)) = \mathcal{O}(n)$.
\cref{rgl:rtwo,rgl:rthree} must be applied on every pair of vertices and take time $\sum_{v,w \in V}\mathcal{O}(d(v) + d(w)) = \mathcal{O}(n^2)$
we assume that checking for termination as we defined in \cref{def:reduced} takes only constant time in each step.
In the worst case, in each iteration, only one rule will be applied and one vertex reduced. 
Hence, we have to do at most $n$ iterations.
Therefore, the graph can be reduced in time $\mathcal{O}(n^3)$.
\end{proof}

\noindent By utilizing all the previous results, we are now finally ready to prove \cref{thm:central}: 

\centraltheo*

\begin{proof}
    Let \G be the plane input graph and $G'=(V',E')$ be the graph obtained by the exhaustive application of \cref{rgl:rone,rgl:rtwo,rgl:rthree}.
    As none of our rules change the size of a possible solution $D' \subseteq V'$ in $G'$, we know by \cref{lemma:correctnessone}, \cref{lemma:correctnesstwo} and \cref{lemma:correctnessthree} that $G'$ has a \sdom of size $k$ if and only if $G$ has an sds set of size $k$.
    Furthermore, by \cref{lemma:runtime}, the preprocessing procedure runs in polynomial time.
    
    By taking the size of each region proven in \cref{lemma:outside}, the total number of regions in a maximal \dreg~(\cref{lemma:numRegions}) and the number of vertices that can lay outside of any region (\cref{lemma:outside}), we obtain the following bound:
    
    \begin{equation}
         87 \cdot (3k - 6) + 97 \cdot k  < \kernelsize \cdot k
    \end{equation}
    \noindent If $\abs{V(G')} > \kernelsize \cdot k$ we replace $G'$ by one single vertex $v$, which is trivially a \emph{no}-instance, because $v$ has no witness to form an sds.

\end{proof}

%% file: fig/tikz/neighborhoods-single-vertex.tikz
\begin{tikzpicture}[scale=1.6, thick]
	\begin{pgfonlayer}{nodelayer}
		\node [style=BLACK, label={above:$v$}] (0) at (1, 2) {};
		\node [style=NONE] (1) at (4.25, 2.75) {};
		\node [style=NONE] (2) at (4.25, 0) {};
		\node [style=none] (3) at (5.25, 1.75) {};
		\node [style=none] (4) at (5.75, 2.75) {};
		\node [style=none] (5) at (5.25, 3.75) {};
		\node [style=none] (6) at (5.25, 1) {};
		\node [style=none] (7) at (5.75, 0) {};
		\node [style=none] (8) at (5.25, -1) {};
		\node [style=NTHR] (9) at (2, 4) {};
		\node [style=NTWO] (10) at (1.25, 0.25) {};
		\node [style=NTWO] (11) at (3, 3.75) {};
		\node [style=NTWO] (12) at (0.25, -0.5) {};
		\node [style=NTHR] (14) at (-0.5, 3.75) {};
		\node [style=NTHR] (15) at (-1.5, 2.25) {};
		\node [style=none] (16) at (-2, 4.25) {};
		\node [style=none] (17) at (4.75, 4.25) {};
		\node [style=none] (18) at (4.75, -1) {};
		\node [style=none] (19) at (-2, -1) {};
		\node [style=none] (21) at (-1.5, 4.75) {$N(v)$};
	\end{pgfonlayer}
	\begin{pgfonlayer}{edgelayer}
		\draw (0) to (1);
		\draw (0) to (2);
		\draw (4.center) to (1);
		\draw (3.center) to (1);
		\draw (1) to (5.center);
		\draw (6.center) to (2);
		\draw (7.center) to (2);
		\draw (8.center) to (2);
		\draw (1) to (2);
		\draw (9) to (0);
		\draw (0) to (11);
		\draw (11) to (1);
		\draw (10) to (0);
		\draw (10) to (2);
		\draw (0) to (12);
		\draw (12) to (2);
		\draw (0) to (15);
		\draw (12) to (15);
		\draw (14) to (15);
		\draw (14) to (0);
		\draw [style=EDGEDASHED] (16.center) to (17.center);
		\draw [style=EDGEDASHED] (17.center) to (18.center);
		\draw [style=EDGEDASHED] (18.center) to (19.center);
		\draw [style=EDGEDASHED] (19.center) to (16.center);
	\end{pgfonlayer}
\end{tikzpicture}

%% file: fig/tikz/neighborhoods-two-vertices.tikz
\begin{tikzpicture}
	\begin{pgfonlayer}{nodelayer}
		\node [style=BLACK, label={above:$v$}] (0) at (-5, 0) {};
		\node [style=BLACK, label={below:$w$}] (2) at (6, 0) {};
		\node [style=BLACK] (3) at (3, 3) {};
		\node [style=BLACK] (4) at (3, -3.5) {};
		\node [style=BLACK] (5) at (-2, 3) {};
		\node [style=BLACK] (6) at (-2, -3.5) {};
		\node [style=BLACK] (7) at (-7, -1) {};
		\node [style=BLACK] (8) at (-7, 0) {};
		\node [style=BLACK] (10) at (-2.75, -1.75) {};
		\node [style=BLACK] (11) at (-3.25, 2.5) {};
		\node [style=BLACK] (12) at (-3.5, -2.5) {};
		\node [style=BLACK] (13) at (-7, 1) {};
		\node [style=NONE] (14) at (-7, 1) {};
		\node [style=NONE] (15) at (-7, 0) {};
		\node [style=NONE] (16) at (-7, -1) {};
		\node [style=NTWO] (17) at (-2.5, 1.75) {};
		\node [style=NTWO] (18) at (-3.25, 2.5) {};
		\node [style=NTWO] (19) at (-2.75, -1.75) {};
		\node [style=NTWO] (20) at (-3.5, -2.5) {};
		\node [style=NONE] (21) at (-2, 3) {};
		\node [style=NONE] (22) at (-2, -3.5) {};
		\node [style=NTHR] (23) at (-7, -1) {};
		\node [style=NTHR] (24) at (-7, 0) {};
		\node [style=NTHR] (25) at (-7, 1) {};
		\node [style=NTHR] (26) at (-2.5, 0.75) {};
		\node [style=NTHR] (27) at (-2.5, -0.75) {};
		\node [style=NTHR] (28) at (2.75, 0.75) {};
		\node [style=NTHR] (29) at (2.75, -0.75) {};
		\node [style=NTWO] (30) at (6, 2.25) {};
		\node [style=NTWO] (31) at (5.25, 2) {};
		\node [style=NTWO] (32) at (7, 2.5) {};
		\node [style=NTWO] (33) at (8, 3) {};
		\node [style=NTWO] (34) at (5.25, 2) {};
		\node [style=NONE] (35) at (3, 3) {};
		\node [style=NONE] (36) at (3, -3.5) {};
		\node [style=none] (37) at (-1, 4) {};
		\node [style=none] (38) at (-2, 4.5) {};
		\node [style=none] (39) at (-3, 4) {};
		\node [style=none] (40) at (2, 4) {};
		\node [style=none] (41) at (3, 4.5) {};
		\node [style=none] (42) at (4, 4) {};
		\node [style=none] (43) at (2, -4.5) {};
		\node [style=none] (44) at (3, -5) {};
		\node [style=none] (45) at (4, -4.5) {};
		\node [style=none] (46) at (-1, -4.5) {};
		\node [style=none] (47) at (-2, -5) {};
		\node [style=none] (48) at (-3, -4.5) {};
		\node [style=NTHR] (49) at (7.25, -1.25) {};
		\node [style=NTHR] (50) at (7.75, 0.25) {};
		\node [style=NTHR] (51) at (7.5, 1) {};
		\node [style=NTHR] (52) at (7.75, -0.5) {};
		\node [style=NONE] (53) at (-6.25, 3) {};
		\node [style=none] (54) at (-5.25, 4) {};
		\node [style=none] (55) at (-6.25, 4.25) {};
		\node [style=none] (56) at (-7.5, 4) {};
		\node [style=none] (57) at (8.75, 3.5) {};
		\node [style=none] (59) at (8.75, -4) {};
		\node [style=none] (60) at (-8.5, -4) {};
		\node [style=none] (61) at (-8.5, 3.5) {};
		\node [style=none] (63) at (7.25, 4.25) {$N(v,w)$};
		\node [style=none] (64) at (2.75, 1.5) {$z$};
	\end{pgfonlayer}
	\begin{pgfonlayer}{edgelayer}
		\draw (5) to (0);
		\draw (6) to (0);
		\draw (2) to (3);
		\draw (2) to (4);
		\draw (5) to (3);
		\draw (6) to (4);
		\draw (0) to (10);
		\draw (0) to (12);
		\draw (12) to (6);
		\draw (10) to (6);
		\draw (11) to (0);
		\draw (11) to (5);
		\draw (8) to (0);
		\draw (7) to (0);
		\draw (8) to (7);
		\draw (13) to (0);
		\draw (23) to (20);
		\draw (0) to (26);
		\draw (26) to (17);
		\draw (27) to (0);
		\draw (27) to (19);
		\draw (21) to (17);
		\draw (17) to (0);
		\draw (26) to (28);
		\draw (29) to (27);
		\draw (29) to (2);
		\draw (28) to (2);
		\draw (27) to (28);
		\draw (3) to (34);
		\draw (2) to (34);
		\draw (30) to (2);
		\draw (32) to (2);
		\draw (33) to (2);
		\draw (30) to (3);
		\draw (32) to (3);
		\draw (33) to (3);
		\draw (38.center) to (21);
		\draw (37.center) to (21);
		\draw (21) to (39.center);
		\draw (35) to (40.center);
		\draw (41.center) to (35);
		\draw (42.center) to (35);
		\draw (48.center) to (22);
		\draw (22) to (47.center);
		\draw (46.center) to (22);
		\draw (43.center) to (36);
		\draw (44.center) to (36);
		\draw (45.center) to (36);
		\draw (49) to (2);
		\draw (2) to (52);
		\draw (50) to (2);
		\draw (51) to (2);
		\draw (26) to (27);
		\draw (0) to (53);
		\draw (54.center) to (53);
		\draw (55.center) to (53);
		\draw (53) to (56.center);
		\draw [style=EDGEDASHED] (57.center) to (59.center);
		\draw [style=EDGEDASHED] (59.center) to (60.center);
		\draw [style=EDGEDASHED] (61.center) to (60.center);
		\draw [style=EDGEDASHED] (61.center) to (57.center);
	\end{pgfonlayer}
\end{tikzpicture}

%% file: fig/tikz/ntwoimportant.tikz
\begin{tikzpicture}
	\begin{pgfonlayer}{nodelayer}
		\node [style=BLACK] (0) at (-7.5, 0.75) {};
		\node [style=BLACK] (1) at (-1.5, 0.75) {};
		\node [style=NTHR] (2) at (-4.5, 0.75) {};
		\node [style=DOM] (4) at (-4.5, 2) {};
		\node [style=NONE] (5) at (-4.5, 3.25) {};
		\node [style=none] (6) at (-1, 3.75) {};
		\node [style=none] (7) at (-1, 0) {};
		\node [style=none] (8) at (-8, 0) {};
		\node [style=none] (9) at (-8, 3.75) {};
		\node [style=DOM] (10) at (-3.25, 4.75) {};
		\node [style=none] (12) at (-7.5, 1.5) {$v$};
		\node [style=none] (15) at (-1.5, 1.75) {$w$};
		\node [style=BLACK] (16) at (-4.5, 5) {};
		\node [style=BLACK] (18) at (-3.25, 6) {};
		\node [style=BLACK] (20) at (-2, 5) {};
		\node [style=none] (22) at (-2.75, 4.25) {$d_1$};
		\node [style=none] (24) at (-3.75, 2.5) {$d_2$};
		\node [style=none] (25) at (-6.75, 4.5) {$N(v,w)$};
		\node [style=BLACK] (26) at (1.5, 0.75) {};
		\node [style=BLACK] (27) at (7.5, 0.75) {};
		\node [style=NTHR] (28) at (4.5, 0.75) {};
		\node [style=DOM] (30) at (4.5, 3.25) {};
		\node [style=none] (31) at (8, 3.75) {};
		\node [style=none] (32) at (8, 0) {};
		\node [style=none] (33) at (1, 0) {};
		\node [style=none] (34) at (1, 3.75) {};
		\node [style=BLACK] (35) at (5.75, 4.75) {};
		\node [style=none] (36) at (1.5, 1.5) {$v$};
		\node [style=none] (37) at (7.5, 2) {$w$};
		\node [style=none] (41) at (4.5, 2.5) {$d_1$};
		\node [style=none] (43) at (2.25, 4.5) {$N(v,w)$};
		\node [style=none] (44) at (4.5, 1.5) {$d_2$};
		\node [style=none] (45) at (0, 6.25) {};
		\node [style=none] (46) at (0, -1) {};
	\end{pgfonlayer}
	\begin{pgfonlayer}{edgelayer}
		\draw (0) to (2);
		\draw (2) to (1);
		\draw (1) to (4);
		\draw (4) to (0);
		\draw (2) to (4);
		\draw (4) to (5);
		\draw [in=60, out=180] (5) to (0);
		\draw [in=120, out=0] (5) to (1);
		\draw [style=EDGEDASHED] (7.center)
			 to (8.center)
			 to (9.center)
			 to (6.center)
			 to cycle;
		\draw [style=EDGE] (5) to (10);
		\draw (18) to (10);
		\draw (20) to (10);
		\draw (10) to (16);
		\draw (26) to (28);
		\draw (28) to (27);
		\draw [in=60, out=180] (30) to (26);
		\draw [in=120, out=0] (30) to (27);
		\draw [style=EDGEDASHED] (32.center)
			 to (33.center)
			 to (34.center)
			 to (31.center)
			 to cycle;
		\draw [style=EDGE] (30) to (35);
		\draw [style=EDGEDASHED] (45.center) to (46.center);
	\end{pgfonlayer}
\end{tikzpicture}

%% file: fig/tikz/region-example.tikz
\begin{tikzpicture}
	\begin{pgfonlayer}{nodelayer}
		\node [style=BLACK] (0) at (-11.75, -0.25) {};
		\node [style=BLACK] (1) at (-12.5, -2.25) {};
		\node [style=BLACK] (2) at (-8.75, -5) {};
		\node [style=BLACK] (3) at (-10, -1.5) {};
		\node [style=BLACK] (5) at (-7.75, 1) {};
		\node [style=DOM, label={above:$d_1$}] (6) at (-4.75, -2.75) {};
		\node [style=BLACK] (7) at (-9.75, -0.5) {};
		\node [style=BLACK] (9) at (-5, 0) {};
		\node [style=DOM, label={above:$d_2$}] (10) at (-11.75, -0.25) {};
		\node [style=BLACK] (11) at (-8.25, -2.5) {};
		\node [style=BLACK] (12) at (-6, -1.75) {};
		\node [style=BLACK] (13) at (-8.5, -0.5) {};
		\node [style=DOM, label={above:$d_3$}] (14) at (-7.25, -1.25) {};
		\node [style=BLACK] (15) at (-7.5, 2.5) {};
		\node [style=BLACK] (16) at (-4, -4.25) {};
		\node [style=BLACK] (17) at (-2.25, -3.25) {};
		\node [style=BLACK] (18) at (-2.25, -1.5) {};
		\node [style=BLACK] (19) at (-1.25, -2) {};
		\node [style=BLACK] (20) at (-5.75, -5.25) {};
		\node [style=BLACK] (21) at (-11, -2) {};
		\node [style=BLACK] (22) at (-7.25, -3.25) {};
		\node [style=BLACK] (23) at (-7.25, -4) {};
		\node [style=BLACK] (25) at (1, 0) {};
		\node [style=DOM, label={above:$d_2$}] (33) at (1, 0) {};
		\node [style=DOM] (38) at (5.5, -1) {};
		\node [style=DOM] (39) at (8, -2.5) {};
		\node [style=none] (42) at (5, 0) {$d_3$};
		\node [style=none] (43) at (8.5, -1.75) {$d_1$};
		\node [style=none] (44) at (0, 3.5) {};
		\node [style=none] (45) at (0, -5.25) {};
	\end{pgfonlayer}
	\begin{pgfonlayer}{edgelayer}
		\draw [style=FILLPURPLE] (6.center)
			 to (11.center)
			 to (3.center)
			 to (0.center)
			 to (1.center)
			 to (2.center)
			 to [bend right] cycle;
		\draw (2) to (3);
		\draw [style=FILLGREEN] (7.center)
			 to (0.center)
			 to (5.center)
			 to (14.center)
			 to (13.center)
			 to cycle;
		\draw [style=FILLDARKBLUE] (9.center)
			 to (5.center)
			 to (10.center)
			 to (15.center)
			 to cycle;
		\draw (2) to (11);
		\draw [style=FILLDARKBLUE] (14) to (12);
		\draw (12) to (11);
		\draw (13) to (5);
		\draw (7) to (5);
		\draw [style=FILLDARKBLUEINVISIBLE] (6.center)
			 to [bend right=75, looseness=1.25] (9.center)
			 to [in=75, out=-45] cycle;
		\draw (6) to (18);
		\draw (19) to (6);
		\draw (17) to (6);
		\draw (16) to (6);
		\draw (16) to (17);
		\draw (19) to (18);
		\draw (6) to (20);
		\draw (20) to (16);
		\draw [bend left=45, looseness=1.25] (9) to (6);
		\draw [style=FILLDARKBLUEINVISIBLE] (14.center)
			 to (6.center)
			 to [bend right=15] cycle;
		\draw [style=FILLDARKBLUEINVISIBLE] (14.center)
			 to (6.center)
			 to [bend right=15] cycle;
		\draw [style=FILLDARKBLUEINVISIBLE] (14.center)
			 to (6.center)
			 to [bend right=15] cycle;
		\draw (14) to (12);
		\draw (6) to (12);
		\draw (21) to (10);
		\draw (2) to (21);
		\draw (23) to (6);
		\draw (22) to (6);
		\draw (22) to (2);
		\draw (23) to (2);
		\draw (33) to (38);
		\draw (38) to (39);
		\draw [bend right, looseness=0.25] (33) to (39);
		\draw [bend right=60] (39) to (33);
		\draw [style=EDGEDASHED] (44.center) to (45.center);
	\end{pgfonlayer}
\end{tikzpicture}

%% file: fig/tikz/simple-region-example.tikz
\begin{tikzpicture}
	\begin{pgfonlayer}{nodelayer}
		\node [style=BLACK, label={above:$v$}] (0) at (-3.25, 0) {};
		\node [style=BLACK, label={above:$w$}] (2) at (3, 0) {};
		\node [style=NONE] (3) at (0, 4.25) {};
		\node [style=NONE] (4) at (0, -2.25) {};
		\node [style=NTWO] (5) at (0, 3.25) {};
		\node [style=NTWO] (6) at (0, -1.5) {};
		\node [style=NTHR] (7) at (0, 2.25) {};
		\node [style=NTHR] (8) at (0, 1.5) {};
		\node [style=NTHR] (9) at (0, 0.75) {};
		\node [style=NTHR] (10) at (0, 0) {};
		\node [style=NTHR] (11) at (0, -0.75) {};
		\node [style=none] (12) at (1, 5) {};
		\node [style=none] (13) at (0, 5.5) {};
		\node [style=none] (14) at (-1, 5) {};
		\node [style=none] (15) at (0.75, -3) {};
		\node [style=none] (16) at (0, -3.5) {};
		\node [style=none] (17) at (-0.75, -3) {};
		\node [style=none] (18) at (-4, 4.75) {};
		\node [style=none] (19) at (3.75, 4.75) {};
		\node [style=none] (20) at (3.75, -2.75) {};
		\node [style=none] (21) at (-4, -2.75) {};
		\node [style=none] (22) at (-2.5, 5.75) {$N(v,w)$};
	\end{pgfonlayer}
	\begin{pgfonlayer}{edgelayer}
		\draw [style=FILLBLUE] (0.center)
			 to [bend right=15] (4.center)
			 to [bend right=15, looseness=0.75] (2.center)
			 to [bend right=15] (3.center)
			 to [bend right=15] cycle;
		\draw (5) to (0);
		\draw (7) to (0);
		\draw (8) to (0);
		\draw (9) to (0);
		\draw (10) to (0);
		\draw (8) to (9);
		\draw (5) to (7);
		\draw (5) to (3);
		\draw (6) to (4);
		\draw (6) to (0);
		\draw (6) to (2);
		\draw (11) to (2);
		\draw (10) to (2);
		\draw (11) to (0);
		\draw (9) to (2);
		\draw (8) to (2);
		\draw (2) to (7);
		\draw (5) to (2);
		\draw (7) to (8);
		\draw (10) to (11);
		\draw (3) to (13.center);
		\draw (3) to (12.center);
		\draw (3) to (14.center);
		\draw (16.center) to (4);
		\draw (4) to (15.center);
		\draw (17.center) to (4);
		\draw [style=EDGEDASHED] (18.center) to (21.center);
		\draw [style=EDGEDASHED] (20.center) to (21.center);
		\draw [style=EDGEDASHED] (20.center) to (19.center);
		\draw [style=EDGEDASHED] (19.center) to (18.center);
	\end{pgfonlayer}
\end{tikzpicture}

%% file: fig/tikz/obtainingKernel.tikz
\begin{tikzpicture}
	\begin{pgfonlayer}{nodelayer}
		\node [style=BLACK] (0) at (-5.5, 1.5) {};
		\node [style=BLACK] (1) at (-6.25, -0.5) {};
		\node [style=BLACK] (2) at (-2.5, -3.25) {};
		\node [style=BLACK] (3) at (-3.75, 0.25) {};
		\node [style=BLACK] (5) at (-1.5, 2.75) {};
		\node [style=DOM, label={above:$d_1$}] (6) at (1.5, -1) {};
		\node [style=BLACK] (7) at (-3.5, 1.25) {};
		\node [style=BLACK] (9) at (1.25, 1.75) {};
		\node [style=DOM, label={above:$d_2$}] (10) at (-5.5, 1.5) {};
		\node [style=BLACK] (11) at (-2, -0.75) {};
		\node [style=BLACK] (12) at (0.25, 0) {};
		\node [style=BLACK] (13) at (-2.25, 1.25) {};
		\node [style=DOM, label={above:$d_3$}] (14) at (-1, 0.5) {};
		\node [style=BLACK] (15) at (-1.25, 4.25) {};
		\node [style=BLACK] (16) at (2.25, -2.5) {};
		\node [style=BLACK] (17) at (4, -1.5) {};
		\node [style=BLACK] (18) at (4, 0.25) {};
		\node [style=BLACK] (19) at (5, -0.25) {};
		\node [style=BLACK] (20) at (0.5, -3.5) {};
		\node [style=BLACK] (21) at (-4.75, -0.25) {};
		\node [style=BLACK] (22) at (-1, -1.5) {};
		\node [style=BLACK] (23) at (-1, -2.25) {};
		\node [style=none] (24) at (-6.5, -2.5) {};
		\node [style=none] (25) at (-3.75, -1) {};
		\node [style=none] (26) at (3.5, 1.25) {};
		\node [style=none] (27) at (5.75, 0) {};
		\node [style=none] (28) at (4.75, -2.5) {};
		\node [style=none] (29) at (1.75, -4) {};
		\node [style=none] (30) at (-1, -3.5) {};
		\node [style=none] (31) at (1.75, -1.75) {};
		\node [style=none] (32) at (2.75, 0.25) {};
		\node [style=none] (33) at (4, -4.5) {};
		\node [style=none] (34) at (3.25, -2.75) {};
		\node [style=none] (35) at (3.25, -2.75) {};
		\node [style=NTWO] (37) at (4.75, -5.5) {2};
		\node [style=NTWO] (39) at (-7.25, -2.75) {1};
		\node [style=NTWO] (40) at (-6, 4.5) {3};
		\node [style=none] (41) at (-5.25, 4.25) {};
		\node [style=none] (42) at (-2, 3.25) {};
		\node [style=none] (43) at (-2.5, 1.75) {};
		\node [style=none] (44) at (-5, 0.5) {};
		\node [style=none] (45) at (-0.25, 0) {};
		\node [style=none] (46) at (2, 1) {};
		\node [style=none] (47) at (-3.5, 4.75) {};
		\node [style=none] (48) at (-2.75, 4.75) {$...$};
		\node [style=none] (49) at (-5, 3.25) {};
		\node [style=none] (50) at (0.75, -0.5) {};
		\node [style=none] (54) at (8.75, -5) {};
		\node [style=none] (55) at (8.75, -5.75) {};
		\node [style=none] (56) at (-9, -6) {};
		\node [style=none] (57) at (-9, -5) {};
		\node [style=none] (58) at (0, -7.75) {};
		\node [style=none] (59) at (0, -6.75) {};
		\node [style=TEXTHIGH, text width=7.5cm] (60) at (0, -9.5) {\psdom has a linear kernel of size at most $\kernelsize k$  [\Cref{thm:central}]};
		\node [style=TEXTHIGH, text width=5cm] (61) at (10.5, -2.75) {$\abs{V\setminus(V(\mathfrak{R}) \cup D} \leq 97 \abs{D}$: [\Cref{lemma:outside}]};
		\node [style=TEXTHIGH, text width=12cm] (62) at (-4.75, 8.5) {There exists maximal \dreg $\mathfrak{R}$ with $\abs{R} \leq 3* \abs{D}-6$ [\Cref{lemma:numRegions}]};
		\node [style=none] (63) at (-9.75, 0.75) {};
		\node [style=none] (64) at (-8, -2) {};
		\node [style=none] (65) at (-6, 7.25) {};
		\node [style=none] (66) at (-6, 5.5) {};
		\node [style=none] (67) at (7.25, -4.25) {};
		\node [style=none] (68) at (5.75, -5) {};
		\node [style=TEXTHIGH, text width=5cm] (69) at (-12.75, 1.75) {$\abs{V(R)} \leq 87$ [\Cref{lemma:inside}]};
		\node [style=TEXTHIGHGRAY, text width=4cm] (70) at (-16.75, -1.5) {$\abs{N_2(v,w) \cap V(R)} \leq 24$ [\Cref{lemma:ntwocover}]};
		\node [style=TEXTHIGHGRAY, text width=4cm] (71) at (-14.25, -5.75) {$\abs{N_3(v,w) \cap V(R)} \leq 56$ [\Cref{lemma:rtwosr}]};
		\node [style=TEXTHIGHGRAY, text width=4cm] (72) at (-17, 5) {$ \abs{N_1(v,w) \cap V(R)} \leq 4 $ [\Cref{lemma:nonecover}]};
		\node [style=none] (73) at (-15.5, 3.5) {};
		\node [style=none] (74) at (-14.25, 2.75) {};
		\node [style=none] (75) at (-15, -0.25) {};
		\node [style=none] (76) at (-15, 0.75) {};
		\node [style=none] (77) at (-11.75, -4.5) {};
		\node [style=none] (78) at (-11.75, 0.5) {};
	\end{pgfonlayer}
	\begin{pgfonlayer}{edgelayer}
		\draw [style=FILLPURPLE] (6.center)
			 to (11.center)
			 to (3.center)
			 to (0.center)
			 to (1.center)
			 to (2.center)
			 to [bend right] cycle;
		\draw (2) to (3);
		\draw [style=FILLGREEN] (7.center)
			 to (0.center)
			 to (5.center)
			 to (14.center)
			 to (13.center)
			 to cycle;
		\draw [style=FILLDARKBLUE] (9.center)
			 to (5.center)
			 to (10.center)
			 to (15.center)
			 to cycle;
		\draw (2) to (11);
		\draw [style=FILLDARKBLUE] (14) to (12);
		\draw (12) to (11);
		\draw (13) to (5);
		\draw (7) to (5);
		\draw [style=FILLDARKBLUEINVISIBLE] (6.center)
			 to [bend right=75, looseness=1.25] (9.center)
			 to [in=75, out=-45] cycle;
		\draw (6) to (18);
		\draw (19) to (6);
		\draw (17) to (6);
		\draw (16) to (6);
		\draw (16) to (17);
		\draw (19) to (18);
		\draw (6) to (20);
		\draw (20) to (16);
		\draw [bend left=45, looseness=1.25] (9) to (6);
		\draw [style=FILLDARKBLUEINVISIBLE] (14.center)
			 to (6.center)
			 to [bend right=15] cycle;
		\draw [style=FILLDARKBLUEINVISIBLE] (14.center)
			 to (6.center)
			 to [bend right=15] cycle;
		\draw [style=FILLDARKBLUEINVISIBLE] (14.center)
			 to (6.center)
			 to [bend right=15] cycle;
		\draw (14) to (12);
		\draw (6) to (12);
		\draw (21) to (10);
		\draw (2) to (21);
		\draw (23) to (6);
		\draw (22) to (6);
		\draw (22) to (2);
		\draw (23) to (2);
		\draw [style=simple dashed directed edge not thick] (24.center) to (25.center);
		\draw [style=EDGEDASHED] (32.center) to (26.center);
		\draw [style=EDGEDASHED, bend left=60, looseness=0.75] (26.center) to (27.center);
		\draw [style=EDGEDASHED, in=30, out=-75, looseness=0.75] (27.center) to (28.center);
		\draw [style=EDGEDASHED] (28.center) to (29.center);
		\draw [style=EDGEDASHED, in=-120, out=-165, looseness=1.25] (29.center) to (30.center);
		\draw [style=EDGEDASHED] (30.center) to (31.center);
		\draw [style=EDGEDASHED, in=-135, out=30] (31.center) to (32.center);
		\draw [style=simple dashed directed edge not thick] (33.center) to (35.center);
		\draw [style=simple dashed directed edge not thick] (41.center) to (42.center);
		\draw [style=simple dashed directed edge not thick] (41.center) to (43.center);
		\draw [style=simple dashed directed edge not thick] (41.center) to (44.center);
		\draw [style=simple dashed directed edge not thick, bend right] (41.center) to (45.center);
		\draw [style=simple dashed directed edge not thick] (41.center) to (47.center);
		\draw [style=simple dashed directed edge not thick, bend right=15, looseness=0.75] (41.center) to (46.center);
		\draw [style=simple dashed directed edge not thick, bend right, looseness=1.50] (41.center) to (50.center);
		\draw [style=EDGEDASHED] (57.center) to (56.center);
		\draw [style=EDGEDASHED, in=-135, out=-30, looseness=0.25] (56.center) to (55.center);
		\draw [style=EDGEDASHED] (54.center) to (55.center);
		\draw [style=simple dashed directed edge] (59.center) to (58.center);
		\draw [style=EDGE] (65.center) to (66.center);
		\draw [style=EDGE] (63.center) to (64.center);
		\draw [style=EDGE] (67.center) to (68.center);
		\draw [style=BLUE] (73.center) to (74.center);
		\draw [style=BLUE] (75.center) to (76.center);
		\draw [style=BLUE] (77.center) to (78.center);
	\end{pgfonlayer}
\end{tikzpicture}

%% file: fig/tikz/ruleOne.tikz
\begin{tikzpicture}
	\begin{pgfonlayer}{nodelayer}
		\node [style=none] (0) at (0, 0) {$\rightarrow$};
		\node [style=BLACK] (3) at (-3.5, 0) {};
		\node [style=NONE] (5) at (-3.5, 2) {};
		\node [style=NONE] (6) at (-4.75, 2) {};
		\node [style=NONE] (7) at (-5.5, 1) {};
		\node [style=NTWO] (8) at (-5.5, 0) {};
		\node [style=NTHR] (9) at (-5.5, -1) {};
		\node [style=NTHR] (10) at (-4.75, -2) {};
		\node [style=NTHR] (11) at (-3.5, -2) {};
		\node [style=NTHR] (12) at (-2.25, -2) {};
		\node [style=NTHR] (13) at (-2.5, 1) {};
		\node [style=NTHR] (14) at (-1.75, 0) {};
		\node [style=NTHR] (15) at (-1.75, -1) {};
		\node [style=none] (16) at (-3, 2.75) {};
		\node [style=none] (17) at (-3.5, 3) {};
		\node [style=none] (18) at (-4, 2.75) {};
		\node [style=none] (19) at (-4.75, 2.75) {};
		\node [style=none] (20) at (-5.5, 2.25) {};
		\node [style=none] (21) at (-5.25, 2.75) {};
		\node [style=none] (22) at (-6.5, 1) {};
		\node [style=none] (23) at (-6.25, 1.5) {};
		\node [style=none] (24) at (-6.25, 0.5) {};
		\node [style=BLACK] (25) at (3.5, 0) {};
		\node [style=NONE] (26) at (3.5, 2) {};
		\node [style=NONE] (27) at (2.25, 2) {};
		\node [style=NONE] (28) at (1.5, 1) {};
		\node [style=BRGRAY] (29) at (1.5, 0) {};
		\node [style=BRGRAY] (30) at (1.5, -1) {};
		\node [style=BRGRAY] (31) at (2.25, -2) {};
		\node [style=BRGRAY] (32) at (3.5, -2) {};
		\node [style=BRGRAY] (33) at (4.75, -2) {};
		\node [style=NTHR, label={above:$v'$}] (34) at (4.5, 1) {};
		\node [style=BRGRAY] (35) at (5.25, 0) {};
		\node [style=BRGRAY] (36) at (5.25, -1) {};
		\node [style=none] (37) at (4, 2.75) {};
		\node [style=none] (38) at (3.5, 3) {};
		\node [style=none] (39) at (3, 2.75) {};
		\node [style=none] (40) at (2.25, 2.75) {};
		\node [style=none] (41) at (1.5, 2.25) {};
		\node [style=none] (42) at (1.75, 2.75) {};
		\node [style=none] (43) at (0.5, 1) {};
		\node [style=none] (44) at (0.75, 1.5) {};
		\node [style=none] (45) at (0.75, 0.5) {};
		\node [style=none] (46) at (-3.25, 0.75) {$v$};
		\node [style=none] (48) at (3.75, 0.75) {$v$};
		\node [style=none] (49) at (-6, 3.5) {$G$};
		\node [style=none] (51) at (1, 3.5) {$G'$};
	\end{pgfonlayer}
	\begin{pgfonlayer}{edgelayer}
		\draw (3) to (7);
		\draw (3) to (6);
		\draw (3) to (5);
		\draw (6) to (5);
		\draw (6) to (7);
		\draw (8) to (7);
		\draw (8) to (3);
		\draw (8) to (9);
		\draw (10) to (3);
		\draw (11) to (3);
		\draw (12) to (3);
		\draw (10) to (9);
		\draw (10) to (11);
		\draw (11) to (12);
		\draw [bend right] (10) to (12);
		\draw (13) to (3);
		\draw (14) to (3);
		\draw (15) to (3);
		\draw (14) to (15);
		\draw (19.center) to (6);
		\draw (6) to (20.center);
		\draw (6) to (21.center);
		\draw (5) to (18.center);
		\draw (16.center) to (5);
		\draw (17.center) to (5);
		\draw (24.center) to (7);
		\draw (7) to (22.center);
		\draw (23.center) to (7);
		\draw (25) to (28);
		\draw (25) to (27);
		\draw (25) to (26);
		\draw (27) to (26);
		\draw (27) to (28);
		\draw [style=ULTRAGRAY] (29) to (28);
		\draw [style=ULTRAGRAY] (29) to (25);
		\draw [style=ULTRAGRAY] (29) to (30);
		\draw [style=ULTRAGRAY] (31) to (25);
		\draw [style=ULTRAGRAY] (32) to (25);
		\draw [style=ULTRAGRAY] (33) to (25);
		\draw [style=ULTRAGRAY] (31) to (30);
		\draw [style=ULTRAGRAY] (31) to (32);
		\draw [style=ULTRAGRAY] (32) to (33);
		\draw [style=ULTRAGRAY, bend right] (31) to (33);
		\draw [style=GRAY] (34) to (25);
		\draw [style=ULTRAGRAY] (35) to (25);
		\draw [style=ULTRAGRAY] (36) to (25);
		\draw [style=ULTRAGRAY] (35) to (36);
		\draw (40.center) to (27);
		\draw (27) to (41.center);
		\draw (27) to (42.center);
		\draw (26) to (39.center);
		\draw (37.center) to (26);
		\draw (38.center) to (26);
		\draw (45.center) to (28);
		\draw (28) to (43.center);
		\draw (44.center) to (28);
	\end{pgfonlayer}
\end{tikzpicture}

%% file: fig/tikz/ruletwo.tikz
\begin{tikzpicture}
	\begin{pgfonlayer}{nodelayer}
		\node [style=BLACK] (0) at (-13.75, 5.5) {};
		\node [style=BLACK] (1) at (-3.75, 5.5) {};
		\node [style=NTHR] (2) at (-14, 3.75) {};
		\node [style=NTHR] (3) at (-10, 4.25) {};
		\node [style=NTHR] (4) at (-9.5, 5.25) {};
		\node [style=NTHR] (5) at (-9.25, 7) {};
		\node [style=NONE] (6) at (-11.75, 7.75) {};
		\node [style=NTHR] (7) at (-6.25, 7.75) {};
		\node [style=NTHR] (8) at (-8, 4.25) {};
		\node [style=NTHR] (9) at (-7.5, 6) {};
		\node [style=NONE] (10) at (-15.25, 8) {};
		\node [style=NTWO] (11) at (-13.75, 8) {};
		\node [style=NTWO] (12) at (-2, 6.75) {};
		\node [style=NONE] (13) at (-2.75, 8.25) {};
		\node [style=none] (14) at (-3.25, 6.25) {$w$};
		\node [style=none] (15) at (-14.25, 6.5) {$v$};
		\node [style=NTHR] (16) at (-5.75, 3.25) {};
		\node [style=NTHR] (17) at (-11.5, 3.25) {};
		\node [style=none] (18) at (-2.25, 8.75) {};
		\node [style=none] (19) at (-2.75, 9) {};
		\node [style=none] (20) at (-3.25, 8.75) {};
		\node [style=none] (21) at (-11, 7.75) {};
		\node [style=none] (22) at (-11, 8.25) {};
		\node [style=none] (23) at (-11.5, 8.5) {};
		\node [style=none] (24) at (-15.75, 8.5) {};
		\node [style=none] (25) at (-15.25, 8.75) {};
		\node [style=none] (26) at (-14.75, 8.5) {};
		\node [style=NTHR] (27) at (-12.25, 2.25) {};
		\node [style=NTHR] (28) at (-5.25, 2.25) {};
		\node [style=NTHR] (58) at (-6.5, 5.25) {};
		\node [style=NTHR] (59) at (-7, 7.25) {};
		\node [style=BLACK] (60) at (3, 5.75) {};
		\node [style=BLACK] (61) at (13, 5.75) {};
		\node [style=BRGRAY] (62) at (2.75, 4) {};
		\node [style=BRGRAY] (63) at (6.75, 4.5) {};
		\node [style=BRGRAY] (64) at (7.25, 5.5) {};
		\node [style=BRGRAY] (65) at (7.5, 7.25) {};
		\node [style=NONE] (66) at (5, 8) {};
		\node [style=BRGRAY] (67) at (10.5, 8) {};
		\node [style=BRGRAY] (68) at (8.75, 4.5) {};
		\node [style=BRGRAY] (69) at (9.25, 6.25) {};
		\node [style=NONE] (70) at (1.5, 8.25) {};
		\node [style=BRGRAY] (71) at (3, 8.25) {};
		\node [style=BRGRAY] (72) at (14.75, 7) {};
		\node [style=NONE] (73) at (14, 8.5) {};
		\node [style=none] (74) at (13.5, 6.5) {$w$};
		\node [style=none] (75) at (2.5, 6.75) {$v$};
		\node [style=BRGRAY] (76) at (11, 3.5) {};
		\node [style=BRGRAY] (77) at (5.25, 3.5) {};
		\node [style=none] (78) at (14.5, 9) {};
		\node [style=none] (79) at (14, 9.25) {};
		\node [style=none] (80) at (13.5, 9) {};
		\node [style=none] (81) at (5.75, 8) {};
		\node [style=none] (82) at (5.75, 8.5) {};
		\node [style=none] (83) at (5.25, 8.75) {};
		\node [style=none] (84) at (1, 8.75) {};
		\node [style=none] (85) at (1.5, 9) {};
		\node [style=none] (86) at (2, 8.75) {};
		\node [style=BRGRAY] (87) at (4.5, 2.5) {};
		\node [style=BRGRAY] (88) at (11.5, 2.5) {};
		\node [style=BRGRAY] (89) at (10.25, 5.5) {};
		\node [style=BRGRAY] (90) at (9.75, 7.5) {};
		\node [style=BLACK] (91) at (1.5, 4.75) {};
		\node [style=BLACK] (92) at (14.25, 4.5) {};
		\node [style=none] (93) at (1.5, 5.5) {$v'$};
		\node [style=none] (94) at (14.5, 5.25) {$w'$};
		\node [style=BLACK] (95) at (6.5, 6.75) {};
		\node [style=BLACK] (96) at (9.75, 5) {};
		\node [style=none] (97) at (6.75, 7.5) {$y$};
		\node [style=none] (98) at (9.75, 5.75) {$y'$};
		\node [style=DOM] (104) at (-13.5, -5.75) {};
		\node [style=NONE] (105) at (-11.5, -3.5) {};
		\node [style=NONE] (106) at (-15, -3.25) {};
		\node [style=NTWO] (107) at (-13.5, -3.25) {};
		\node [style=none] (108) at (-14, -4.75) {$v$};
		\node [style=none] (109) at (-10.75, -3.5) {};
		\node [style=none] (110) at (-10.75, -3) {};
		\node [style=none] (111) at (-11.25, -2.75) {};
		\node [style=none] (112) at (-15.5, -2.75) {};
		\node [style=none] (113) at (-15, -2.5) {};
		\node [style=none] (114) at (-14.5, -2.75) {};
		\node [style=BLACK] (115) at (-3, -9.5) {};
		\node [style=NTWO] (116) at (-1.5, -8) {};
		\node [style=NONE] (117) at (-2.25, -6.5) {};
		\node [style=none] (118) at (-2.75, -8.5) {$w$};
		\node [style=none] (119) at (-1.75, -6) {};
		\node [style=none] (120) at (-2.25, -5.75) {};
		\node [style=none] (121) at (-2.75, -6) {};
		\node [style=NTHR] (122) at (-14, -7.75) {};
		\node [style=NTHR] (123) at (-11.75, -9) {};
		\node [style=NTHR] (124) at (-10.75, -9.5) {};
		\node [style=NTHR] (125) at (-10, -8.25) {};
		\node [style=NTHR] (126) at (-13.25, -9.25) {};
		\node [style=NTHR] (127) at (-12, -10.75) {};
		\node [style=NTHR] (128) at (-10.5, -8.75) {};
		\node [style=DOM] (129) at (-8.75, -9.75) {};
		\node [style=NTHR] (130) at (-9, -12.75) {};
		\node [style=NTHR] (131) at (-9.25, -11.75) {};
		\node [style=NTHR] (132) at (-15.75, -7) {};
		\node [style=NTHR] (133) at (-11, -4.25) {};
		\node [style=NTHR] (134) at (-9.25, -4.25) {};
		\node [style=NTHR] (135) at (-9, -5) {};
		\node [style=NTHR] (136) at (-7, -5.75) {};
		\node [style=NTHR] (137) at (-7.75, -6.5) {};
		\node [style=NTHR] (138) at (-9.75, -6.75) {};
		\node [style=NTHR] (139) at (-10, -7.5) {};
		\node [style=BLACK] (140) at (-5.75, -4.5) {};
		\node [style=DOM] (141) at (-5.75, -4.5) {};
		\node [style=NTHR] (142) at (-6.75, -7.25) {};
		\node [style=NTHR] (143) at (-6.5, -8.25) {};
		\node [style=NTHR] (144) at (-6.75, -9) {};
		\node [style=NTHR] (145) at (-5.5, -6.5) {};
		\node [style=NTHR] (146) at (-4.5, -7.25) {};
		\node [style=none] (147) at (0, 6) {$\rightarrow$};
		\node [style=none] (148) at (0, -7.75) {$\rightarrow$};
		\node [style=none] (149) at (-16.5, -2.5) {};
		\node [style=none] (150) at (-17.25, -10.75) {};
		\node [style=none] (151) at (-9.75, -10.5) {};
		\node [style=none] (152) at (-9, -8.5) {};
		\node [style=none] (153) at (-6.25, -6) {};
		\node [style=none] (154) at (-7, -2.75) {};
		\node [style=none] (156) at (-6.5, -13.5) {};
		\node [style=none] (157) at (-15.25, -1) {$N(v)$};
		\node [style=DOM] (158) at (4.5, -5.75) {};
		\node [style=NONE] (159) at (6.5, -3.5) {};
		\node [style=NONE] (160) at (3, -3.25) {};
		\node [style=BRGRAY] (161) at (4.5, -3.25) {};
		\node [style=none] (162) at (4, -4.75) {$v$};
		\node [style=none] (163) at (7.25, -3.5) {};
		\node [style=none] (164) at (7.25, -3) {};
		\node [style=none] (165) at (6.75, -2.75) {};
		\node [style=none] (166) at (2.5, -2.75) {};
		\node [style=none] (167) at (3, -2.5) {};
		\node [style=none] (168) at (3.5, -2.75) {};
		\node [style=BLACK] (169) at (15, -9.5) {};
		\node [style=NTWO] (170) at (15.75, -8) {};
		\node [style=NONE] (171) at (15.25, -7) {};
		\node [style=none] (172) at (15.25, -8.5) {$w$};
		\node [style=none] (173) at (15.75, -6.5) {};
		\node [style=none] (174) at (15.25, -6.25) {};
		\node [style=none] (175) at (14.75, -6.5) {};
		\node [style=BRGRAY] (176) at (4, -7.75) {};
		\node [style=BRGRAY] (177) at (6.25, -9) {};
		\node [style=BRGRAY] (178) at (7.25, -9.5) {};
		\node [style=NTHR] (179) at (8, -8.25) {};
		\node [style=BRGRAY] (180) at (4.75, -9.25) {};
		\node [style=NTHR] (181) at (6, -10.75) {};
		\node [style=BRGRAY] (182) at (7.5, -8.75) {};
		\node [style=DOM] (183) at (9.25, -9.75) {};
		\node [style=NTHR] (184) at (9, -12.75) {};
		\node [style=NTHR] (185) at (8.75, -11.75) {};
		\node [style=BRGRAY] (186) at (2.25, -7) {};
		\node [style=BRGRAY] (187) at (7, -4.25) {};
		\node [style=NTHR] (188) at (8.75, -4.25) {};
		\node [style=BRGRAY] (189) at (9, -5) {};
		\node [style=NTHR] (190) at (11, -5.75) {};
		\node [style=BRGRAY] (191) at (10.25, -6.5) {};
		\node [style=BRGRAY] (192) at (8.25, -6.75) {};
		\node [style=BRGRAY] (193) at (8, -7.5) {};
		\node [style=BLACK] (194) at (12.25, -4.5) {};
		\node [style=DOM] (195) at (12.25, -4.5) {};
		\node [style=NTHR] (196) at (11.25, -7.25) {};
		\node [style=NTHR] (197) at (11.5, -8.25) {};
		\node [style=NTHR] (198) at (11.25, -9) {};
		\node [style=NTHR] (199) at (12.5, -6.75) {};
		\node [style=NTHR] (200) at (13.5, -7.5) {};
		\node [style=none] (202) at (1.5, -2.5) {};
		\node [style=none] (203) at (0.75, -10.75) {};
		\node [style=none] (204) at (8.25, -10.5) {};
		\node [style=none] (205) at (9, -8.5) {};
		\node [style=none] (206) at (11.75, -6) {};
		\node [style=none] (207) at (11, -2.75) {};
		\node [style=none] (208) at (11.5, -13.5) {};
		\node [style=none] (209) at (2.75, -1) {$N(v)$};
		\node [style=none] (210) at (-5.5, -3.5) {$d_1$};
		\node [style=none] (211) at (-8.25, -10.5) {$d_2$};
		\node [style=BLACK] (212) at (2, -5.25) {};
		\node [style=none] (213) at (2, -4.5) {$v'$};
		\node [style=none] (214) at (12.5, -3.5) {$d_1$};
		\node [style=none] (215) at (9.75, -10.5) {$d_2$};
		\node [style=none] (216) at (-2, 7) {};
		\node [style=none] (217) at (15, 1) {(1)};
		\node [style=none] (218) at (15, -13) {(2)};
	\end{pgfonlayer}
	\begin{pgfonlayer}{edgelayer}
		\draw [style=BLUE] (0) to (4);
		\draw [bend right=15] (0) to (3);
		\draw [bend left] (0) to (10);
		\draw (0) to (2);
		\draw (0) to (11);
		\draw (6) to (0);
		\draw (11) to (6);
		\draw [bend left=15, looseness=1.25] (0) to (5);
		\draw (3) to (8);
		\draw [bend left=15] (9) to (4);
		\draw [bend left, looseness=0.75] (5) to (7);
		\draw [bend left=15, looseness=1.25] (1) to (8);
		\draw [in=165, out=30, looseness=0.75] (9) to (1);
		\draw [bend right, looseness=0.75] (1) to (7);
		\draw [bend left] (12) to (1);
		\draw [bend left, looseness=1.25] (1) to (13);
		\draw [bend left=45, looseness=1.25] (13) to (12);
		\draw (17) to (0);
		\draw [bend right=15, looseness=1.25] (16) to (1);
		\draw (17) to (16);
		\draw (18.center) to (13);
		\draw (19.center) to (13);
		\draw (13) to (20.center);
		\draw (6) to (21.center);
		\draw (22.center) to (6);
		\draw (23.center) to (6);
		\draw (10) to (26.center);
		\draw (10) to (25.center);
		\draw (10) to (24.center);
		\draw [bend right=15] (0) to (27);
		\draw (27) to (28);
		\draw [bend right=15, looseness=1.25] (28) to (1);
		\draw (28) to (16);
		\draw (17) to (27);
		\draw [style=BLUE] (58) to (1);
		\draw [style=BLUE] (58) to (4);
		\draw [style=EDGE, bend right=15, looseness=0.75] (5) to (59);
		\draw [style=EDGE, bend left, looseness=0.75] (59) to (1);
		\draw [style=ULTRAGRAY] (60) to (64);
		\draw [style=ULTRAGRAY, bend right=15] (60) to (63);
		\draw [bend left] (60) to (70);
		\draw [style=ULTRAGRAY] (60) to (62);
		\draw [style=ULTRAGRAY] (60) to (71);
		\draw (66) to (60);
		\draw [style=ULTRAGRAY] (71) to (66);
		\draw [style=ULTRAGRAY, bend left=15, looseness=1.25] (60) to (65);
		\draw [style=ULTRAGRAY] (63) to (68);
		\draw [style=ULTRAGRAY, bend left=15] (69) to (64);
		\draw [style=ULTRAGRAY, bend left, looseness=0.75] (65) to (67);
		\draw [style=ULTRAGRAY, bend left=15, looseness=1.25] (61) to (68);
		\draw [style=ULTRAGRAY, in=165, out=30, looseness=0.75] (69) to (61);
		\draw [style=ULTRAGRAY, bend right, looseness=0.75] (61) to (67);
		\draw [style=ULTRAGRAY, bend left] (72) to (61);
		\draw [bend left, looseness=1.25] (61) to (73);
		\draw [style=ULTRAGRAY, bend left=45, looseness=1.25] (73) to (72);
		\draw [style=ULTRAGRAY] (77) to (60);
		\draw [style=ULTRAGRAY, bend right=15, looseness=1.25] (76) to (61);
		\draw [style=ULTRAGRAY] (77) to (76);
		\draw (78.center) to (73);
		\draw (79.center) to (73);
		\draw (73) to (80.center);
		\draw (66) to (81.center);
		\draw (82.center) to (66);
		\draw (83.center) to (66);
		\draw (70) to (86.center);
		\draw (70) to (85.center);
		\draw (70) to (84.center);
		\draw [style=ULTRAGRAY, bend right=15] (60) to (87);
		\draw [style=ULTRAGRAY] (87) to (88);
		\draw [style=ULTRAGRAY, bend right=15, looseness=1.25] (88) to (61);
		\draw [style=ULTRAGRAY] (88) to (76);
		\draw [style=ULTRAGRAY] (77) to (87);
		\draw [style=ULTRAGRAY] (89) to (61);
		\draw [style=ULTRAGRAY] (89) to (64);
		\draw [style=ULTRAGRAY, bend right=15, looseness=0.75] (65) to (90);
		\draw [style=ULTRAGRAY, bend left, looseness=0.75] (90) to (61);
		\draw [style=ULTRAGRAY] (91) to (60);
		\draw [style=EDGE] (61) to (92);
		\draw [style=EDGE] (60) to (91);
		\draw [style=EDGE, in=180, out=0, looseness=1.25] (60) to (95);
		\draw [style=EDGE, in=180, out=0, looseness=1.50] (96) to (61);
		\draw [style=EDGE, in=0, out=180, looseness=0.75] (96) to (95);
		\draw [bend left] (104) to (106);
		\draw (104) to (107);
		\draw (105) to (104);
		\draw (107) to (105);
		\draw (105) to (109.center);
		\draw (110.center) to (105);
		\draw (111.center) to (105);
		\draw (106) to (114.center);
		\draw (106) to (113.center);
		\draw (106) to (112.center);
		\draw [bend left] (116) to (115);
		\draw [bend left, looseness=1.25] (115) to (117);
		\draw [bend left=45, looseness=1.25] (117) to (116);
		\draw (119.center) to (117);
		\draw (120.center) to (117);
		\draw (117) to (121.center);
		\draw [style=EDGE] (125) to (104);
		\draw [style=EDGE] (104) to (124);
		\draw [style=EDGE] (104) to (123);
		\draw [style=EDGE] (122) to (104);
		\draw [style=EDGE] (127) to (104);
		\draw [style=EDGE] (104) to (126);
		\draw [style=EDGE] (128) to (104);
		\draw [style=EDGE] (125) to (129);
		\draw [style=EDGE] (129) to (127);
		\draw [style=EDGE] (124) to (128);
		\draw [style=EDGE] (123) to (127);
		\draw [style=EDGE] (127) to (126);
		\draw [style=EDGE] (126) to (122);
		\draw [style=EDGE, in=195, out=-135, looseness=1.25] (104) to (131);
		\draw [style=EDGE] (131) to (115);
		\draw [style=EDGE, in=180, out=-135, looseness=1.50] (104) to (130);
		\draw [style=EDGE, bend right=15] (130) to (115);
		\draw [style=EDGE, in=180, out=0] (132) to (104);
		\draw [style=EDGE] (104) to (133);
		\draw [style=EDGE] (134) to (104);
		\draw [style=EDGE] (135) to (104);
		\draw [style=EDGE] (136) to (104);
		\draw [style=EDGE] (104) to (137);
		\draw [style=EDGE] (137) to (104);
		\draw [style=EDGE] (138) to (104);
		\draw [style=EDGE] (139) to (104);
		\draw [style=EDGE] (133) to (134);
		\draw [style=EDGE] (134) to (135);
		\draw [style=EDGE] (136) to (135);
		\draw [style=EDGE] (136) to (137);
		\draw [style=EDGE] (137) to (138);
		\draw [style=EDGE] (138) to (139);
		\draw [style=EDGE] (125) to (139);
		\draw [style=EDGE] (134) to (140);
		\draw [style=EDGE] (136) to (140);
		\draw [style=EDGE, in=15, out=120, looseness=1.25] (115) to (140);
		\draw [style=EDGE, in=165, out=0] (142) to (115);
		\draw [style=EDGE] (143) to (115);
		\draw [style=EDGE] (144) to (115);
		\draw [style=EDGE] (129) to (115);
		\draw [style=EDGE] (129) to (144);
		\draw [style=EDGE] (129) to (143);
		\draw [style=EDGE, bend left] (129) to (142);
		\draw [style=EDGE] (115) to (146);
		\draw [style=EDGE] (141) to (146);
		\draw [style=EDGE] (145) to (141);
		\draw [style=EDGE, bend right=15] (145) to (115);
		\draw [style=EDGEDASHED, in=45, out=-105] (154.center) to (153.center);
		\draw [style=EDGEDASHED, in=60, out=-135] (153.center) to (152.center);
		\draw [style=EDGEDASHED, in=75, out=-105] (152.center) to (151.center);
		\draw [style=EDGEDASHED, in=45, out=-105, looseness=0.50] (151.center) to (156.center);
		\draw [style=EDGEDASHED, in=315, out=-150, looseness=0.50] (156.center) to (150.center);
		\draw [style=EDGEDASHED, in=90, out=-90] (149.center) to (150.center);
		\draw [style=EDGEDASHED, in=90, out=90, looseness=0.25] (149.center) to (154.center);
		\draw [bend left] (158) to (160);
		\draw [style=ULTRAGRAY] (158) to (161);
		\draw (159) to (158);
		\draw [style=ULTRAGRAY] (161) to (159);
		\draw (159) to (163.center);
		\draw (164.center) to (159);
		\draw (165.center) to (159);
		\draw (160) to (168.center);
		\draw (160) to (167.center);
		\draw (160) to (166.center);
		\draw [bend left] (170) to (169);
		\draw [bend left, looseness=1.25] (169) to (171);
		\draw [bend left=45, looseness=1.25] (171) to (170);
		\draw (173.center) to (171);
		\draw (174.center) to (171);
		\draw (171) to (175.center);
		\draw [style=EDGE] (179) to (158);
		\draw [style=ULTRAGRAY] (158) to (178);
		\draw [style=ULTRAGRAY] (158) to (177);
		\draw [style=ULTRAGRAY] (176) to (158);
		\draw [style=EDGE] (181) to (158);
		\draw [style=ULTRAGRAY] (158) to (180);
		\draw [style=ULTRAGRAY] (182) to (158);
		\draw [style=EDGE] (179) to (183);
		\draw [style=EDGE] (183) to (181);
		\draw [style=ULTRAGRAY] (178) to (182);
		\draw [style=ULTRAGRAY] (177) to (181);
		\draw [style=ULTRAGRAY] (181) to (180);
		\draw [style=ULTRAGRAY] (180) to (176);
		\draw [style=EDGE, in=195, out=-135, looseness=1.25] (158) to (185);
		\draw [style=EDGE] (185) to (169);
		\draw [style=EDGE, in=180, out=-135, looseness=1.50] (158) to (184);
		\draw [style=EDGE, bend right=15] (184) to (169);
		\draw [style=ULTRAGRAY, in=180, out=0] (186) to (158);
		\draw [style=ULTRAGRAY] (158) to (187);
		\draw [style=EDGE] (188) to (158);
		\draw [style=ULTRAGRAY] (189) to (158);
		\draw [style=EDGE] (190) to (158);
		\draw [style=EDGE] (158) to (191);
		\draw [style=ULTRAGRAY] (191) to (158);
		\draw [style=ULTRAGRAY] (192) to (158);
		\draw [style=ULTRAGRAY] (193) to (158);
		\draw [style=ULTRAGRAY] (187) to (188);
		\draw [style=ULTRAGRAY] (188) to (189);
		\draw [style=ULTRAGRAY] (190) to (189);
		\draw [style=ULTRAGRAY] (190) to (191);
		\draw [style=ULTRAGRAY] (191) to (192);
		\draw [style=ULTRAGRAY] (192) to (193);
		\draw [style=ULTRAGRAY] (179) to (193);
		\draw [style=EDGE] (188) to (194);
		\draw [style=EDGE] (190) to (194);
		\draw [style=EDGE, in=15, out=120, looseness=1.25] (169) to (194);
		\draw [style=EDGE, in=165, out=0] (196) to (169);
		\draw [style=EDGE] (197) to (169);
		\draw [style=EDGE] (198) to (169);
		\draw [style=EDGE] (183) to (169);
		\draw [style=EDGE] (183) to (198);
		\draw [style=EDGE] (183) to (197);
		\draw [style=EDGE, bend left] (183) to (196);
		\draw [style=EDGE] (169) to (200);
		\draw [style=EDGE] (195) to (200);
		\draw [style=EDGE] (199) to (195);
		\draw [style=EDGE, bend right=15] (199) to (169);
		\draw [style=EDGEDASHED, in=45, out=-105] (207.center) to (206.center);
		\draw [style=EDGEDASHED, in=60, out=-135] (206.center) to (205.center);
		\draw [style=EDGEDASHED, in=75, out=-105] (205.center) to (204.center);
		\draw [style=EDGEDASHED, in=45, out=-105, looseness=0.50] (204.center) to (208.center);
		\draw [style=EDGEDASHED, in=315, out=-150, looseness=0.50] (208.center) to (203.center);
		\draw [style=EDGEDASHED, in=105, out=-90, looseness=0.50] (202.center) to (203.center);
		\draw [style=EDGEDASHED, in=90, out=90, looseness=0.25] (202.center) to (207.center);
		\draw [style=EDGE] (212) to (158);
	\end{pgfonlayer}
\end{tikzpicture}

%% file: fig/tikz/rulethree.tikz
\begin{tikzpicture}
	\begin{pgfonlayer}{nodelayer}
		\node [style=DOM] (0) at (-14.25, 0) {};
		\node [style=DOM] (1) at (-8.25, 0) {};
		\node [style=NONE] (2) at (-11.25, 3.5) {};
		\node [style=NTWO] (3) at (-11.25, 2.75) {};
		\node [style=NTHR] (4) at (-11.25, 1.75) {};
		\node [style=NTHR] (5) at (-11.25, 0.75) {};
		\node [style=NTHR] (8) at (-11.25, -0.5) {};
		\node [style=NTHR] (9) at (-11.25, -1.5) {};
		\node [style=NTWO] (10) at (-11.25, -2.5) {};
		\node [style=NONE] (11) at (-11.25, -3.25) {};
		\node [style=none] (12) at (-7.5, 0) {$\rightarrow$};
		\node [style=DOM] (13) at (-6.75, 0) {};
		\node [style=BLACK] (14) at (-0.75, 0) {};
		\node [style=NONE] (15) at (-3.75, 3.5) {};
		\node [style=BRGRAY] (16) at (-3.75, 2.75) {};
		\node [style=BRGRAY] (17) at (-3.75, 1.5) {};
		\node [style=BRGRAY] (21) at (-3.75, -0.5) {};
		\node [style=BRGRAY] (22) at (-3.75, -1.5) {};
		\node [style=BRGRAY] (23) at (-3.75, -2.5) {};
		\node [style=NONE] (24) at (-3.75, -3.25) {};
		\node [style=BLACK] (25) at (-3.75, 0.75) {};
		\node [style=BLACK] (26) at (-3.75, -0.75) {};
		\node [style=none] (27) at (-14.5, 0.75) {$v$};
		\node [style=none] (28) at (-8.25, 0.75) {$w$};
		\node [style=none] (31) at (-0.5, 0.75) {$w$};
		\node [style=none] (32) at (-6.75, 0.75) {$v$};
		\node [style=none] (33) at (-3.25, 1.5) {$y$};
		\node [style=none] (34) at (-3.25, -1.5) {$y'$};
		\node [style=none] (35) at (-3, 3.75) {};
		\node [style=none] (36) at (-3.75, 4.25) {};
		\node [style=none] (37) at (-4.5, 3.75) {};
		\node [style=none] (38) at (-10.5, 3.75) {};
		\node [style=none] (39) at (-11.25, 4.25) {};
		\node [style=none] (40) at (-12, 3.75) {};
		\node [style=none] (41) at (-12, -3.75) {};
		\node [style=none] (42) at (-10.5, -3.75) {};
		\node [style=none] (43) at (-11.25, -4.25) {};
		\node [style=none] (44) at (-4.5, -3.75) {};
		\node [style=none] (45) at (-3.75, -4.25) {};
		\node [style=none] (46) at (-3, -3.75) {};
		\node [style=BLACK] (47) at (0.75, 0) {};
		\node [style=BLACK] (48) at (6.75, 0) {};
		\node [style=NONE] (49) at (3.75, 3.5) {};
		\node [style=NTWO] (50) at (3.75, 1.5) {};
		\node [style=DOM] (52) at (3.75, 0) {};
		\node [style=NTHR] (54) at (3.75, -1.5) {};
		\node [style=NONE] (56) at (3.75, -3.25) {};
		\node [style=none] (57) at (7.5, 0) {$\rightarrow$};
		\node [style=BLACK] (58) at (8.25, 0) {};
		\node [style=BLACK] (59) at (14.25, 0) {};
		\node [style=NONE] (60) at (11.25, 3.5) {};
		\node [style=BRGRAY] (61) at (11.25, 1.5) {};
		\node [style=BRGRAY] (63) at (11.25, 0) {};
		\node [style=BRGRAY] (64) at (11.25, -1.5) {};
		\node [style=NONE] (66) at (11.25, -3.25) {};
		\node [style=DOM] (67) at (11.25, 0.25) {};
		\node [style=none] (69) at (0.5, 0.75) {$v$};
		\node [style=none] (70) at (6.75, 0.75) {$w$};
		\node [style=none] (71) at (14.5, 0.75) {$w$};
		\node [style=none] (72) at (8.25, 0.75) {$v$};
		\node [style=none] (74) at (11.75, 0.75) {$y$};
		\node [style=none] (75) at (12, 3.75) {};
		\node [style=none] (76) at (11.25, 4.25) {};
		\node [style=none] (77) at (10.5, 3.75) {};
		\node [style=none] (78) at (4.5, 3.75) {};
		\node [style=none] (79) at (3.75, 4.25) {};
		\node [style=none] (80) at (3, 3.75) {};
		\node [style=none] (81) at (3, -3.75) {};
		\node [style=none] (82) at (4.5, -3.75) {};
		\node [style=none] (83) at (3.75, -4.25) {};
		\node [style=none] (84) at (10.5, -3.75) {};
		\node [style=none] (85) at (11.25, -4.25) {};
		\node [style=none] (86) at (12, -3.75) {};
		\node [style=none] (87) at (0, 5) {};
		\node [style=none] (88) at (0, -5) {};
		\node [style=none] (89) at (0, -5) {};
		\node [style=none] (90) at (7.5, 5) {$G[R\setminus\partial R] \cong P_3$};
		\node [style=none] (91) at (-7.5, 5) {$G[R\setminus\partial R] \ncong P_3$};
		\node [style=none] (92) at (-12, 2.75) {};
		\node [style=none] (93) at (-10.5, 2.75) {};
		\node [style=none] (94) at (-12, -2.5) {};
		\node [style=none] (95) at (-10.5, -2.5) {};
		\node [style=none] (96) at (-10.5, 2.75) {};
		\node [style=none] (97) at (3, 2) {};
		\node [style=none] (98) at (5, 2) {};
		\node [style=none] (99) at (5, -2) {};
		\node [style=none] (100) at (3, -2) {};
		\node [style=none] (101) at (-13.5, 3.75) {$R \setminus \partial R$};
		\node [style=none] (102) at (1.25, 3.25) {$R \setminus \partial R$};
		\node [style=none] (103) at (-12.5, 2.25) {};
		\node [style=none] (104) at (2.5, 1.25) {};
		\node [style=none] (105) at (2.5, 1.25) {};
		\node [style=none] (106) at (-13.5, 3) {};
		\node [style=none] (107) at (1.25, 2.5) {};
		\node [style=none] (108) at (4.25, 2.75) {$P_1$};
		\node [style=none] (109) at (4.25, 0.75) {$P_2$};
		\node [style=none] (110) at (4.25, -0.75) {$P_3$};
		\node [style=none] (111) at (-10.75, 4.5) {$b$};
		\node [style=none] (112) at (11.75, 4.5) {$b$};
		\node [style=none] (113) at (4.25, 4.5) {$b$};
		\node [style=none] (114) at (-3.25, 4.5) {$b$};
		\node [style=none] (115) at (4.25, -4.25) {$b'$};
		\node [style=none] (116) at (11.75, -4.25) {$b'$};
		\node [style=none] (117) at (-3.25, -4.25) {$b'$};
		\node [style=none] (118) at (-10.75, -4.25) {$b'$};
	\end{pgfonlayer}
	\begin{pgfonlayer}{edgelayer}
		\draw [style=FILLBLUE] (11.center)
			 to [in=240, out=15] (1.center)
			 to [in=-15, out=120] (2.center)
			 to [in=60, out=-165] (0.center)
			 to [in=165, out=-60] cycle;
		\draw (3) to (0);
		\draw (0) to (4);
		\draw (5) to (1);
		\draw (0) to (5);
		\draw (4) to (1);
		\draw (1) to (8);
		\draw (8) to (0);
		\draw (0) to (9);
		\draw (9) to (1);
		\draw (10) to (0);
		\draw (1) to (10);
		\draw (3) to (1);
		\draw (10) to (11);
		\draw (3) to (2);
		\draw (10) to (9);
		\draw (3) to (4);
		\draw (5) to (4);
		\draw [style=FILLBLUE] (13.center)
			 to [in=165, out=-60] (24.center)
			 to [in=240, out=15] (14.center)
			 to [in=-15, out=120] (15.center)
			 to [in=60, out=-165] cycle;
		\draw [style=ULTRAGRAY] (16) to (13);
		\draw [style=ULTRAGRAY] (13) to (17);
		\draw [style=ULTRAGRAY] (17) to (14);
		\draw [style=ULTRAGRAY] (14) to (21);
		\draw [style=ULTRAGRAY] (21) to (13);
		\draw [style=ULTRAGRAY] (13) to (22);
		\draw [style=ULTRAGRAY] (22) to (14);
		\draw [style=ULTRAGRAY] (23) to (13);
		\draw [style=ULTRAGRAY] (14) to (23);
		\draw [style=ULTRAGRAY] (16) to (14);
		\draw (23) to (24);
		\draw (16) to (15);
		\draw [style=ULTRAGRAY] (23) to (22);
		\draw [style=ULTRAGRAY] (16) to (17);
		\draw [bend right=15] (25) to (14);
		\draw [bend left=15] (14) to (26);
		\draw [bend right=15] (26) to (13);
		\draw [bend left=15] (13) to (25);
		\draw [style=FILLGREEN] (40.center) to (2);
		\draw [style=FILLGREEN] (2) to (39.center);
		\draw [style=FILLGREEN] (38.center) to (2);
		\draw [style=FILLGREEN] (37.center) to (15);
		\draw [style=FILLGREEN] (36.center) to (15);
		\draw [style=FILLGREEN] (35.center) to (15);
		\draw [style=FILLGREEN] (43.center) to (11);
		\draw [style=FILLGREEN] (41.center) to (11);
		\draw [style=FILLGREEN] (42.center) to (11);
		\draw [style=FILLGREEN] (24) to (44.center);
		\draw [style=FILLGREEN] (24) to (45.center);
		\draw [style=FILLGREEN] (46.center) to (24);
		\draw [style=FILLBLUE] (56.center)
			 to [in=240, out=15] (48.center)
			 to [in=-15, out=120] (49.center)
			 to [in=60, out=-165] (47.center)
			 to [in=165, out=-60] cycle;
		\draw (50) to (47);
		\draw (52) to (48);
		\draw (47) to (52);
		\draw (47) to (54);
		\draw (54) to (48);
		\draw (50) to (48);
		\draw (50) to (49);
		\draw [style=FILLBLUE] (58.center)
			 to [in=165, out=-60] (66.center)
			 to [in=240, out=15] (59.center)
			 to [in=-15, out=120] (60.center)
			 to [in=60, out=-165] cycle;
		\draw [style=ULTRAGRAY] (61) to (58);
		\draw [style=ULTRAGRAY] (59) to (63);
		\draw [style=ULTRAGRAY] (63) to (58);
		\draw [style=ULTRAGRAY] (58) to (64);
		\draw [style=ULTRAGRAY] (64) to (59);
		\draw [style=ULTRAGRAY] (61) to (59);
		\draw [style=ULTRAGRAY] (61) to (60);
		\draw [bend right=15] (67) to (59);
		\draw [bend left=15] (58) to (67);
		\draw [style=FILLGREEN] (80.center) to (49);
		\draw [style=FILLGREEN] (49) to (79.center);
		\draw [style=FILLGREEN] (78.center) to (49);
		\draw [style=FILLGREEN] (77.center) to (60);
		\draw [style=FILLGREEN] (76.center) to (60);
		\draw [style=FILLGREEN] (75.center) to (60);
		\draw [style=FILLGREEN] (83.center) to (56);
		\draw [style=FILLGREEN] (81.center) to (56);
		\draw [style=FILLGREEN] (82.center) to (56);
		\draw [style=FILLGREEN] (66) to (84.center);
		\draw [style=FILLGREEN] (66) to (85.center);
		\draw [style=FILLGREEN] (86.center) to (66);
		\draw [style=EDGEDASHED] (87.center) to (89.center);
		\draw [style=EDGE] (52) to (54);
		\draw [style=ULTRAGRAY] (64) to (63);
		\draw [style=EDGE] (52) to (50);
		\draw [style=ULTRAGRAY] (63) to (61);
		\draw [style=EDGEDASHED] (92.center) to (96.center);
		\draw [style=EDGEDASHED] (96.center) to (95.center);
		\draw [style=EDGEDASHED] (94.center) to (92.center);
		\draw [style=EDGEDASHED] (94.center) to (95.center);
		\draw [style=EDGEDASHED] (97.center) to (98.center);
		\draw [style=EDGEDASHED] (98.center) to (99.center);
		\draw [style=EDGEDASHED] (99.center) to (100.center);
		\draw [style=EDGEDASHED] (97.center) to (100.center);
		\draw [style=simple directed edge, bend right, looseness=1.25] (106.center) to (103.center);
		\draw [style=simple directed edge, bend right] (107.center) to (105.center);
	\end{pgfonlayer}
\end{tikzpicture}

%% file: fig/tikz/maxntwo.tikz
\begin{tikzpicture}
	\begin{pgfonlayer}{nodelayer}
		\node [style=DOM] (0) at (-4, 0) {};
		\node [style=NONE] (1) at (-2, 3.75) {};
		\node [style=NONE] (2) at (4, 3.75) {};
		\node [style=DOM] (3) at (6, 0) {};
		\node [style=NONE] (4) at (4, -2) {};
		\node [style=NONE] (5) at (-2, -2) {};
		\node [style=none] (6) at (-4.5, 0.5) {$v$};
		\node [style=none] (7) at (6, 0.75) {$w$};
		\node [style=none] (9) at (4, 5) {$u_2$};
		\node [style=none] (10) at (-2, 5) {$u_1$};
		\node [style=none] (11) at (-2, -3.25) {$u_3$};
		\node [style=none] (12) at (4, -3.25) {$u_4$};
		\node [style=NTWO] (13) at (-3.25, 2) {};
		\node [style=NTWO] (14) at (-2.5, 1.75) {};
		\node [style=NTWO] (15) at (-1.75, 1.5) {};
		\node [style=NTWO] (16) at (-1, 1.25) {};
		\node [style=NTWO] (17) at (0, 1) {};
		\node [style=none] (19) at (-1.5, 4.25) {};
		\node [style=none] (20) at (-2, 4.5) {};
		\node [style=none] (21) at (-2.5, 4.25) {};
		\node [style=none] (22) at (3.5, 4.25) {};
		\node [style=none] (23) at (4, 4.5) {};
		\node [style=none] (24) at (4.5, 4.25) {};
		\node [style=none] (25) at (-2, -2.75) {};
		\node [style=none] (26) at (-2.5, -2.5) {};
		\node [style=none] (27) at (-1.5, -2.5) {};
		\node [style=none] (28) at (4, -2.75) {};
		\node [style=none] (29) at (4.5, -2.5) {};
		\node [style=none] (30) at (3.5, -2.5) {};
	\end{pgfonlayer}
	\begin{pgfonlayer}{edgelayer}
		\draw (1) to (2);
		\draw (5) to (4);
		\draw [style=EDGEDASHED] (3) to (1);
		\draw [style=FILLGREENINVISBLE, bend right, looseness=1.25] (1) to (0);
		\draw [style=FILLGREENINVISBLE, bend right, looseness=0.75] (0) to (1);
		\draw [style=FILLGREENINVISBLE, bend right, looseness=1.25] (1) to (0);
		\draw [style=FILLGREENINVISBLE, bend right=60, looseness=1.25] (0) to (1);
		\draw [style=FILLBLUE] (0.center)
			 to [in=-135, out=15, looseness=1.50] (17.center)
			 to [in=-45, out=75, looseness=1.25] (1.center)
			 to [bend right, looseness=1.25] cycle;
		\draw [style=FILLBLUE] (0.center)
			 to [bend right=45, looseness=1.50] (2.center)
			 to [bend left, looseness=1.50] cycle;
		\draw [style=FILLBLUE] (5.center)
			 to [bend right, looseness=0.75] (0.center)
			 to cycle;
		\draw [style=FILLBLUE] (2.center)
			 to [bend right, looseness=0.75] (3.center)
			 to cycle;
		\draw [style=FILLBLUE] (4.center)
			 to [bend left, looseness=0.75] (3.center)
			 to cycle;
		\draw [style=FILLBLUE] (5.center)
			 to [bend left=15, looseness=0.50] (3.center)
			 to [bend left=15, looseness=0.50] cycle;
		\draw (0) to (13);
		\draw (13) to (1);
		\draw (0) to (14);
		\draw (14) to (1);
		\draw (16) to (0);
		\draw (16) to (1);
		\draw [style=EDGEDASHED] (0) to (4);
		\draw (21.center) to (1);
		\draw (20.center) to (1);
		\draw (19.center) to (1);
		\draw (23.center) to (2);
		\draw (24.center) to (2);
		\draw (22.center) to (2);
		\draw (25.center) to (5);
		\draw (27.center) to (5);
		\draw (26.center) to (5);
		\draw (30.center) to (4);
		\draw (28.center) to (4);
		\draw (29.center) to (4);
		\draw (14) to (15);
		\draw (0) to (15);
		\draw (15) to (1);
		\draw (15) to (16);
	\end{pgfonlayer}
\end{tikzpicture}

%% file: fig/tikz/nthreeinside.tikz
\begin{tikzpicture}
	\begin{pgfonlayer}{nodelayer}
		\node [style=BLACK] (0) at (-12, 3.5) {};
		\node [style=BLACK] (1) at (-2, 3.5) {};
		\node [style=NONE] (2) at (-10, 6.5) {};
		\node [style=NONE] (3) at (-4, 6.5) {};
		\node [style=NONE] (4) at (-4, 0.5) {};
		\node [style=NONE] (5) at (-10, 0.5) {};
		\node [style=none] (6) at (-1.5, 4.25) {$w$};
		\node [style=none] (7) at (-12.75, 4.25) {$v$};
		\node [style=DOM] (8) at (-7, 5) {};
		\node [style=DOM] (9) at (-7, 3.25) {};
		\node [style=DOM] (10) at (-7, 1) {};
		\node [style=NTHR] (12) at (-8.5, 6) {};
		\node [style=NTHR] (13) at (-9.75, 5.5) {};
		\node [style=NTHR] (14) at (-8.5, 5.25) {};
		\node [style=NTHR] (15) at (-10, 4.5) {};
		\node [style=none] (16) at (-3.5, 7) {};
		\node [style=none] (17) at (-4, 7.25) {};
		\node [style=none] (18) at (-4.5, 7) {};
		\node [style=none] (19) at (-9.5, 7) {};
		\node [style=none] (20) at (-10, 7.25) {};
		\node [style=none] (21) at (-10.5, 7) {};
		\node [style=none] (22) at (-10.5, 0) {};
		\node [style=none] (23) at (-10, -0.25) {};
		\node [style=none] (24) at (-9.5, 0) {};
		\node [style=none] (25) at (-4.5, 0) {};
		\node [style=none] (26) at (-4, -0.25) {};
		\node [style=none] (27) at (-3.5, 0) {};
		\node [style=none] (28) at (-6.75, 8.5) {\textbf{Case 0.1}: Maximal 6 simple regions};
		\node [style=BLACK] (29) at (1, -10) {};
		\node [style=BLACK] (30) at (11, -10) {};
		\node [style=NONE] (31) at (3, -7) {};
		\node [style=NONE] (32) at (9, -7) {};
		\node [style=NONE] (33) at (9, -13) {};
		\node [style=NONE] (34) at (3, -13) {};
		\node [style=none] (35) at (11.5, -9.25) {$w$};
		\node [style=none] (36) at (0.25, -7.75) {$v$};
		\node [style=none] (44) at (9.5, -6.5) {};
		\node [style=none] (45) at (9, -6.25) {};
		\node [style=none] (46) at (8.5, -6.5) {};
		\node [style=none] (47) at (3.5, -6.5) {};
		\node [style=none] (48) at (3, -6.25) {};
		\node [style=none] (49) at (2.5, -6.5) {};
		\node [style=none] (50) at (2.5, -13.5) {};
		\node [style=none] (51) at (3, -13.75) {};
		\node [style=none] (52) at (3.5, -13.5) {};
		\node [style=none] (53) at (8.5, -13.5) {};
		\node [style=none] (54) at (9, -13.75) {};
		\node [style=none] (55) at (9.5, -13.5) {};
		\node [style=none] (56) at (6, -5) {\textbf{Case 1}: At most 4 vertices};
		\node [style=BLACK] (57) at (-12, -7.5) {};
		\node [style=BLACK] (58) at (-2, -7.5) {};
		\node [style=NONE] (59) at (-10, -4.5) {};
		\node [style=NONE] (60) at (-4, -4.5) {};
		\node [style=NONE] (61) at (-4, -13) {};
		\node [style=NONE] (62) at (-10, -13) {};
		\node [style=none] (63) at (-1.5, -6.75) {$w$};
		\node [style=none] (64) at (-12.75, -6.75) {$v$};
		\node [style=none] (72) at (-3.5, -4) {};
		\node [style=none] (73) at (-4, -3.75) {};
		\node [style=none] (74) at (-4.5, -4) {};
		\node [style=none] (75) at (-9.5, -4) {};
		\node [style=none] (76) at (-10, -3.75) {};
		\node [style=none] (77) at (-10.5, -4) {};
		\node [style=none] (78) at (-10.5, -13.5) {};
		\node [style=none] (79) at (-10, -13.75) {};
		\node [style=none] (80) at (-9.5, -13.5) {};
		\node [style=none] (81) at (-4.5, -13.5) {};
		\node [style=none] (82) at (-4, -13.75) {};
		\node [style=none] (83) at (-3.5, -13.5) {};
		\node [style=none] (84) at (-6.75, -2.75) {\textbf{Case 2 and 3}: 14 regions};
		\node [style=none] (112) at (6, 8.5) {\textbf{Case 0.2}: At most 11 simple regions};
		\node [style=NTHR] (113) at (3, -11.75) {};
		\node [style=NTHR] (114) at (9.25, -8.5) {};
		\node [style=NTHR] (115) at (4.75, -10) {};
		\node [style=none] (116) at (8.25, -7.75) {$v'$};
		\node [style=none] (117) at (4.75, -9.25) {$y$};
		\node [style=none] (118) at (4, -11.25) {$w'$};
		\node [style=none] (119) at (-6.75, 6) {$d_1$};
		\node [style=none] (120) at (-6.75, 4.25) {$d_2$};
		\node [style=none] (121) at (-6.75, 2) {$d_3$};
		\node [style=BLACK] (122) at (-7.25, -7) {};
		\node [style=BLACK] (123) at (-6, -10.5) {};
		\node [style=BLACK] (124) at (-4.25, -6.5) {};
		\node [style=BLACK] (125) at (-4.5, -7.75) {};
		\node [style=BLACK] (126) at (-5.25, -9.25) {};
		\node [style=BLACK] (127) at (-4.25, -11) {};
		\node [style=BLACK] (128) at (-2, -7.5) {};
		\node [style=NTHR] (138) at (7.5, -9.5) {};
		\node [style=none] (139) at (7, -9) {$y'$};
		\node [style=BRGRAY] (140) at (3, -7.75) {};
		\node [style=BRGRAY] (141) at (4.5, -7.5) {};
		\node [style=BRGRAY] (142) at (8.75, -11) {};
		\node [style=BRGRAY] (143) at (9.25, -10.25) {};
		\node [style=BRGRAY] (144) at (6, -10.75) {};
		\node [style=BRGRAY] (145) at (5.75, -8.25) {};
		\node [style=BRGRAY] (146) at (8, -11.75) {};
		\node [style=BRGRAY] (147) at (6.5, -7.75) {};
		\node [style=BRGRAY] (148) at (3.5, -8.75) {};
		\node [style=BRGRAY] (149) at (6.75, -12) {};
		\node [style=BRGRAY] (150) at (3.25, -9.5) {};
		\node [style=NTHR] (164) at (-9.5, -11) {};
		\node [style=none] (165) at (-8.75, -11.25) {$v'$};
		\node [style=BLACK] (166) at (1, 3.75) {};
		\node [style=BLACK] (167) at (11, 3.75) {};
		\node [style=NONE] (168) at (3, 6.75) {};
		\node [style=NONE] (169) at (9, 6.75) {};
		\node [style=NONE] (170) at (9, -1.75) {};
		\node [style=NONE] (171) at (3, -1.75) {};
		\node [style=none] (172) at (11.5, 4.5) {$w$};
		\node [style=none] (173) at (0.25, 4.5) {$v$};
		\node [style=none] (174) at (9.5, 7.25) {};
		\node [style=none] (175) at (9, 7.5) {};
		\node [style=none] (176) at (8.5, 7.25) {};
		\node [style=none] (177) at (3.5, 7.25) {};
		\node [style=none] (178) at (3, 7.5) {};
		\node [style=none] (179) at (2.5, 7.25) {};
		\node [style=none] (180) at (2.5, -2.25) {};
		\node [style=none] (181) at (3, -2.5) {};
		\node [style=none] (182) at (3.5, -2.25) {};
		\node [style=none] (183) at (8.5, -2.25) {};
		\node [style=none] (184) at (9, -2.5) {};
		\node [style=none] (185) at (9.5, -2.25) {};
		\node [style=BLACK] (186) at (7, 5.5) {};
		\node [style=BLACK] (187) at (7, 4) {};
		\node [style=BLACK] (188) at (4.5, 2.25) {};
		\node [style=BLACK] (189) at (4.75, 0.25) {};
		\node [style=BLACK] (190) at (11, 3.75) {};
		\node [style=none] (191) at (6.25, 5.25) {$u_1$};
		\node [style=none] (192) at (5.75, 0) {$u_2'$};
		\node [style=none] (193) at (5.75, 2.25) {$u_1'$};
		\node [style=none] (194) at (6.25, 3.75) {$u_2$};
	\end{pgfonlayer}
	\begin{pgfonlayer}{edgelayer}
		\draw [bend left=15] (1) to (4);
		\draw (4) to (5);
		\draw [bend left=15] (5) to (0);
		\draw [bend left=15] (0) to (2);
		\draw (2) to (3);
		\draw [bend left=15] (3) to (1);
		\draw [style=FILLBLUE] (8.center)
			 to (0.center)
			 to [bend left=60, looseness=1.25] cycle;
		\draw [style=FILLBLUE] (0.center)
			 to [bend left=15] (9.center)
			 to [bend right=345] cycle;
		\draw [style=FILLBLUE] (10.center)
			 to [bend right=15] (0.center)
			 to [bend right=15] cycle;
		\draw [style=FILLBLUE] (1.center)
			 to [bend right=15] (8.center)
			 to cycle;
		\draw [style=FILLBLUE] (1.center)
			 to [bend right=15] (9.center)
			 to [bend right=15] cycle;
		\draw [style=FILLBLUE] (1.center)
			 to [bend left=15, looseness=0.75] (10.center)
			 to [bend left=15] cycle;
		\draw [style=FILLGREENINVISBLE] (12) to (8);
		\draw [style=FILLGREENINVISBLE] (8) to (13);
		\draw [style=FILLGREENINVISBLE] (13) to (8);
		\draw [bend right, looseness=0.75] (8) to (13);
		\draw (8) to (14);
		\draw (16.center) to (3);
		\draw (3) to (17.center);
		\draw (3) to (18.center);
		\draw (2) to (20.center);
		\draw (19.center) to (2);
		\draw (2) to (21.center);
		\draw (24.center) to (5);
		\draw (5) to (23.center);
		\draw (22.center) to (5);
		\draw (25.center) to (4);
		\draw (4) to (26.center);
		\draw (27.center) to (4);
		\draw [bend left=15] (30) to (33);
		\draw (33) to (34);
		\draw [bend left=15] (34) to (29);
		\draw [bend left=15] (29) to (31);
		\draw (31) to (32);
		\draw [bend left=15] (32) to (30);
		\draw (44.center) to (32);
		\draw (32) to (45.center);
		\draw (32) to (46.center);
		\draw (31) to (48.center);
		\draw (47.center) to (31);
		\draw (31) to (49.center);
		\draw (52.center) to (34);
		\draw (34) to (51.center);
		\draw (50.center) to (34);
		\draw (53.center) to (33);
		\draw (33) to (54.center);
		\draw (55.center) to (33);
		\draw [bend left=15] (58) to (61);
		\draw (61) to (62);
		\draw [bend left=15, looseness=1.25] (62) to (57);
		\draw [bend left=15] (57) to (59);
		\draw (59) to (60);
		\draw [bend left=15] (60) to (58);
		\draw (72.center) to (60);
		\draw (60) to (73.center);
		\draw (60) to (74.center);
		\draw (59) to (76.center);
		\draw (75.center) to (59);
		\draw (59) to (77.center);
		\draw (80.center) to (62);
		\draw (62) to (79.center);
		\draw (78.center) to (62);
		\draw (81.center) to (61);
		\draw (61) to (82.center);
		\draw (83.center) to (61);
		\draw [bend right=15] (29) to (115);
		\draw [bend left=15] (115) to (30);
		\draw (30) to (114);
		\draw (113) to (29);
		\draw [style=FILLBLUE] (124.center)
			 to (122.center)
			 to (125.center)
			 to [in=180, out=360] (128.center)
			 to cycle;
		\draw [style=FILLBLUE] (126.center)
			 to (123.center)
			 to (127.center)
			 to (128.center)
			 to cycle;
		\draw [style=FILLBLUE] (57.center)
			 to [bend right=345] (124.center)
			 to [bend left=330] cycle;
		\draw [style=FILLBLUE] (57.center)
			 to [bend left=345, looseness=1.25] (126.center)
			 to [bend right=345] cycle;
		\draw [style=FILLBLUE] (57.center)
			 to (122.center)
			 to [bend left=15, looseness=0.25] cycle;
		\draw [style=FILLBLUE] (57.center)
			 to [bend left=60] (128.center)
			 to [in=60, out=135] cycle;
		\draw [style=FILLBLUE] (128.center)
			 to [bend left, looseness=0.75] (57.center)
			 to [bend right=75, looseness=0.25] cycle;
		\draw [style=FILLBLUE] (57.center)
			 to (125.center)
			 to [bend left=15, looseness=0.50] cycle;
		\draw [style=FILLBLUE] (57.center)
			 to [bend right=60, looseness=0.75] (127.center)
			 to [bend right=330] cycle;
		\draw [style=FILLBLUE] (123.center)
			 to [bend right=345] (57.center)
			 to [bend right] cycle;
		\draw [style=FILLBLUE] (57.center)
			 to [bend right=75, looseness=1.75] (128.center)
			 to [bend left=90, looseness=1.75] cycle;
		\draw [style=EDGE, bend left, looseness=0.75] (0) to (14);
		\draw [style=EDGE, bend right=15] (13) to (0);
		\draw [style=EDGE] (0) to (15);
		\draw [style=EDGE] (15) to (8);
		\draw [style=ULTRAGRAY] (140) to (29);
		\draw [style=ULTRAGRAY] (141) to (29);
		\draw [style=ULTRAGRAY] (143) to (30);
		\draw [style=ULTRAGRAY] (30) to (142);
		\draw [style=ULTRAGRAY] (142) to (143);
		\draw [style=ULTRAGRAY] (141) to (140);
		\draw [style=ULTRAGRAY, bend right=345] (144) to (29);
		\draw [style=ULTRAGRAY] (144) to (143);
		\draw [style=ULTRAGRAY] (145) to (141);
		\draw [style=ULTRAGRAY] (145) to (30);
		\draw [style=ULTRAGRAY, bend right] (146) to (30);
		\draw [style=ULTRAGRAY] (30) to (147);
		\draw [style=ULTRAGRAY] (147) to (141);
		\draw [style=ULTRAGRAY] (142) to (146);
		\draw [style=ULTRAGRAY] (148) to (29);
		\draw [style=ULTRAGRAY] (148) to (141);
		\draw [style=ULTRAGRAY] (146) to (149);
		\draw [style=ULTRAGRAY, bend right=45] (149) to (30);
		\draw [style=ULTRAGRAY] (29) to (150);
		\draw [style=ULTRAGRAY] (150) to (148);
		\draw [style=ULTRAGRAY] (150) to (148);
		\draw [style=EDGE, bend left=15, looseness=1.25] (164) to (57);
		\draw [bend left=15] (167) to (170);
		\draw (170) to (171);
		\draw [bend left=15] (171) to (166);
		\draw [bend left=15] (166) to (168);
		\draw (168) to (169);
		\draw [bend left=15] (169) to (167);
		\draw (174.center) to (169);
		\draw (169) to (175.center);
		\draw (169) to (176.center);
		\draw (168) to (178.center);
		\draw (177.center) to (168);
		\draw (168) to (179.center);
		\draw (182.center) to (171);
		\draw (171) to (181.center);
		\draw (180.center) to (171);
		\draw (183.center) to (170);
		\draw (170) to (184.center);
		\draw (185.center) to (170);
		\draw [style=FILLBLUE] (167.center)
			 to [bend left=315] (166.center)
			 to [bend left=45, looseness=1.25] cycle;
		\draw [style=FILLBLUE] (186.center)
			 to (167.center)
			 to [in=345, out=135, looseness=0.75] cycle;
		\draw [style=FILLBLUE] (166.center)
			 to [in=165, out=15] (167.center)
			 to [bend left=330, looseness=0.75] cycle;
		\draw [style=FILLBLUE] (190)
			 to (187.center)
			 to [bend right=15] (167);
		\draw [style=FILLBLUE] (190.center)
			 to [in=345, out=-165, looseness=0.75] (166.center)
			 to [in=-150, out=-30, looseness=0.50] cycle;
		\draw [style=FILLBLUE] (166.center)
			 to [bend right=15] (188.center)
			 to cycle;
		\draw [style=FILLBLUE] (166.center)
			 to [bend right=45] (190.center)
			 to [in=315, out=-135, looseness=1.25] cycle;
		\draw [style=FILLBLUE] (189.center)
			 to (166.center)
			 to [bend right=15] cycle;
		\draw [style=FILLBLUE] (166.center)
			 to [bend right=60, looseness=1.75] (190.center)
			 to [bend left=60, looseness=2.00] cycle;
	\end{pgfonlayer}
\end{tikzpicture}

%% file: fig/tikz/kernelSize.tikz
\begin{tikzpicture}
	\begin{pgfonlayer}{nodelayer}
		\node [style=DOM] (0) at (0, 0) {};
		\node [style=NONE] (1) at (-3, 2) {};
		\node [style=NONE] (2) at (-3, -2) {};
		\node [style=NONE] (3) at (-7, -2) {};
		\node [style=NONE] (4) at (-7, 2) {};
		\node [style=DOM] (5) at (-10, -1.5) {};
		\node [style=NONE] (6) at (1.75, 3.25) {};
		\node [style=NONE] (7) at (3.5, 2.25) {};
		\node [style=none] (10) at (0, 0.75) {$v$};
		\node [style=NTWO] (11) at (-1, -1.5) {};
		\node [style=NTWO] (12) at (1.75, -3.5) {};
		\node [style=NTWO] (13) at (-0.5, -2) {};
		\node [style=NTWO] (15) at (-1.5, 2.25) {};
		\node [style=NTWO] (16) at (-1.5, 1.5) {};
		\node [style=none] (21) at (-7.25, 3) {};
		\node [style=none] (22) at (-8, 2.75) {};
		\node [style=none] (23) at (-8.25, 2) {};
		\node [style=none] (24) at (-7, -3) {};
		\node [style=none] (25) at (-7.75, -2.75) {};
		\node [style=none] (26) at (-6.25, -2.75) {};
		\node [style=none] (27) at (-2.75, -2.75) {};
		\node [style=none] (28) at (-3.75, 2.75) {};
		\node [style=none] (29) at (-2.25, 2.75) {};
		\node [style=none] (30) at (-3, 3) {};
		\node [style=none] (31) at (-3.5, -2.75) {};
		\node [style=none] (34) at (-6.5, 1.25) {$n_2$};
		\node [style=none] (35) at (-3.5, 1.5) {$n_1$};
		\node [style=none] (36) at (-3, -1.25) {$n_3$};
		\node [style=none] (37) at (-7, -1.25) {$n_4$};
		\node [style=none] (38) at (-4.75, -0.25) {$\mathbf{R_1}$};
		\node [style=none] (39) at (3.25, 3.5) {$\mathbf{R_2}$};
		\node [style=none] (40) at (6.75, -1.25) {};
		\node [style=none] (41) at (7.25, -1.75) {$\mathbf{R_l}$};
		\node [style=NONE] (42) at (5, 4) {};
		\node [style=NONE] (43) at (2.75, 5) {};
		\node [style=DOM] (44) at (4.75, 6) {};
		\node [style=none] (45) at (5.5, 5.75) {};
		\node [style=none] (46) at (5.5, 6.5) {};
		\node [style=none] (47) at (4.25, 6.5) {};
		\node [style=none] (48) at (2.5, 5.75) {};
		\node [style=none] (49) at (1.75, 5.5) {};
		\node [style=none] (50) at (1.75, 4.75) {};
		\node [style=none] (51) at (1.75, 3.75) {};
		\node [style=none] (52) at (1.25, 3.5) {};
		\node [style=none] (53) at (1.25, 3) {};
		\node [style=none] (54) at (5.75, 4.25) {};
		\node [style=none] (55) at (5.75, 3.75) {};
		\node [style=none] (56) at (5.25, 3.25) {};
		\node [style=none] (57) at (4.5, 2.5) {};
		\node [style=none] (58) at (4.5, 2) {};
		\node [style=none] (59) at (4.25, 1.5) {};
		\node [style=none] (60) at (6.25, -2.25) {};
		\node [style=none] (61) at (-2.75, 5.25) {};
		\node [style=none] (62) at (5.5, -3.25) {};
		\node [style=none] (63) at (1.75, -7.25) {};
		\node [style=none] (64) at (-8.75, -4.25) {};
		\node [style=none] (65) at (-11.5, -1) {};
		\node [style=none] (66) at (-8.5, 4) {};
		\node [style=NTWO] (67) at (2.5, -4.25) {};
		\node [style=NTWO] (68) at (3.25, -4.75) {};
		\node [style=NTWO] (69) at (4.25, -0.25) {};
		\node [style=NTWO] (70) at (2.75, 1) {};
		\node [style=NTWO] (71) at (3.25, 0.5) {};
		\node [style=NTHR] (72) at (0, -2.5) {};
	\end{pgfonlayer}
	\begin{pgfonlayer}{edgelayer}
		\draw [style=FILLDARKBLUE] (1.center)
			 to (4.center)
			 to (5.center)
			 to (3.center)
			 to (2.center)
			 to (0.center)
			 to cycle;
		\draw [style=FILLDARKBLUEINVISIBLE] (7) to (0);
		\draw [style=FILLDARKBLUEINVISIBLE] (0) to (6);
		\draw [style=FILLDARKBLUE] (43.center)
			 to (44.center)
			 to (42.center)
			 to (7.center)
			 to [in=15, out=-135] (0.center)
			 to [bend left=15] (6.center)
			 to cycle;
		\draw [style=FILLBLUE] (15.center)
			 to (1.center)
			 to (0.center)
			 to cycle;
		\draw [style=EDGE] (16) to (0);
		\draw [style=EDGE] (1) to (16);
		\draw [style=FILLBLUE] (13.center)
			 to [in=0, out=-165] (2.center)
			 to (0.center)
			 to [in=30, out=-120, looseness=1.25] cycle;
		\draw [style=FILLBLUE] (2) to (0);
		\draw [style=EDGE] (2) to (11);
		\draw [style=FILLBLUE] (2.center)
			 to [bend left=330, looseness=1.25] (68.center)
			 to [bend right] (0.center)
			 to (12.center)
			 to cycle;
		\draw (22.center) to (4);
		\draw (4) to (21.center);
		\draw (4) to (23.center);
		\draw (24.center) to (3);
		\draw (3) to (25.center);
		\draw (26.center) to (3);
		\draw (27.center) to (2);
		\draw (1) to (28.center);
		\draw (29.center) to (1);
		\draw (30.center) to (1);
		\draw (31.center) to (2);
		\draw (0) to (11);
		\draw [style=EDGEDASHED] (0) to (40.center);
		\draw (45.center) to (44);
		\draw (44) to (46.center);
		\draw (47.center) to (44);
		\draw (43) to (48.center);
		\draw (49.center) to (43);
		\draw (50.center) to (43);
		\draw (51.center) to (6);
		\draw (52.center) to (6);
		\draw (53.center) to (6);
		\draw (57.center) to (7);
		\draw (58.center) to (7);
		\draw (59.center) to (7);
		\draw (56.center) to (42);
		\draw (55.center) to (42);
		\draw (54.center) to (42);
		\draw [style=EDGEDASHED] (0) to (60.center);
		\draw [style=EDGEDASHED] (66.center) to (61.center);
		\draw [style=EDGEDASHED] (66.center)
			 to (65.center)
			 to (64.center)
			 to (63.center)
			 to (62.center)
			 to (61.center);
		\draw (67) to (12);
		\draw [bend left=15, looseness=1.25] (67) to (2);
		\draw [bend right] (67) to (0);
		\draw [style=FILLBLUE] (69.center)
			 to (7.center)
			 to [bend left=15] (0.center)
			 to cycle;
		\draw (71) to (0);
		\draw (70) to (0);
		\draw (71) to (7);
		\draw (70) to (7);
		\draw [style=EDGE, in=0, out=-75, looseness=0.75] (0) to (72);
	\end{pgfonlayer}
\end{tikzpicture}

%% file: pages/content/outlook/open-q-further-research.tex
\chapter{Open Questions and Further Research}\label{ch:closing}

\vspace*{-50pt}

\begin{figure}[ht]
        \includegraphics[width=0.35\textwidth, right]{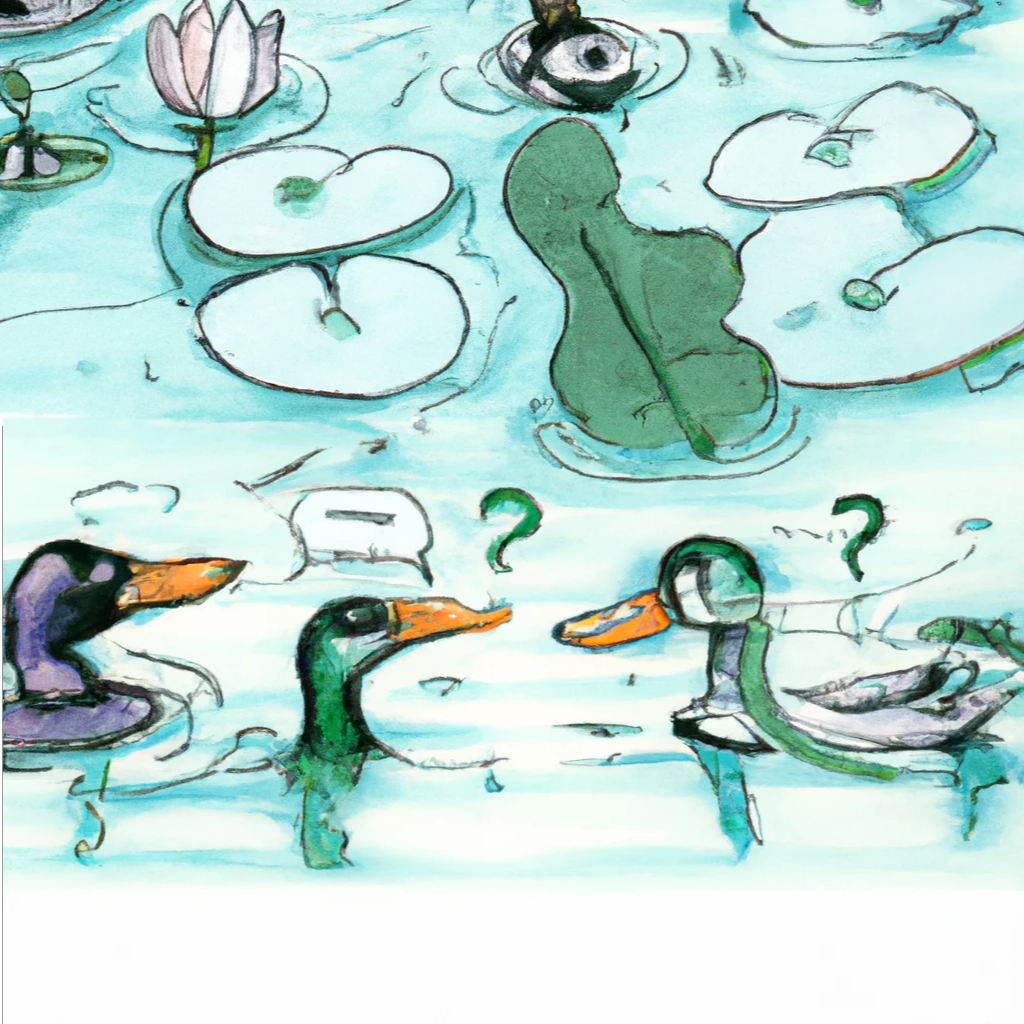}
        \captionsetup{textformat=empty,labelformat=blank}
        \caption[Generated with Dalle-E. Knowledge Cutoff 09-2022]{Generated with Dall-E. \url{https://labs.openai.com/}. ``more ducks asking further questions and research topics''}
\end{figure}

\epigraph{\itshape ``Peter does what he usually does when he doesn’t know what to do next: he gives up.''}{Qualityland, \textit{Marc-Uwe Kling}}

Now that all ducks are happy again, you still have some questions in mind.
After you have shown a linear kernel of size $\kernelsize \cdot k$ and, when parameterized by the solution size, that it does probably not exist for \textit{split graphs} and \textit{bipartite} graphs - including \textit{triangle-free} and \textit{chordal} graphs - you still have some open questions worth to be investigated.

\noindent \textbf{Improving Kernel~}
The constant of the kernel size $\kernelsize \cdot k$ is very high and the usage of an exponential time algorithm yields a total running time of  $\mathcal{O}(2^{\kernelsize k} \cdot poly(n))$, which is too large for practical applications.
It would be interesting to improve them and one idea would be the following:
\cref{rgl:rtwo} uses the fact that $N(v,w)$ can be dominated by at most four vertices: $v$, $w$ and two neighbors as a witness.
Observe that if $d(v,w) \leq 3$, this witness might be shared in an sds, because choosing one single witness on the path from $v$ to $w$ is sufficient to semitotally dominate $N(v,w)$.
Therefore $\Dvw$, $\Dv$ and $\Dw$ could be redefined to contain sets of size at most two (instead of three) which will improve the reduction. 
Note that in our analysis, the poles of a region $R$ in the \dreg $\mathfrak{R}$ satisfy $d(v,w) \leq 3$ by definition.
Therefore, requiring $d(v,w) \leq 3$ would be ok and we can still assume that \cref{rgl:rtwo} has been applied for any $R \in \mathfrak{R}$.
$d(v,w)$ can be calculated in linear time and would not blow up our runtime.

\noindent \textbf{Experimental Results~}
Our bound and its large constants refer to the worst-case scenario which tells us little about the performance applied to real-world instances.
Already Alber et al.~\cite{Alber2004} noticed that these kinds of reduction rules behave very well in nature and showed in an experimental setting for \pdom that on average more than $79\%$ of the vertices and $88\%$ of the edges have been removed by their reduction rules from a sample set of random planar graphs with up to $4000$ vertices. 
As our reduction rules differ from those given in \cite{Alber2004}, it would be interesting to see, how much random graphs are reduced using our preprocessing algorithm. \cref{rgl:rone,rgl:rtwo,rgl:rthree}.

\noindent \textbf{Other Open Problems~}
The classical complexities for \textit{dually chordal} and \textit{tolerance} graphs have already been asked for by Galby et al.~\cite{Galby2020} and are still open.
Furthermore, it would be interesting to complement the parameterized complexities on the hard classes we have started in this work.
open are those for \textit{circle}, \textit{chordal bipartite} and \textit{undirected path} graphs.
We think that \sdoms on \textit{chordal bipartite} graphs is at least \WONEhs-hard when parameterized by solution size, but we have been unable to prove it.
Furthermore, the reduction for \textit{circle} graphs~\cite{Kloks2021} to show \NP-completeness depends on the input size, but \doms and \tdoms have already been shown to be \WONEhs-hard on \textit{circle} graphs~\cite{Bousquet2012}. 
Maybe these reductions can be adjusted to show \WONEhs-hardness for \sdoms as well.
Last but not least, there exists an fpt algorithm for \doms on \textit{undirected path} graphs~\cite{Figueiredo2022}. 
Does it transfer to \sdoms as well?